\documentclass[twoside,12pt]{article}
\usepackage{graphicx}
\usepackage{amssymb}
\def\R{{\mathbb R}}
\def\E{{\mathbb E}}
\def\N{{\mathbb N}}
\def\S{{\mathbb S}}
\def\Z{{\mathbb Z}}
\def\T{{\mathbb T}}
\def\Lat{{\mathbb L}}
\def\supp{{\mbox{\rm supp}\, }}
\def\C{{\mathbb C}}
\def\Q{{\mathbb Q}}
\def\kasten{$~~\mbox{\hfil\vrule height6pt width5pt depth-1pt}$ }
\newtheorem{theorem}{Theorem}[section]

\newtheorem{proposition}[theorem]{Proposition}
\newtheorem{corollary}[theorem]{Corollary}

\newtheorem{lemma}[theorem]{Lemma}
\newtheorem{remark}[theorem]{Remark}

\begin{document}
\pagestyle{myheadings} \markboth{ S. Albeverio, H. Gottschalk and
M. W. Yoshida }{Particle systems in the GCE, scaling and QFT} \thispagestyle{empty}
\begin{center}
 {\Large \bf Systems of classical particles in the grand canonical ensemble, scaling limits and
 quantum field theory}

\

\noindent Sergio Albeverio and Hanno Gottschalk\\
\vspace{.25cm}
{\small Institut f\"ur angewandte Mathematik, \\ Rheinische
Friedrich-Wilhelms-Universit\"at Bonn,\\ Wegelerstr. 6, D-53115
Bonn, Germany\\ e-mail: albeverio@uni-bonn.de /
gottscha@wiener.iam.uni-bonn.de}

\

Minoru W. Yoshida\\
\vspace{.25cm}
{\small Department of Mathematics and Systems Engeneering\\
The University of Electrocommunications 1-5-1\\
Chofugaku, Tokyo 182-8585, Japan\\
e-mail: yoshida@se.uec.ac.jp}
\end{center}
\vspace{.25cm} {\noindent \small {\bf Abstract.} Euclidean quantum
fields obtained as solutions of stochastic partial pseudo
differential equations driven by a Poisson white noise have paths
given by locally integrable functions. This makes it possible to
define a class of ultra-violet finite local interactions for these
models (in any space-time dimension). The corresponding
interacting Euclidean quantum fields can be identified with
systems of classical "charged" particles in the grand canonical
ensemble with an interaction given by a nonlinear energy density
of the "static field" generated by the particles' charges via a
"generalized Poisson equation". A new definition of some
well-known systems of statistical mechanics is given by
formulating the related field theoretic local interactions. The
infinite volume limit of such systems is discussed for models with
trigonometric interactions using a representation of such models
as Widom-Rowlinson models associated with a (formal) Potts models
at imaginary temperature. The infinite volume correlation
functional of such Potts models can be constructed by a cluster
expansion. This leads to the construction of extremal Gibbs
measures with trigonometric interactions in the low-density high
temperature (LD-HT) regime. For Poissonian models with certain
trigonometric interactions an extension of the well-known relation
between the (massive) sine-Gordon model and the Yukawa particle
gas connecting characteristic- and correlation functionals is
given and used to derive infinite volume measures for interacting
Poisson quantum field models through an alternative route. The
continuum limit of the particle systems under consideration is
also investigated and the formal analogy with the scaling limit of
renormalization group theory is pointed out. In some simple cases
the question of (non-) triviality of the continuum limits is
clarified. }

\vspace{.25cm} {\small \noindent {\bf Keywords}: {\it Euclidean
quantum field theory, Poisson random fields, local interactions,
particle  systems in the grand canonical ensemble, correlation
functionals, Potts- and Widom-Rowlinson models, cluster expansion,
extremal Gibbs measures,
continuum limit of particle systems, sine-Gordon model.}}
 \section{Introduction}
Strong  connections between classical statistical mechanics and
quantum field theory have been established in the framework of
Euclidean quantum field theory (EQFT), see e.g.
\cite{AFHKL,GJ,Si}. In particular this applies to the
approximation of Euclidean quantum fields by lattice spin systems
\cite{GJ,Si}, the representation as a gas of interacting random
walks \cite{AFHKL,FFS,Sy}, or the connection of quantum field
models with trigonometric interaction (e.g. the sine-Gordon model)
with the gas of particles interacting through Yukawa-- or Coulomb
forces \cite{AHK1,AHK2,DL,F,FP,FS}. In this way, cluster
expansions or correlation inequalities from classical particle- or
ferromagnetic spin systems have been applied to the solution of
the infra-red problem in Euclidean quantum field theory.

Basically,  all these constructions concern models of quantum
fields given by a classical Euclidean action functional
$S(X)=S^0(X)+\beta V_{\Lambda}(X)$ with the free term
$S^0(X)={1\over 2}\int_{\R^d}\left[|\nabla X|^2+m^2X^2\right] dx$
and the interaction term $V_\Lambda(X)$ being an additive
functional in the infra-red regularizer $\Lambda\subseteq\R^d$ of
local type, i.e. $V_\Lambda(X) =\int_\Lambda v(X)\, dx$ for some
function $v:\R\to\R$. Inserting this into a heuristic path
integral of Feynman type, one gets the well-known heuristic
formula for the vacuum expectation values of the relativistic
quantized field continued to imaginary times (Schwinger functions)
as
\begin{equation}
\label{1.1eqa} S_{\Lambda,n}(y_1,\ldots,y_n)={1\over
Z_\Lambda}\int X(y_1)\cdots X(y_n)\, e^{-S^0(X)-\beta
V_\Lambda(X)}\, {\cal D} X\, ,
\end{equation}
where $y_1, \ldots,y_n\in\R^d$, $\beta\in\R$, $m^2{>\atop (-)}0$.
While the path integral itself makes sense -- $e^{-S^0(X)}{\cal
D}X/Z_\emptyset $ can be identified with the Gaussian measure with
covariance operator $(-\Delta+m^2)^{-1}$, i.e. the Nelson free
field measure (for $m^2=0$, $d=1$ the Wiener measure) -- it is
difficult to define the interaction term $V_\Lambda(X)$, since the
field configurations $X$ in the support of the Gaussian measure
are functions only if $d=1$: If $d\geq 2$ the field configurations
in the integral (\ref{1.1eqa}) generically are distributions and
expressions as $v(X)$ are ill-defined. This also limits the EQFT
approach essentially to space-time dimension $d=1$ or $d=2$, where
$v(X)$ for polynomial, trigonometric or exponential $v$ can be
regularized by Wick-ordering (for the construction of the
$\phi^4$-model in $d=3$ dimensions see \cite{GJ2}).

In the present paper we suggest to replace Nelson's measure
$e^{-S_0(X)}{\cal D}X/Z_\emptyset$ in (\ref{1.1eqa}) by a
convoluted Poisson noise measure \cite{AGW1}. Since Nelson's
measure can be seen as a convoluted Gaussian white noise measure,
from a mathematical point of view it is natural to generalize
Equation (\ref{1.1eqa}) to Poisson path space measures.
Furthermore, given the fact that convoluted Poisson white noise
measures have support on locally integrable field configurations
$X$, for a certain class of functions $v$ we can define potentials
$V_\Lambda(X)$ without any ultra-violet renormalization (not even
Wick-ordering) independently of the dimension $d\geq 2$. As we
will show in Section 4, the Euclidean quantum field models
obtained in this way can be identified with systems of classical
continuous and interacting particles in the grand canonical
ensemble. Since at least in principle the perturbed Gaussian free
field models can be recovered from the related interacting
"Poissonian" quantum field models by a scaling limit of the
associated particle system, we can consider the above replacement
as a new approximation of EQFTs by systems of statistical
mechanics. Also, properties of EQFT, as e.g. Euclidean invariance,
are preserved in the infinite volume limit $\Lambda\uparrow\R^d$.
In this sense this new approximation takes care of important
structural aspects of EQFT (which are violated e.g. by the lattice
approximation, discussed e.g. in \cite{AFHKL,GJ,Si}).

Another motivation for our suggestion is the constructive approach
to quantized gauge type fields developed in [3-6,10-12]: The basic
framework in these references is the one of covariant stochastic
partial (pseudo) differential equations driven by noise not
necessarily of the Gaussian type, in contrast to Nelson's
Euclidean approach \cite{N1,N2} which can be considered in the
framework of stochastic partial (pseudo) differential equations of
the Gaussian type. This approach started in the study of
quaternionian vector \cite{AH1,AH2,AH3,AIK} and scalar models
\cite{AGW1,AGW2,AW}, it has then extended to much more general
fields, see \cite{AGW4,BGL,GL1,GL2,GS,Jo}. In these cases the
axiomatic framework for the relativistic fields to be
accommodated, when possibly constructed, is the concept of quantum
fields with indefinite metric \cite{MS,St}. In fact, analytic
continuation for these models from Euclidean imaginary time to
relativistic real time is possible and the modified Wightman
axioms \cite{MS} for quantum fields with indefinite metric can be
verified explicitly \cite{AGW1,AGW4,AIK,BGL}. In particular,
fields with interesting scattering behavior have been found in
this class of models, also in the physical space-time dimension
$4$, cf. \cite{AG,AGW3,AGW4,Jo}. Therefore the connection with
relativistic quantum field theory does not get lost, if we replace
the Nelson's measure by a convoluted Poisson noise measure.

An alternative way to describe the main attitude  of this paper is
to say that  a systematic discussion is given how to introduce
perturbations of the basic (indefinite metric) Euclidean quantum
fields to construct other such fields. In analogy with the
standard constructive approach, this is achieved by constructing
Gibbs type measures for a bounded region of space-time (finite
volume) and then removing this restriction in the sense of a
thermodynamic limit. The main result of this paper consists in
showing that such an approach indeed can be developed and yields
at the same time interesting new relations with models of
classical statistical mechanics. Some results of this work have
been announced in \cite{AGY}.

Let us finally describe the content of the single  sections of
this paper: In Section 2 the basic notions of generalized white
noise convoluted generalized white noise are recalled. It is also
described, how the corresponding random fields lead, by analytic
continuation of their moment functions (Schwinger functions), to
relativistic Wightman functions satisfying all axioms of an
indefinite metric quantum field theory. Some special Green's
functions used to perform the convolution are discussed and the
scattering behavior of the associated quantum field models is
recalled. Finally, we show that the lattice approximation of the
Euclidean noise fields canonically leads to the notion of a
generalized white noise.

In Section 3, path properties of convoluted Poisson  noise (CPN)
are discussed and exploited to construct ultra-violet finite,
local interactions: In Section 3.2 we recall that pure Poisson
noise has paths in the space of locally finite "marked"
configurations and hence convolution with an integrable kernel
leads to fields with paths which are locally integrable
(independently of the dimension $d\geq 2$), cf. Section 3.3. In
Section 3.4 we then define the interaction term $V_\Lambda$ for
any $v$ measurable s.t. $|v(t)|\leq a+b|t|$ for some $a,b>0$.

Section 4 is devoted to the connection between the  particle
systems in a grand canonical ensemble (GCE) and quantum fields
defined by convoluted Poisson white noise with interaction.
Theorem \ref{4.1theo} shows the stability (in the statistical
mechanics sense) of the field theoretic interaction potential for
the associated system of classical, continuous particles.

In Section 5 several models of statistical mechanics  are looked
upon as systems of classical particles associated (in the sense of
Section 4) with convoluted, interacting Poisson white noise. In
particular the cases of a gas of hard spheres, particle systems
with potentials of stochastic geometry or pair potentials which
are positive definite fit into this framework.

Section 6 is the technical core  of this work. We give a complete
solution of the problem of taking the infinite volume limit of the
models of quantum fields resp. statistical mechanics in the
low-density high-temperature regime (LD-HT) and trigonometric
interactions (cf. Section 6.1 for the definition of the
interaction). This is presumably one of the first cluster
expansion for a continuous particle system for an interaction that
is not a pair-interaction\footnote{It is probably known to some
experts that the cluster expansion for the standard Potts model at
positive temperature leads to a construction of the ordinary
Widom-Rowlinson model in the LD-HT regime (corresponding to
"exponential" interactions for systems of particles with only
positive charges \cite{Go2}). But neither have the details been
worked out, nor has the flexibility of this method in connection
with "charged" or "marked" particles been realized.}. The strategy
is to represent such a model as the projection of a (formal) Potts
model at imaginary temperature to one of its components
(Widom-Rowlinson model), cf. Section 6.2. Even though such formal
Potts models are only represented as complex valued measures on
the space of locally finite configurations with a extra mark
indicating the "component" and cannot be interpreted in terms of
statistical mechanics, the standard cluster expansion \cite{Ru}
for their correlation functionals goes through (Section 6.3). The
projection to the first component then defines the correlation
functional of the system with trigonometric interaction. Using
standard arguments \cite{KK,Le} one can then reconstruct the
associated infinite volume measure. Verification of Ruelle
equations in Section 6.4 then implies that such measures are
Gibbs. Cluster properties of correlation functionals in the
infinite volume follow from the cluster expansion and imply
ergodicity of the translation group and hence extremality of the
Gibbs state (Section 6.5). The case of trigonometric interactions
is analyzed in Section 6.6: We extend the previously known
connection for the massive resp. massless sine-Gordon model and
Yukawa resp. Coulomb gas models ("duality transformation").

The continuum (scaling) limit  of interacting convoluted Poisson
noise (with infra-red cut-off) is discussed  in Section 7. We
start with a rather general discussion of scaling limits for
Poisson models and the relation with "renormalization group
methods". The case of trigonometric interactions with ultra-violet
cut-off is then analyzed with the related, ultra-violet
regularized, perturbed free field. Triviality without an
ultra-violet cut-off and without renormalization is shown in
Section 7.3. In Section 7.4 the scaling limit for the
$d=2$-dimensional sine-Gordon model without ultra-violet cut-off
and with a coupling constant renormalization is established in the
sense of formal power series.

\section{Generalized white noise and convoluted generalized white noise}

In this section we introduce our notation and recall some results of \cite{AGW1}

\

{\noindent \it 2.1 Generalized white noise}

\noindent For $d\geq 1$ we identify the $d$-dimensional Euclidean space-time with $\R^d$, by $\cdot$ / $|.|$ we denote
the Euclidean scalar product / norm and $E(d)$ stands for the group of Euclidean transformations on $\R^d$.
The space ${\cal S}$ is the space of real valued fast falling test functions on $\R^d$ endowed with the Schwartz topology. By
${\cal S}'$ we denote it's topological dual space (space of tempered distributions). Let ${\cal B}({\cal S}')$ be the
Borel $\sigma$-algebra on ${\cal S}'$, i.e. the $\sigma$-algebra generated by the open (in the weak topology) subsets of
${\cal S}'$. Then, $({\cal S}',B({\cal S}'))$ is a measurable space.

A (tempered) \underline{random field} over $\R^d$ by definition is a mapping from ${\cal S}$ into the space of real valued random variables
on some probability space $X:{\cal S}\to{\rm RV}(\Omega,{\cal B},P)$ such that (i) $X$ is linear $P$-a.s. and (ii) $f_n\to f$ in ${\cal S}$
$\Rightarrow$ $X(f_n)\stackrel{\cal L}{\to}X(f)$ where $\stackrel{\cal L}{\to}$ means convergence in probability law. Two processes $X_j$, $j=1,2$, on probability spaces
$(\Omega_j,{\cal B}_j,P_j)$, $j=1,2$, are called \underline{equivalent in law} if $P_1\{X_1(f_1)\in B_1,\ldots, X_1(f_n)\in B_n\}=P_2\{X_2(f_1)\in B_1,\ldots, X_2(f_n)\in B_n\}$
$\forall n\in\N$, $f_1,\ldots,f_n\in{\cal S}$ and $B_1,\ldots,B_n\in{\cal B}(\R)$, where ${\cal B}(\R)$ stands for the Borel sigma-algebra\footnote{The sigma-algebra generated by the open subsets.} on $\R$.

By Minlos' theorem \cite{M} there is a one-to-one correspondence (up to equivalence in law) between
tempered random fields and the \underline{characteristic functionals}  (i.e. continuous,
normalized and positive definite functionals)  ${\cal C}:{\cal S}\to\C$ given by ${\cal C}(f)=\E_P[e^{iX(f)}]$.
Furthermore, $X$ can be realized as a \underline{coordinate process}, i.e. there exists an unique probability measure
$P^X$ on $({\cal S}',{\cal B}({\cal S}'))$ such that for the random field $X_c(f)(\omega)=\langle \omega,f\rangle=\omega(f)$ $\forall \omega\in{\cal S}'$
and $f\in{\cal S}$ and $\E_P[e^{iX(f)}]=\E_{P^X}[e^{iX_c(f)}]$ $\forall f\in{\cal S}$. In the following we drop the subscript $c$
and we adopt the general rule that a random field $X$ on the probability space $({\cal S}',{\cal B}({\cal S}'),P^X)$ always is the coordinate process.

Let $\psi:\R\to\C$ be a \underline{L\'evy-characteristic}, i.e. a continuous, conditionally
positive definite function (for $t_j\in\R$, $z_j\in\C$, $j=1,\ldots,n$ s.t. $\sum_{j=1}^nz_j=0$ we have
$\sum_{l,j=1}^n\psi(t_l-t_j){\bar z}_lz_j\geq 0$) such that $\psi(0)=0$. We set
\begin{equation}
\label{2.1eqa}
{\cal C}_F(f)=e^{\int_{\R^d}\psi(f)\,dx}~~\forall f\in{\cal S}
\end{equation}
and we get from Theorem 6 p. 283 of \cite{GV} that ${\cal C}_F$ is a characteristic functional.
The associated random field $F$ is called a \underline{generalized white noise}. $F$ has infinitely divisible probability law,
is invariant in law under Euclidean transformations and for $f,h\in {\cal S}$ such that $\supp f\cap\supp h=\emptyset$
$F(f)$ and $F(h)$ are independent random variables.

Provided $\psi$ is $C^1$-differentiable at $0$, one can derive the following representation for $\psi$ (cf. \cite{BF})
\begin{equation}
\label{2.2eqa}
\psi(t)=iat-{\sigma^2\over 2}t^2+z\int_{\R}\left(e^{ist}-1\right)dr(s)~.
\end{equation}
Here $a\in\R$, $z,\sigma^2\in[0,\infty)$ and $r$ is a probability measure on $\R$ such that $r\{0\}=0$. The representation (\ref{2.2eqa}) is unique (for $z>0$).
Using notions which are slightly different from the standard definitions, we call $r$ the \underline{L\'evy measure} of $\psi$ and
$z$ is called the \underline{activity}. The first term in (\ref{2.2eqa}) is called \underline{deterministic part},
the second one the \underline{Gaussian part} and the third one the \underline{Poisson part}.

Inserting (\ref{2.2eqa}) into (\ref{2.1eqa}) we see that $F$ can be written as the sum of
independent deterministic (i.e. constant), Gaussian and Poisson parts which are uniquely determined by $\psi$.

\

{\noindent \it 2.2 Convoluted generalized white noise}

\noindent Let $L:{\cal S}'\to{\cal S}'$ be a symmetric, Euclidean invariant linear operator.
For reasons which will become transparent in Section 4, we call an equation of the type
$L\xi=\eta$, $\eta\in{\cal S}'$, a \underline{generalized Poisson equation}\footnote{Set $L=-\Delta$
and $\eta$ a signed measure to obtain the Poisson equation of electrostatics.} (GPE). Suppose that
$L$ is continuously invertible\footnote{Here we only deal with GPEs leading to short range static felds.}
by a Green's function $G\in{\cal S}'$, i.e. $G*\omega=L^{-1}\omega$ $\forall \omega\in{\cal S}'$. Then the stochastic GPE
\begin{equation}
\label{2.3eqa}
LX=F
\end{equation}
has a pathwhise solution $X=G*F$ and $X$ is called a \underline{convoluted generalized} \underline{white noise}.

If the L\'evy measure $r$ of $F$  has moments of all orders, then the Schwinger functions
\begin{equation}
\label{2.4eqa}
S_n(f_1\otimes\cdots\otimes f_n)=\E_{P^X}\left[\prod_{l=1}^nX(f_l)\right],~~f_1,\ldots,f_n\in{\cal S}
\end{equation}
exist  and can be calculated explicitly. They fulfill the requirements of temperedness, symmetry, invariance, Hermiticity
and clustering of the Osterwalder--Schrader axioms \cite{OS}. In general they do not fulfill the axiom of reflection
positivity, cf. \cite{AGW1,GS} for some counter examples (but we also note that the question is not yet
completely settled in the general case). Nevertheless, if $G$ has a representation of the form
\begin{equation}
\label{2.5eqa}
G=\int_{0}^\infty C_m\, d\rho(m^2), ~~~ \int_{0}^\infty {d|\rho|(m^2)\over m^2}<\infty\, ,
\end{equation}
for some (signed) measure $\rho$ and $C_m$ the covariance function of Nelson's free field of mass $m$, then
the Schwinger functions (\ref{2.4eqa}) can be analytically continued to a sequence of Wightman functions
which fulfill all Wightman's axioms \cite{SW} except (possibly) for positivity \cite{AGW1}. The Wightman functions however
fulfill the Hilbert space structure condition of Morchio and Strocchi \cite{MS} and therefore can be
considered as vacuum expectation values of a local, relativistic quantum field with indefinite metric \cite{AGW2}.

\

{\noindent \it 2.3 Some special Green's functions}

\noindent The Green's functions $G=G_\alpha$ associated with the partial pseudo differential operators
$L_\alpha=(-\Delta+m^2_0)^{\alpha}$, $m_0>0$, $0<\alpha\leq 1/2$, are of particular interest, since for
$F$ a purely Gaussian white noise, $X$ is a generalized free field \cite{Gr}, in particular, $X$ is reflection positive \cite{OS,Si} (cf. item (i) below).
In the special case $\alpha=1/2$, $X$ is Nelson's free field of mass $m_0>0$ \cite{N2}. We give a list of the properties
of the kernels $G_\alpha$ in the following
\begin{proposition}
\label{2.1prop}
For $m_0>0$ and $\alpha\in (0,1]$ let $G_\alpha=G_{\alpha,m_0}$ be the Green's function of
$(-\Delta+m_0^2)^\alpha$. Then

\noindent (i) $G_\alpha$ has a representation (\ref{2.5eqa}) with
\begin{equation}
\label{2.6eqa}
d\rho_\alpha(m^2)=\sin (\pi\alpha) 1_{\{m^2>m_0^2\}}(m^2){dm^2\over (m^2-m_0^2)^\alpha}~~~~0<\alpha<1
\end{equation}
and $\rho_1(dm^2)=\delta(m^2-m_0^2)dm^2$;

\noindent (ii) $G_\alpha\in L^1(\R^d,dx)$ and $G_\alpha$ is smooth on $\R^d\setminus\{0\}$;

\noindent (iii) $G_\alpha(x) >0\, \forall x\in\R^d\setminus\{0\}$;

\noindent (iv) $\exists \, C>0$ such that $G_\alpha(x)\leq Ce^{-m_0|x|}$ $\forall x\in\R^d: |x|>1$;

\noindent (v) For $\lambda >0$, $G_{\alpha,m_0}(\lambda x)=\lambda^{2\alpha-d}G_{\alpha,\lambda m_0}(x)$.

\noindent (vi) $|G_{\alpha m_0}(x)|< c_\alpha(d)\,|x|^{-(d-2\alpha)}$ for $x\in\R^d\setminus\{0\}$, where $0<c_\alpha(d)<\infty$, for $d\geq 2$, $0<\alpha<1$, can
be chosen optimal as in Eq. (\ref{2.8eqa}) below.
\end{proposition}
\noindent {\bf Proof.} All properties hold for $C_{m_0}=G_{1,m_0}$, cf. \cite{GJ} p. 126. The representation (i)
has been established in \cite{AGW1}, Section 6. (iii) now follows from the fact that $\rho_\alpha$ is a positive
measure and $C_m(x)>0$ $\forall x\not=0$. (iv) follows from (i) and the related property of $C_m$, $m\geq m_0$. (v) is an
consequence of the representation
\begin{equation}
\label{2.7eqa}
G_{\alpha,m_0}(x)=(2\pi)^{-d}\int_{\R^d}{e^{ik\cdot x}\over (|k|^2+m_0^2)^\alpha}\,dk
\end{equation}
where the integral has to be understood in the sense of Fourier transform of a
tempered distribution.
(ii) follows from (iv) and (vi);
Smoothness of $G_\alpha$ for $x\not=0$ follows from the fact that by (i)
$G_\alpha$ can be represented as a Fourier-Laplace transform and therefore is real analytic for
such $x$. The same argument (using also the "mass-gap" in (i)) also shows that that partial derivatives of $G$ are in $L^1(\R^d\setminus B_1(0),dx)$.

 Finally it remains to prove (vi): Let $\lambda=|x|$ and $\hat e\in\R^d$, $|\hat e|=1$. By rotation invariance of the
 $G_{\alpha,m_0}$ we get using (i), (v) and the residuum theorem
 \begin{eqnarray}
 \label{2.8eqa}
 G_{\alpha,m_0}(x)&=&\lambda^{-(d-2\alpha)}\sin(\pi\alpha)\int_{\lambda^2m_0^2}^\infty C_m(\hat e)\, {dm^2\over (m^2-m_0^2)^\alpha}\nonumber \\
 &=&\lambda^{-(d-2\alpha)}\gamma_\alpha(d)\int_0^\infty\int_0^\infty {e^{-\sqrt{t^2+m^2+\lambda^2m_0^2}}\over
 \sqrt{t^2+m^2+\lambda^2m_0^2}}\, t^{d-2} dt \, {dm^2\over m^{2\alpha}}\nonumber\\
 &<&\lambda^{-(d-2\alpha)}\gamma_\alpha(d)\int_0^\infty\int_0^\infty {e^{-\sqrt{t^2+m^2}}\over
 \sqrt{t^2+m^2}}\, t^{d-2} dt \, {dm^2\over m^{2\alpha}} \, .
 \end{eqnarray}
 We have set $\gamma_\alpha(d)=\mbox{Vol}(S^{d-2})\sin(\pi\alpha)/4\pi$. Here the right hand side multiplied with $\lambda^{d-2\alpha}$ defines the constants $c_\alpha(d)$ and it is clear from the calculation that
 these constants are optimal for $\lambda \to 0$. For $d> 2$ it is obvious, that the integrals converge. For $d=2$, $0<\alpha<1$, the inner integral has a
 logarithmic singularity at $m=0$. This singularity multiplied with $m^{-2\alpha}$ is however $dm^2$-integrable and thus $c_\alpha(d)<\infty$ also in this case.
\kasten

\begin{remark}
\label{2.1rem}
{\rm (i) As Prop. \ref{2.1prop} (v) shows, $G_{\alpha,0}\not\in L^1(\R^d,dx)$ but $G_{\alpha,0}\in L^1_{\rm loc}(\R^d,dx)$.

\noindent (ii) For $m_0>0$, $d/4\geq \alpha>0$ we have $G_\alpha\not\in L^2(\R^d,dx)$ since
$\int_{\R^d}G_\alpha^2\, dx=G_{2\alpha}(0)=\infty$, see also Prop. \ref{2.1prop} (vi).

\noindent (iii) For $d=1$, $\alpha >1/4$ we have $G_\alpha\in L^2(\R^d,dx)$, in particular this applies to $\alpha=1/2$.
}
\end{remark}
In particular, we can deduce from Proposition \ref{2.1prop} (i) that the Schwinger functions of the
model with $G=G_\alpha$ can be analytically continued to Wightman functions, which have been calculated explicitly
in \cite{AGW1}. From these explicit formulae one can see that for $0<\alpha<1/2$ the mass-shell singularities
of the truncated Wightman functions are of order $\kappa^{-\alpha}$ ($\kappa=k^0-\omega_{m_0}$,
$\omega_m=(|{\bf k}|^2+m^2)^{1/2}$) and hence the model does not
describe scattering particles\footnote{The use of partial pseudo differential operators leads to mass smearing which in some sense
is related to the concept of "infra particles", cf. \cite{Schr}.} . In the most important case $\alpha=1/2$ one can construct incoming and outgoing multi-particle
states using the method of \cite{AG} but the scattering is trivial, since the mass shell singularities of the Wightman functions in momentum space are of
the order $\kappa^{-1/2}$ and are thus too weak
to produce nontrivial scattering (for that one requires order $\kappa^{-1} $). In this sense, the convoluted generalized white noise models can still be
considered to be "free fields", even though higher order truncated Wightman functions do not vanish. But it should also
be noted that such higher order truncated Wightman functions can be decomposed into a superposition of "structure
functions" with non trivial scattering behavior \cite{AG}.

\

{\noindent \it 2.4 Lattice approximation of noise fields and infinitely divisible laws}

\noindent Finally in this section we want to give some heuristic evidence that
it is natural to define the generalized white noise $F$ as in (\ref{2.1eqa}) and (\ref{2.2eqa}): Heuristically
speaking, a noise field is a collection of independent identically distributed (i.i.d.) random variables $\{ F(x)\}_{x\in\R^d}$. To make this
notion precise, we substitute the continuum  $\R^d$ with a lattice $\Lat_n={1\over n}\Z^d$, $n$ odd, of lattice spacing $1/n$
and we consider the limit $n\to \infty$ for i.i.d. random variables $\{F_n(x)\}_{x\in\Lat_n}$. We require
that the distribution of the average of the random variables $F_n(x)$ remains constant in the unit cube $\Lambda_1$
centered at zero, i.e.
\begin{equation}
\label{2.9eqa}
F_1(0)\stackrel{\cal L}{=}\sum_{x\in\Lat_n\cap \Lambda_1}F_n(x)/n^d.
\end{equation}
We remark that $\sharp \Lambda_1\cap \Lat_n=n^d$.  Eq. (\ref{2.9eqa}) can only be fulfilled for $n\in\N$ arbitrary if $F_1(0)$ has infinitely divisible probability law
and thus by Schoenberg's theorem \cite{BF} $\E \left[ e^{itF_1(0)}\right]=e^{\psi(t)}$ for some conditionally
positive definite function $\psi$ and $\E\left[e^{itF_n(x)/n^d}\right]=e^{\psi(t)/n^d}$. Furthermore, (if $\psi$ is $C^1$-differentiable)
a representation (\ref{2.2eqa}) is given by the L\'evy-Khintchine theorem \cite{BF}.
For $f\in{\cal S}$ with compact support we set $\langle F_n,f\rangle=\sum_{x\in\Lat_n}F_n(x)f(x)/n^d$ and we get
\begin{eqnarray}
\label{2.10eqa}
\E\left[e^{i\langle F_n,f\rangle}\right]&=&\prod_{x\in\Lat_n}\E \left[e^{iF_n(x)f(x)/n^d}\right]\nonumber \\
&=& \prod_{ x\in \Lat_n} e^{\psi(f(x))/n^d}\nonumber \\
&=& e^{\sum_{ x\in\Lat_n}\psi(f(x))/n^d}\to e^{\int_{\R^d} \psi(f)\, dx}~~\mbox{as}~n\to\infty
\end{eqnarray}
where the last step shows that the lattice approximation $F_n$ converges to $F$ in law as $n\to\infty$, cf. (\ref{2.1eqa}).
For further information on the lattice approximation see \cite{AW}.

 \section{Path properties of convoluted Poisson white noise and ultra-violet finite local interactions}

 {\noindent \it 3.1 Path properties and quantum field theory}

 \noindent We say that a random field $X$ realized on the probability space $({\cal S}',{\cal B}({\cal S}'),P^X)$
 \underline{has paths in $E$}, where $E\subseteq {\cal S}'$ is a continuously embedded topological
 vector space, if $E$ has $P^X$ inner measure one, i.e. $\sup_{{\cal B}({\cal S}')\ni B\subseteq E}P^X\{X\in B\}=1$.\footnote{Usually one only demands that $\{X\in E\}$ has $P^X$ outer measure $1$, but
 for our considerations we need this stronger formulation.} The path properties of $X$ then are given
 by the general properties of the distributions in $E$, e.g. the property that they can be represented
 as functions.

   The rather irregular paths of Nelson's free field can be considered as the main source
 of problems in constructive quantum field theory. For $d\geq 2$ the paths are contained in
 weighted Sobolev spaces with negative index and have no representation as function spaces \cite{CL,RR,Y}. Consequently,
 energy densities $v(X)$ needed for the construction of local field interactions are ill-defined. In $d=2$ (and partially
 also in $d=3$) local interactions of polynomial, exponential and trigonometric type have been defined
 via regularization of paths and application of a renormalization procedute leading to the definition of
 "Wick-ordered" local interactions $:v(X):$ (for the path properties of these $:v(X):$ see e.g. \cite{Y}). Increasing irregularity of the paths as $d\geq 4$ (in physical terms: increasing
 ultra-violet divergences) so far do not allow an application of these techniques to the physical case $d=4$.

 It is therefore an interesting feature of convoluted Poissson noise (CPN), i.e. a convoluted generalized white noise
 such that the L\'evy characteristic (\ref{2.2eqa}) has only a Poisson part\footnote{With only minor modifications, the considerations
 of this work can be extended to fields which also have a deterministic part.}, that for a large class of convolution kernels $G$
 the paths are given by locally integrable functions and thus some local interactions can be defined without renormalization and
 therefore give ultra-violet finite interactions. This works independently of the space-time dimension $d\geq 2$ (and, of course, also for $d=1$).

 \

 {\noindent \it 3.2 Poisson noise and locally finite marked configurations}

 \noindent Let us first recall a well-known construction, see e.g. \cite{AHHK}: Let $\Lambda_n\subseteq\R^d$ be a monotone
 sequence of compact sets s.t. $\Lambda_n\uparrow \R^d$ as $n\to\infty$ and $\Lambda_0=\emptyset$. For $n\in\N$ let
 $D_n=\Lambda_n\setminus\Lambda_{n-1}$ and we denote the (Lebesgue) volume of $D_n$ by $|D_n|$. Let $(\{N_n\}_{n\in\N},\{Y_n^j\}_{n,j\in\N}
 ,\{S_n^j\}_{j,n\in\N})$ be three families of independent random variables on some proability space
 $(\Omega,{\cal B},P)$ which are distributed as follows: $N_n:\Omega\to\N_0$ has a Poisson law with intensity
 $z|D_n|$, i.e. $P\{N_n=l\}=e^{-z|D_n|}z^l|D_n|^l/l!$, $Y_n^j:\Omega\to\R^d$ has uniform distribution on $D_n$ (i.e.
 $1_{D_n}dx/|D_n|$) and the distribution of $S_n^j:\Omega\to\R$ is given by the L\'evy measure $r$. From now on we will assume that
 $r$ has compact support, $\supp r\subseteq[-c,c]$, for some $c>0$. By ${\cal D}'$ we denote the space
 of (not necessarily tempered) distributions. We define a mapping $\phi:\Omega\to {\cal D}'$ via
 \begin{equation}
 \label{3.1eqa}
 \phi=\sum_{n=1}^\infty \phi_n, ~~~~\phi_n=\sum_{j=1}^{N_n}S_n^j\delta_{Y_n^j}
 \end{equation}
 where $\delta_x$ is the Dirac measure in $X$. Obviously, $\phi$ has range in the space of
 \underline{locally finite marked configurations}, which is defined as the space of (real) signed measures
 $\gamma$ on $\R^d$ such that $\sharp (\supp\gamma\cap\Lambda)<\infty$ for any compact $\Lambda\subseteq \R^d$.
 By $|\gamma|$ we denote the absolute of the signed measure $\gamma$. Let $f$ be a positive measurable function
 on $\R^d$. A signed measure $\gamma$ is called \underline{$f$-finite}, if
 $\int_{\R^d} f\, d|\gamma|<\infty$. We also use the notation $\langle \gamma,f\rangle=\int_{\R^d}f\,d\gamma$ for a (signed)
 measure $\gamma$ on $\R^d$, provided that the integral exists. In particular this is always the case if
 both $f$ and $\gamma$ are nonnegative.
 
 \begin{proposition}
 \label{3.1prop}
 (i) $\phi$ is $P$-a.s. $f$-finite $\forall f\in L^1(\R^d,dx)\cap L^{\infty}(\R^d,dx)$, $f>0$;

  \noindent (ii) In particular, $\phi\in{\cal S}'$ $P$-a.s.. For ${\cal N}$ the exceptional null set, $\phi:(\Omega\setminus
   {\cal N},{\cal B}\cap(\Omega\setminus{\cal N}))\to({\cal S}',{\cal B}({\cal S}'))$ is measurable;

  \noindent (iii) Let $F$ be the Poisson white noise with pure Poisson L\'evy characteristic
  determined by $r$ and $z$ and let $P^F$ be the associated measure on $({\cal S}',{\cal B}({\cal S}'))$
  s.t. $F$ is the coordinate process w.r.t. $P^F$. Then $\phi_*P=P^F$.

  \noindent (iv) Assume that $f$ as above is also continuous. Then $F$ has paths in the space of $f$-finite, locally finite marked configurations, that is an element of ${\cal B}({\cal S}')$.
  \end{proposition}

\noindent The estimates obtained in this proposition actually are not better than those known in the literature. We give a proof for the convenience ofthe reader.

  \noindent {\bf Proof.} (i) Since $[-c,c]\times D_n\ni(s,y)\to \langle s\delta_{y},f\rangle=sf(y)\in\R$ is measurable,
  we get that $\langle\phi_n,f\rangle$ and $\langle |\phi_n|,f\rangle$ are measurable real-valued random variables.
  Since $\langle |\phi|,f\rangle=\sum_{n=1}^\infty\langle|\phi_n|,f\rangle\in[0,\infty]$ converges by monotonicity, the l.h.s. of this
  equation is measurable.

   $\E_P[e^{\langle|\phi|,f\rangle}]<\infty$ implies $P\{\langle|\phi|,f\rangle<\infty\}=1$.
  We can now use the following Laplace transform estimate
  \begin{eqnarray}
  \label{3.2eqa}
  \E_P\left[e^{\langle|\phi|,f\rangle}\right]&=&\lim_{N\to\infty}\E_P\left[e^{\langle |\phi|,1_{\Lambda_N}f\rangle}\right] \nonumber\\
  &=& \lim_{N\to\infty}\prod_{n=1}^N\E_P\left[ e^{\langle|\phi_n|,f\rangle}\right]\nonumber\\
  &=&\lim_{N\to\infty}\prod_{n=1}^N\Bigg[ e^{-z|D_n|}\sum_{l=0}^\infty{z^l|D_n|^l\over l!}\int_{D_n^{\times l}\times[-c,c]^{\times l}}
  \nonumber \\
  &\times& e^{\sum_{j=1}^l|s_n^j|f(y_n^j)}\,{dy_n^1\over |D_n|}\cdots {dy_n^l\over |D_n|}\,dr(s_n^1)\cdots dr(s_n^l)\Bigg]
  \nonumber \\
  &=& \lim_{N\to\infty}\prod_{n=1}^Ne^{z\int_{D_n\times[-c,c]}(e^{|s|f(y)}-1)dydr(s)}\nonumber\\
  &=& e^{z\int_{\R^d}\int_{[-c,c]}(e^{|s|f(y)}-1)dr(s)dy}\nonumber\\
  &\leq& e^{z\int_{\R^d}(e^{cf(y)}-1)dy}\leq e^{zc\|f\|_\infty e^{c\|f\|_\infty}\|f\|_1}<\infty.
  \end{eqnarray}
  Here $\|.\|_p$ denotes the norm of $L^p(\R^d,dx)$, $p\in[1,\infty]$ and the limits in the intermediate steps always exist by
  monotonicity.

(ii) follows immediately, since the choosing $f(x)=1/(1+|x|^2)^d$ shows that $|\phi|$, and hence also $\phi$, is polynomially
bounded $P$-a.s. . To show measurability of $\phi$, by definition of ${\cal B}({\cal S}')$
it suffices to show that $\langle\phi,f\rangle$ is measurable $\forall f\in{\cal S}$ and this
can be proven as in (i).

(iii) By a calculation which is analogous to (\ref{3.2eqa}) one can show that ${\cal C}_F(f)=\E_P[e^{i\langle\phi,f\rangle}]=
\int_{{\cal S}'}e^{i\langle\omega,f\rangle}d\phi_*P(\omega)$ and the statement follows from the uniqueness of $P^F$ which holds by Minlos' theorem.

To show (iv) we first remark that by (iii) the range of $\phi$ is in this set. Thus,
the set of $f$-bounded, locally finite marked configurations has $P^F$ outer measure one.
It remains to show that it is a measurable set. Firstly, the set of locally finite marked configurations $\Gamma$
in ${\cal S}'$ can be written as
\begin{equation}
\label{3.3eqa}
\bigcap_{R\in\Q_+}\bigcup_{n\in\N}\bigcap_{\varepsilon\in \Q_+}\bigcup_{s_1,\ldots,s_n\in \Q \atop y_1,\ldots,y_n\in\Q^d}\bigcap_{h\in\tilde{\cal D}(B_R)\atop \|h\|_\infty <1}\left\{ \omega\in{\cal S}':\left|\omega(h)-\sum_{l=1}^ns_lh(y_l)\right|<\epsilon\right\}
\end{equation}
where $\tilde {\cal D}(B_j)$ is a countable, dense subset of the set of test functions with
support in the ball centered at zero with radius $R$. Thus, $\Gamma$ is measurable. The subset of $f$-finite elements
in $\Gamma$ can be written in manifestly measurable form as
\begin{equation}
\label{3.4eqa}
\bigcup_{C\in\Q_+}\bigcap_{n\in\N}\bigcap_{h\in\tilde {\cal S}\atop\| h\|_\infty \leq 1}\left\{ \omega\in\Gamma:\left|\omega(hf_n)\right|<C\right\}
\end{equation}
where $f_n\in{\cal S}$ is a monotone sequence of positive functions approximating $f$ from below in the local uniform topology and $\tilde {\cal S}$ is a countable, dense subset
of ${\cal S}$. This concludes the proof.
 \kasten

By item (iii) of Prop. \ref{3.1prop} we can identify $\phi$ with $F$ and we therefore drop the notion $\phi$ in the following.

 \

 {\noindent \it 3.3 Path properties of convoluted Poisson noise}

\noindent From the path properties of $F$ we now can deduce the path properties of $X=G*F$ as follows:

 \begin{theorem}
 \label{3.1theo}
 Let $F$ be a Poisson noise with L\'evy measure $r$ of compact support, $\supp \, r\subseteq[-c,c]$, and let $G\in L^1(\R^d,dx)$. Then $X=G*F$
 has paths in $L^1(R^d,g_\epsilon dx)$ where $\epsilon >0$ and $g_\epsilon(x)=1/(1+|x|^2)^{(d+\epsilon)/2}$.
 \end{theorem}
 \noindent {\bf Proof.}
 Let $\Gamma^{|G|*g_\epsilon}$ be the set of $|G|*g_\epsilon$-finite, locally finite marked configurations. One can easily check that $|G|*g_\epsilon$ fulfills the conditions on $f$ in Proposition \ref{3.1prop}. As proven there,
 this set is ${\cal B}({\cal S}')$-measurable. By our general assumptions on $L$, $L^{-1}:{\cal S}'\to{\cal S}'$ is continuous and thus is a measurable
 transformation on $({\cal S}',{\cal B}({\cal S}'))$. Since $P^X=L_*P^F=P^F\circ L^{-1}$, the support of $P^X$ lies in the measurable
 set $L^{-1}\Gamma^{|G|*g_\epsilon}$ and we have to prove that this set lies in $L^1(\R^d,g_\epsilon dx)$. Let $\Lambda_n\uparrow\R^d$, $\Lambda_n\subseteq \R^d$ open and bounded. Let
 furthermore $D_n^l=\Lambda_n\setminus\Lambda_l$ for $n>l$. For $\gamma\in\Gamma^{|G|*g_\epsilon}$, we denote the restriction of $\gamma$ to an open set $A\subseteq \R^d$ by $\gamma_A$. Clearly,
 $G*\gamma_{\Lambda_n}\in L^1(\R^d,g_\epsilon dx)$ since $G$ is in $L^1(\R^d,dx)$ and $\supp \gamma_{\Lambda_n}$ is finite.
 The following estimate shows that $G*\gamma_{\Lambda_n}$ forms a Cauchy sequence in $L^1(\R^d,g_\epsilon\, dx)$. With $\|.\|_{\epsilon,1}$ the $L^1$-norm on that space, we get
 \begin{eqnarray}
 \label{3.5eqa}
 \sup_{n>l}\|G*\gamma_{\Lambda_n}-G*\gamma_{\Lambda_l}\|_{\epsilon,1}&=&\sup_{n>l}\|G*\gamma_{D_n^l}\|_{\epsilon,1}\nonumber\\
 &\leq& \sup_{n>l}\int_{\R^d}|G|*g_\epsilon\, d|\gamma_{D_n^l}| \nonumber\\
 &=& \int_{\R^d\setminus\Lambda_l}|G|*g_\epsilon\, d|\gamma| ~\to 0~\mbox{ as } l\to\infty
 \end{eqnarray}
 since $\gamma$ is $|G|*g_\epsilon$-finite. Also,
\begin{equation}
\label{3.6eqa}
\lim_{n\to\infty}\langle G*\gamma_{\Lambda_n},f\rangle=\lim_{n\to\infty}\langle\gamma_{\Lambda_n},G*f\rangle=\langle\gamma,G*f\rangle=\langle G*\gamma,f\rangle~~\forall f\in{\cal S},
\end{equation}
and by the fact that convergence in $L^1(\R^d,g_\epsilon \, dx)$ implies convergence in ${\cal S}'$, we get that $G*\gamma$ coincides with the limit of $G*\gamma_{\Lambda_n}$
in the Banach space $L^1(\R^d,g_\epsilon \, dx)$.  \kasten

 We remark that by Proposition \ref{2.1prop} the kernels $G_\alpha$ for $0<\alpha\leq 1$ fulfill the requirements of Theorem \ref{3.1theo}.

 In the context of quantum vector fields obtained from SPDEs driven by a Poisson white noise path properties
 have been considered in \cite{AH1,AH2,AH3,GL1,GL2,Ta} where in the latter references it is proven CPN has piecewise smooth paths with discrete singularities.  This has been used to
 define Wilson loop observables or stochastic co-surfaces (for this concept see \cite{AHHK,FFS} and references therein).
Local $L^1$-integrability of paths does not hold for all of these models, since
 the Green's functions for vector-valued fields in many case cannot be represented by locally integrable functions. Nevertheless, most of the analysis of this 
paper would also be possible using the path properties derived in the references given above at the price of more restrictive assumptions on the interactions (to be introduced in the following subsection). 

\

 {\noindent \it 3.4 Definition of local potentials}

 \noindent Having established the path properties of the CPN model, we now want to define
 nonlinear, local interactions. The construction is based on the elementary fact that for a measurable function
 $v:\R\to\R$ such that $|v(t)|\leq a+b|t|$ for some $a,b>0$ the nonlinear transformation $L^1(\R^d,g_\epsilon dx)\ni f\to v(f)\in L^1(\R^d,g_\epsilon dx)$ is well defined.

\begin{theorem}
\label{3.2theo}
Let $v:\R\to\R$ be a measurable function s.t. $|v(t)|\leq a+b|t|$ for some $a,b\geq 0$
and let $X$ be a CPN as in Theorem \ref{3.1theo}. Let $\Lambda\subseteq \R^d$ compact and $\beta \geq 0$.
Then

\noindent (i) $v(X)$ is a random field with paths in $L^1(\R^d,g_\epsilon dx)$, $\epsilon>0$;

\noindent (ii) $V_\Lambda=\langle v(X),1_\Lambda\rangle\in \cap _{p\geq 1} L^p({\cal S}'
,P^X)$;

\noindent (iii)  $e^{-\beta V_\Lambda}\in \cap _{p\geq 1}L^p({\cal S}',P^X)$;

\noindent (iv) Let $\Xi_\Lambda=\Xi(z,\beta,\Lambda)=\E_{P^X}\left[e^{-\beta V_\Lambda}\right]$. Then
\begin{equation}
\label{3.7eqa}
P^{\bar X_\Lambda}={e^{-\beta V_\Lambda}\over \Xi_\Lambda}\,P^X
\end{equation}
defines a probability measure on $({\cal S}',{\cal B}({\cal S}'))$.
\end{theorem}
\noindent {\bf Proof.}
(i) That $X\in L^1(\R^d,g_\epsilon dx)\Rightarrow v(X)\in L^1(\R^d,g_\epsilon dx)$ is elementary. It remains to prove that $\langle v(X),f\rangle$ is
 measurable. To this aim let $v$ be continuous and
$\chi^{\varepsilon}$ be a sequence of Schwartz functions s.t. $\chi^\varepsilon\to \delta_0$ as $\varepsilon\to 0$. Let $\chi_x^{\varepsilon}$ be the
translation of $\chi^\varepsilon$ by $x$. Then $v( X(\chi^{\varepsilon,x}))$ is a random variable. For a fixed random parameter in the set $G*\Gamma^{|G|*g_\epsilon}$
of $P^X$ measure one,
$X(x)$ is a $L^1(\R^d,g_\epsilon dx)$ function in $x$, cf. the proof of Theorem \ref{3.1theo}. For random parameters in the exceptional null set we re-define
$v(X(\chi^{\varepsilon}_x))$ to be zero. Approximating the integral
by a Riemannian sum, we get that also $\int_{\Lambda} v( X(\chi^{\varepsilon}_x))f(x)\, dx$ is measurable  since
the pointwise limit of measurable functions is measurable. Since $X$ is a $L^1(\R^d,g_\epsilon dx)$-function, there exists a subsequence $\varepsilon_n$ s.t. in the limit $\varepsilon\to 0$ $X(\chi^{\varepsilon_n,x})\to X(x)$ $dx$--a.e. and
$v(X(\chi^{\epsilon_n,x}))\to v(X(x))$ $dx$--a.e. for $v$ continous. Consequently, the integral $\int_{\Lambda} v( X(\chi^{\varepsilon_n}_x))f(x)\, dx$ converges to $\langle v(X),f\rangle$ by dominated convergence. Thus, this expression is measurable for continuous $v$. By an approximation
of a measurable $v$ by continuous functions, using the dominated convergence theorem again, we get that $\langle v(X),f\rangle$
is measurable also for $v$ assumed to be only measurable.

(ii) and (iv) follow from (iii) with $v(t)$ replaced with $-|v(t)|$.

(iii) Since $(e^{-\beta V_\Lambda})^p=e^{-p\beta V_\Lambda}$ it suffices to prove the statement for $p=1$.
We note that
\begin{equation}
\label{3.8eqa}
-\beta V_\Lambda =-\beta \left\langle v(X),1_\Lambda\right\rangle \leq \beta b \left\langle |F|,|G|*1_\Lambda\right\rangle+\beta a |\Lambda|
\end{equation}
and that $\beta b |G|*1_\Lambda\in L^1(\R^d,dx)\cap L^\infty (\R^d,dx)$. Thus $\E_{P^X}\left[e^{-\beta V_\Lambda}\right]<\infty$ follows
as in the estimate (\ref{3.2eqa}).
\kasten
\begin{remark}
\label{3.1rem}
{\rm
The growth condition on $v$ in Theorem \ref{3.2theo} can be relaxed in various ways. E.g. obviously for $v$ positive,
(iii) is trivially satisfied and to see that $V_\Lambda<\infty$ $P^X$-a.s. the condition $v(G)\in L^1_{\rm loc}(\R^d,dx)$ would be
sufficient. A refined analysis of this point is postponed to later work (see however the examples of Section 5).
}
\end{remark}
We denote the coordinate process associated to $P^{\bar X_\Lambda}$ by $\bar X_\Lambda$ and we call it
\underline{inter-}\linebreak \underline{acting CPN with infra-red cut-off $\Lambda$}.

\section{The connection with particle systems in the grand canonical
ensemble}
In this section we explain, how models of CPN with local interaction can be interpreted as
systems of interacting classical, continuous particles in the configurational grand canonical ensemble (GCE).

\

{\noindent \it 4.1 Continuous classical particles in the grand canonical ensemble}

\noindent To begin with, we recall some notions of statistical mechanics following \cite{Ru}. Let $(y,p,s)\in\R^d\times\R^d\times[-c,c]$ be
the "coordinates" of a classical point particle of mass $M>0$ in $d$-dimensional Euclidean space. Here $y$ gives the position,
$p$ the momentum and $s$ is an "internal parameter", called \underline{charge}, which is not dynamic, i.e. is not altered by the interaction
with other particles. The classical Hamiltonian of $n$ such particles is given by
\begin{equation}
\label{4.1eqa}
H(y_1,\ldots,y_n;p_1,\ldots,p_n;s_1,\ldots,s_n)=\sum_{l=1}^n{|p_l|^2\over 2M}+U(y_1\ldots,y_n;s_1,\ldots,s_n)
\end{equation}
where $U(y_1,\ldots,y_n;s_1,\ldots,s_n)$ is the potential energy. We assume that there is some {\it a priori} distribution of
the charges $s$ given by a probability measure $r$ with $\supp r\subseteq [-c,c]$.

The GCE at inverse temperature $\beta>0$ with chemical potential $\mu\in\R$ in the finite volume $\Lambda\subseteq \R^d$, $\Lambda$ compact, is given  (up to normalization) by
the following measures on the $n$-particle configuration space
\begin{equation}
\label{4.2eqa}
{1\over n!}\, e^{\beta [n\mu -H(y_1,\ldots,y_n;p_1,\ldots,p_n;s_1,\ldots,s_n)]}dy_1\cdots dy_n\,dp_1\cdots dp_n\,dr(s_1)\cdots dr(s_n)
\end{equation}
where $y_1,\ldots,y_n\in\Lambda$, $s_1,\ldots,s_n\in[-c,c]$. Carrying out the Gaussian integral over the momenta, we pass to the configurational GCE (also abbreviated by GCE in the following)
defined (up to normalization)  through the following measures on spatial $n$-particles configurations ("marked" by charges $s_1,\ldots,s_n$)
\begin{equation}
\label{4.3eqa}
{z^n\over n!} \, e^{-\beta U(y_1,\ldots,y_n;s_1,\ldots,s_n)}dy_1\cdots dy_n\,dr(s_1)\cdots dr(s_n)
\end{equation}
where
\begin{equation}
\label{4.4eqa}
z=e^{\beta \mu}\left( {2\pi M\over\beta}\right)^{d/2}>0
\end{equation}
is the activity of the system\footnote{ By an adaptation of $\mu$ and / or $M$ it is possible to consider $z$ and $\beta$ as independent parameters.}. The functions $e^{-\beta U(y_1,\ldots,y_n;s_1,\ldots,s_n)}$ are called
the \underline{Boltzmann weights} of the system.

\

{\noindent \it 4.2 Interacting Poisson quantum fields and interacting particle systems}

\noindent Identifying $(y_1,\ldots,y_n;s_1,\ldots,s_n)$, $y_j\not=y_l$, $l\not=j$, with $\sum_{l=1}^ns_l\delta_{y_l}$ it
is easy to show (by a calculation analogous to Eq. (\ref{3.2eqa}) ) that in the case $U\equiv 0$ the measure
(\ref{4.3eqa}) can be identified with the Poisson noise $F_\Lambda=1_\Lambda F$ where $F$ is has L\'evy measure $r$
and activity $z$. Thus, $F_\Lambda$ describes a gas of noninteracting particles in the "box" $\Lambda$, see e.g. \cite{Ge,Pre}.
We here want to extend this analogy to the interacting models of the preceding
section:

\begin{itemize}
\item We consider (configurational) GCEs of charged, indistinguishable particles in a finite volume $\Lambda$.

\item The charges of the particles give rise to a \underline{static field}; the field of the unit charge in $y$ is given
by the Green's function $G(x-y)$; the static fields penetrates\footnote{This assumption can be changed by introducing boundary conditions for $L$, cf. Remark \ref{4.1rem} below.} the "walls" of the "box" $\Lambda$.
\item The static field $X$ of a charge configuration $(y_1,\ldots,y_n;s_1,\ldots,s_n)$ $y_j\not=y_l$, $l\not=j$, is obtained by superposition
from the fields of the single particles and is thus given by $\sum_{l=1}^ns_lG(x-y_l)$; equivalently the static field is obtained
as the solution of the generalized Poisson equation $LX=\eta$ with $\eta=\sum_{l=1}^ns_l\delta_{y_l}$ (Fig. \ref{1fig}).
\begin{figure}[htb]
\centerline{\parbox{6cm}{\includegraphics[width=5.9cm]{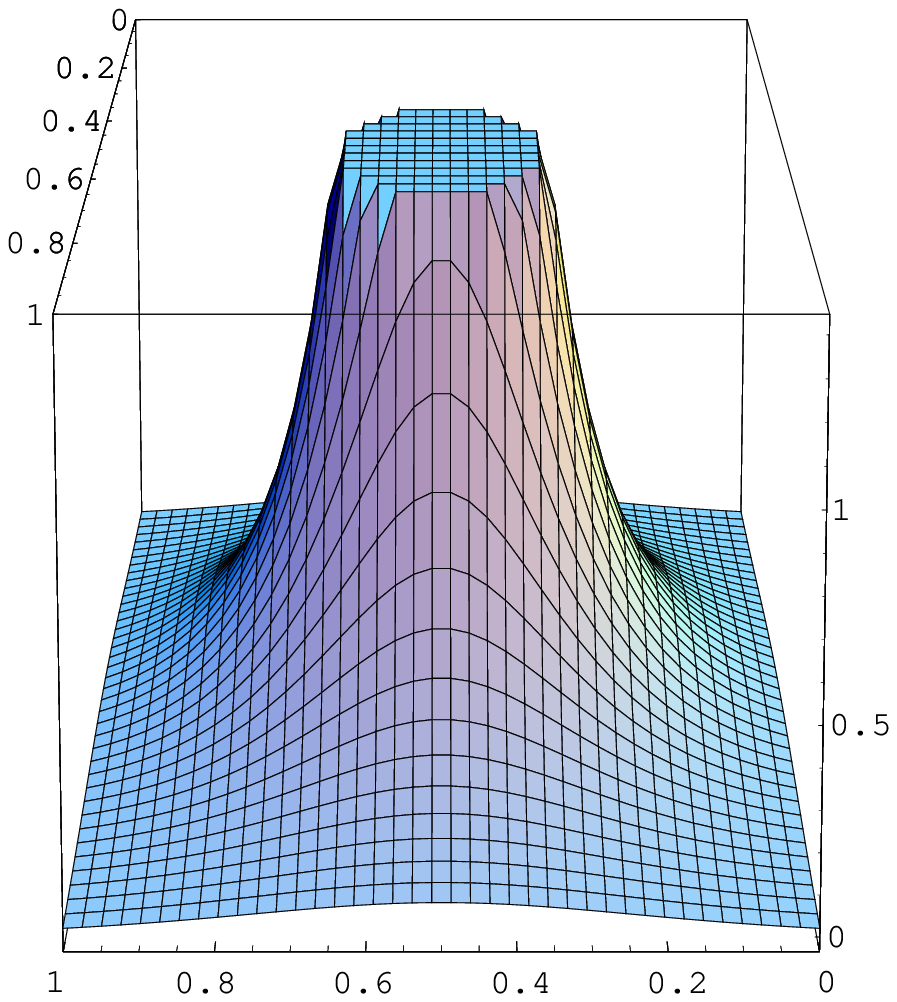}}
\parbox{6cm}{\includegraphics[width=5.9cm]{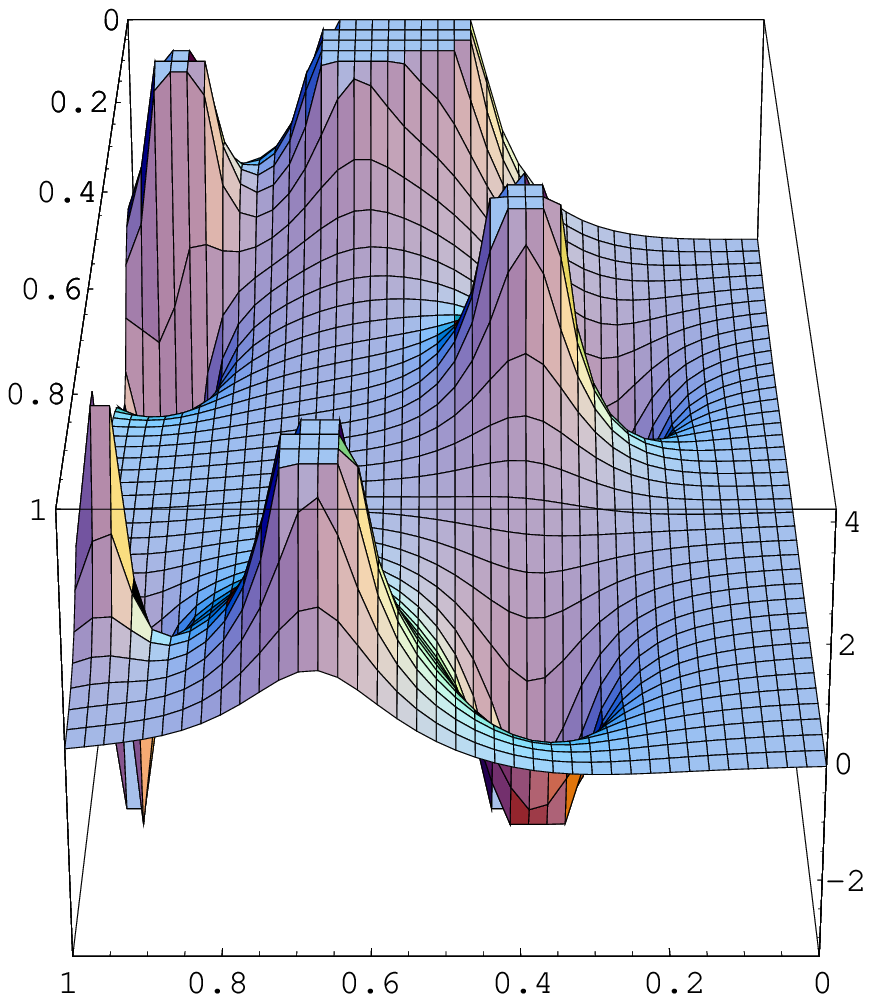}}}
\caption{Field of a unit charge and ten particles with positive and negative charges $\pm 1/\sqrt{10}$, $G(x)=e^{-m_0|x|}/|x|$, $m_0=3$.}
\label{1fig}
\end{figure}
\item The potential energy of the particle configuration $\eta=\sum_{l=1}^ns_l\delta_{y_l}$ is given by a (nonlinear)
 energy density $v:\R\to\R$, $v(0)=0$, of the static field $X$
\begin{eqnarray}
\label{4.5eqa}
U(\eta)=U(y_1,\ldots,y_n;s_1,\ldots,s_n)&=&\int_{\R^d}v\left(\sum_{l=1}^ns_l\, G(x-y_l)\right)\,dx \nonumber \\
&=& \int_{\R^d}v(G*\eta)\, dx\nonumber \\
&=&\int_{\R^d}v(X)\, dx
\end{eqnarray}
\end{itemize}
The \underline{interacting CPN in the finite volume $\Lambda$} is the random field given by the statical
field of the \underline{interacting particle system} in the GCE with potential energy $U$ restricted to the box $\Lambda$.

We remark that the potential $U$ defined in (\ref{4.5eqa}) is Euclidean invariant, provided $G$ is invariant under rotations. Furthermore
$U$ is symmetric under permutations of arguments $(y_1,s_1),\ldots,(y_ns,_n)$.

\

{\noindent \it 4.3 Finite volume vs. infra-red cut-off}

 \noindent Let us now put this into mathematical terms. In particular we want to give sufficient conditions s.t.
 the potential $U$ in (\ref{4.5eqa}) is well-defined and stable.

 Let $F_\Lambda=1_\Lambda F$ be the restriction of $F$ to the compact region $\Lambda$. We set $N_\Lambda=\sharp \supp F_\Lambda$
 and we recall that $N_\Lambda$ is Poisson distributed with intensity $z|\Lambda|$. We have $N_\Lambda<\infty$ $P^F$-a.s. and
 hence $X_\Lambda=G*F_\Lambda\in L^{1}(\R^d,dx)$ $P^F$-a.s. if $G\in L^1(\R^d,dx)$. The crucial observation in (\ref{4.5eqa}) is that
 for the CPN in finite volume $\Lambda$, $X_\Lambda$, we can define local interactions without taking an additional infra-red cut-off as in the usual
 QFT. Throughout the paper we thus distinguish between the techniques of taking an infra-red cut-off (as in Section 3) and restriction of the associated
 particle system to a finite volume. While it seems conceptually clear that both formulations lead to the same system if the infra-red cut-off
 is removed or the infinite volume limit is taken, respectively, this remains to be established mathematically. We now get the counterpart to Theorem \ref{3.2theo}
 using a finite volume instead of an infra-red cut-off:

 \begin{theorem}
 \label{4.1theo}
 Let $F$ be a Poisson noise and $G$, the Green's function of an operator $L$, as in Theorem \ref{3.1theo}.
 Let $X_\Lambda=G*F_\Lambda$. Furthermore, let $v:\R\to\R$ s.t. $|v(t)|\leq b|t|$ for some
 $b>0$ and let $\beta >0$. Then

 \noindent (i) $v(X_\Lambda)$ is a random field with paths in $L^1(\R^d,dx)$;

 \noindent (ii) $\tilde V_\Lambda=\langle v(X_\Lambda),1_{\R^d}\rangle\in\cap_{p\geq 1}L^p({\cal S}',P^{X_\Lambda})$ or, equivalently,
 $U_\Lambda=\linebreak \langle v(G*F_\Lambda),1_{\R^d}\rangle \in\cap_{p\geq 1}L^p({\cal S}',P^F)$;

 \noindent (iii) The potential $U_\Lambda$ is \underline{stable}, i.e. for $U^-_\Lambda$ the negative part of $U_\Lambda$ we have
 $U^-_\Lambda \leq B N_\Lambda$ where $B=c b\, \|G\|_1$;

 \noindent (iv) The \underline{grand partition function}
 \begin{equation}
 \label{4.6eqa}
 \tilde \Xi_\Lambda=\tilde\Xi(z,\beta,\Lambda)=
\E_{P^{X_\Lambda}}\left[ e^{-\beta \tilde V_\Lambda}\right]=\E_{P^F}\left[ e^{-\beta U_\Lambda}\right]
\end{equation}
 is entire analytic in $z$;

\noindent (v) In particular, $e^{-\beta \tilde V_\Lambda}\in \cap _{p\geq 1}L^p({\cal S}',P^{X_\Lambda})$ or,
equivalently, $e^{-\beta U_\Lambda}\in \linebreak \cap_{p\geq 1} L^p({\cal S}',P^F)$;

\noindent (vi) There exist measures on $({\cal S}',{\cal B}({\cal S}'))$ defined by
\begin{equation}
\label{4.7eqa}
P^{\tilde X_\Lambda}={e^{-\beta \tilde V_\Lambda}\over \tilde \Xi_\Lambda}\, P^{X_\Lambda},~~~
P^{\tilde F_\Lambda}={e^{-\beta U_\Lambda}\over \tilde \Xi_\Lambda}\,P^{F_\Lambda}
\end{equation}
related through $L_*P^{\tilde X_\Lambda}=P^{\tilde F_\Lambda}$. Equivalently, the associated
coordinate processes $\tilde X_\Lambda$ and $\tilde F_\Lambda$ fulfill the generalized Poisson equation
$L\tilde X_\Lambda=\tilde F_\Lambda\Leftrightarrow \tilde X_\Lambda =G*\tilde F_\Lambda$.
 \end{theorem}
\noindent {\bf Proof.} (i) That $v(X_\Lambda)$ is a random field can be proven as in Theorem \ref{3.2theo}. That the paths
are in $L^1(\R^d,dx)$ follows from $\langle |v(X_\Lambda)|,1_{\R^d}\rangle\leq b\, \|X_\Lambda\|_1<\infty$.

(ii) follows from (v) with $v$ replaced by $-|v|$. (v) follows from (iv). By \cite{Ru}, Chapter 3, (iv) is a
consequence of the stability of the potential (iii).

To prove (iii) we note that
\begin{eqnarray}
\label{4.8eqa}
U_\Lambda^-&\leq& \left\langle v^-(G*F_\Lambda),1_{\R^d}\right\rangle\nonumber\\
&\leq & b\int_{R^d}|G*F_\Lambda|\, dx \nonumber \\
&\leq& b\int_{R^d}|G|*|F|\, dx=b\, \|G\|_1\int_{\R^d}d|F_\Lambda|
\end{eqnarray}
and $\int_{\R^d}d|F_\Lambda|\leq c \, N_\Lambda$. (vi) now follows from (v), the
fact that $\tilde V_\Lambda =U_\Lambda \circ L$, cf. Eq. (\ref{4.5eqa}),
 and the transformation formula for probablity measures. \kasten

 The conditions of Theorem \ref{4.1theo} on the energy-density $v$ are a little more restrictive
 than those of Theorem \ref{3.2theo}, where e.g. densities of the form $v(t)=\sqrt{|t|}$ are admissible.
 In the framework of Theorem \ref{4.1theo} such potentials can be dealt with at the price of a more technical treatment
 if one e.g. assumes an exponential decay for $G$, since stability is trivial for positive
 potentials.

 We also point out that in the framework of Theorem \ref{4.1theo} we can treat the mass-zero cases (where $G\not\in L^1(\R^d,dx)$, cf Remark \ref{2.1rem} )
 of Prop. \ref{2.1prop} if we demand
 that the (positive) energy density $v$ at $t=0$ tends to zero sufficiently fast,
 e.g. $0\leq v(t)\leq c\, t^\gamma$, for $0\leq |t|\leq \epsilon$, with $\gamma>d/(d-2\alpha)$.

 \begin{remark}
 \label{4.1rem}
 {\rm Most of the constructions presented in Section 3 and 4 can be extended to Riemannian manifolds.
In particular, we can introduce local interactions on compact manifolds without any cut-off.
As a simple example we consider the $d$-dimensional torus $\T^d_l$ of length $l$:
In this case the Green's functions $G=G_{\alpha, m_0}$, $m_0>0$, in Prop. \ref{2.1prop} have to be
modified by introducing periodic boundary conditions
for the Laplacian. Then the translation invariant potential $U$ can be defined by
 $U(y_1,\ldots,y_n;s_1,\ldots,s_n)=\int_{\T_l^d}v(\sum_{j=1}^n s_j G(x-y_j)) \, dx$.
The proof of stability is completely analogue to the one of Theorem \ref{4.1theo}. The infinite volume limit $\T_l^d\to\R^d$
now can be studied as $l\to\infty$.
 }
 \end{remark}

 \

\section{Models of statistical mechanics seen as 'Poisson' quantum fields}
In this section we show that a number of well-known particle systems can be associated
to an interacting CPN -- and hence to a "Poisson", Euclidean QFT -- in the spirit of Theorem \ref{4.1theo} (vi).
Most of the potentials we discuss in this section do not fulfill directly the requirements of Theorem \ref{4.1theo},
however they are known to fulfill the stability condition (e.g. when the potentials are positive
or else by applying well-known criteria, cf. \cite{Ru}). Hence these potentials (with exception of Section 5.1)
can also be used to construct Euclidean quantum field models in our spirit.
Moreover they can be obtained by approximation from potentials in the class considered in Theorem \ref{4.1theo}.

\

{\noindent \it 5.1 The gas of hard spheres}

\noindent  Here we consider a particle system with identical particles carrying a unit charge, hence
we set $r=\delta_1$, the Dirac measure in $1$. Let $B_R=B_R(0)\subseteq \R^d$ be the open ball centered at zero
 with radius $R>0$. We set
 \begin{equation}
 \label{5.1eqa}
 G(x)=1_{B_R}(x)=\left\{ \begin{array}{ll} 1&\mbox{ if }x\in B_R\\
 0&\mbox{ else}\end{array}\right.
 \end{equation}
 and we define
 \begin{equation}
 \label{5.2eqa}
 v^{\rm h.c.}(t)=\left\{\begin{array}{ll} 0&\mbox{ if } t<2\\
 \infty &\mbox{ if } t\geq 2 \end{array}\right.
 \end{equation}
 Then we get for the potential $U$ in Eq. (\ref{4.5eqa})
 \begin{eqnarray}
 \label{5.3eqa}
 U(y_1,\ldots,y_n)&=&\int_{\R^d}v^{\rm h.c.}\left(\sum_{l=1}^nG(x-y_n)\right)\,dx\nonumber\\
 &=& \left\{ \begin{array}{ll} 0 &\mbox{ if }\min_{l,j=1\ldots,n; l\not=j}|y_l-y_j|\geq R\\
 \infty &\mbox{ else}\end{array}\right.
 \end{eqnarray}
Here we did not write out the arguments $s_j\equiv 1$ and  the integral in (\ref{5.3eqa}) is well-defined
as an integral of nonnegative functions with values in $[0,\infty]$. Obviously, on the right hand side of (\ref{5.3eqa})
we have the potential of particles with a hard core of radius $R$ ("gas of hard spheres").

We also note that if we modify (\ref{5.2eqa}) and set $v(t)=0$ if $t<l$ and $v(t)=\infty$ if $t\geq l$, $l\in\N$, $l\geq 2$,
we obtain a system where a non empty intersection of $l$ (and more) balls of radius $R$ is energetically forbidden, but all
configurations without such intersections have zero potential energy. Such systems have pure $l$-point potentials
in the sense of statistical mechanics, cf. \cite{Ru}.

\

{\noindent \it 5.2 Potentials from stochastic geometry}

\begin{figure}[htb]
\centerline{\parbox{6cm}{
\includegraphics[width=5.9cm]{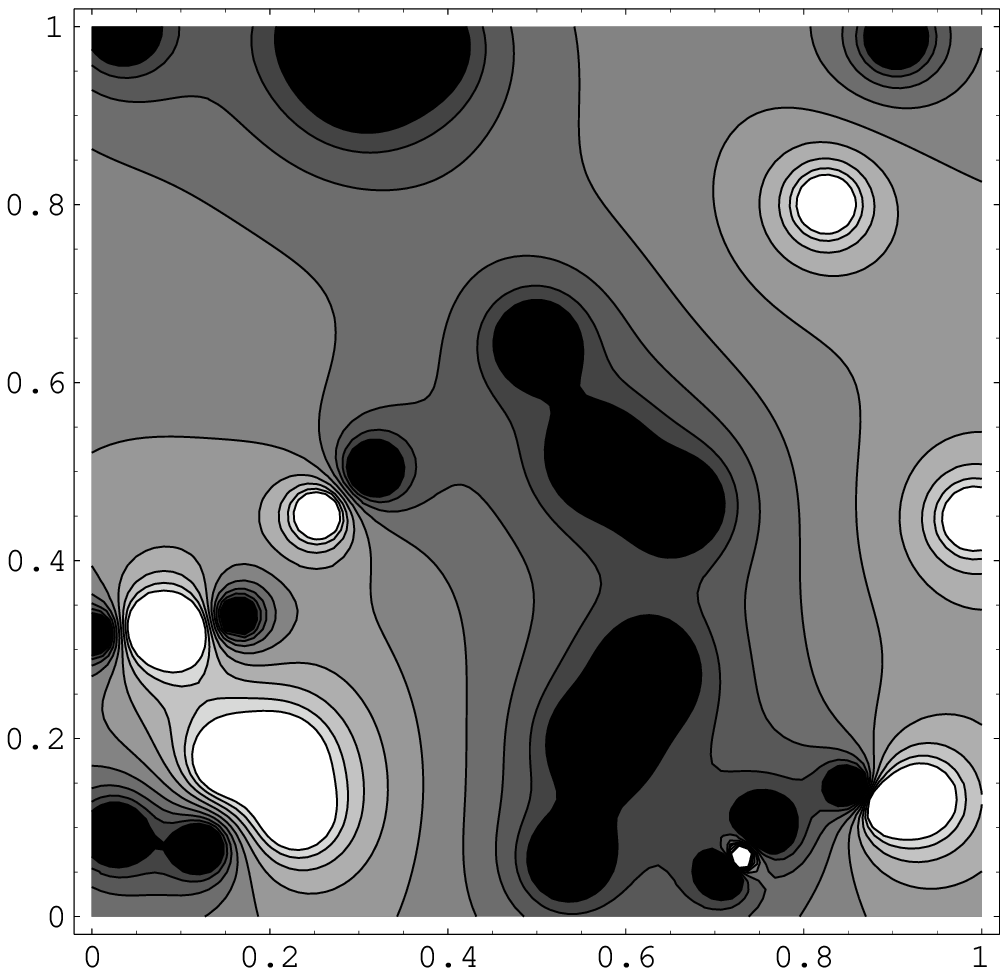}
}
\parbox{6cm}{\hspace{.5cm}
\includegraphics[width=5.9cm]{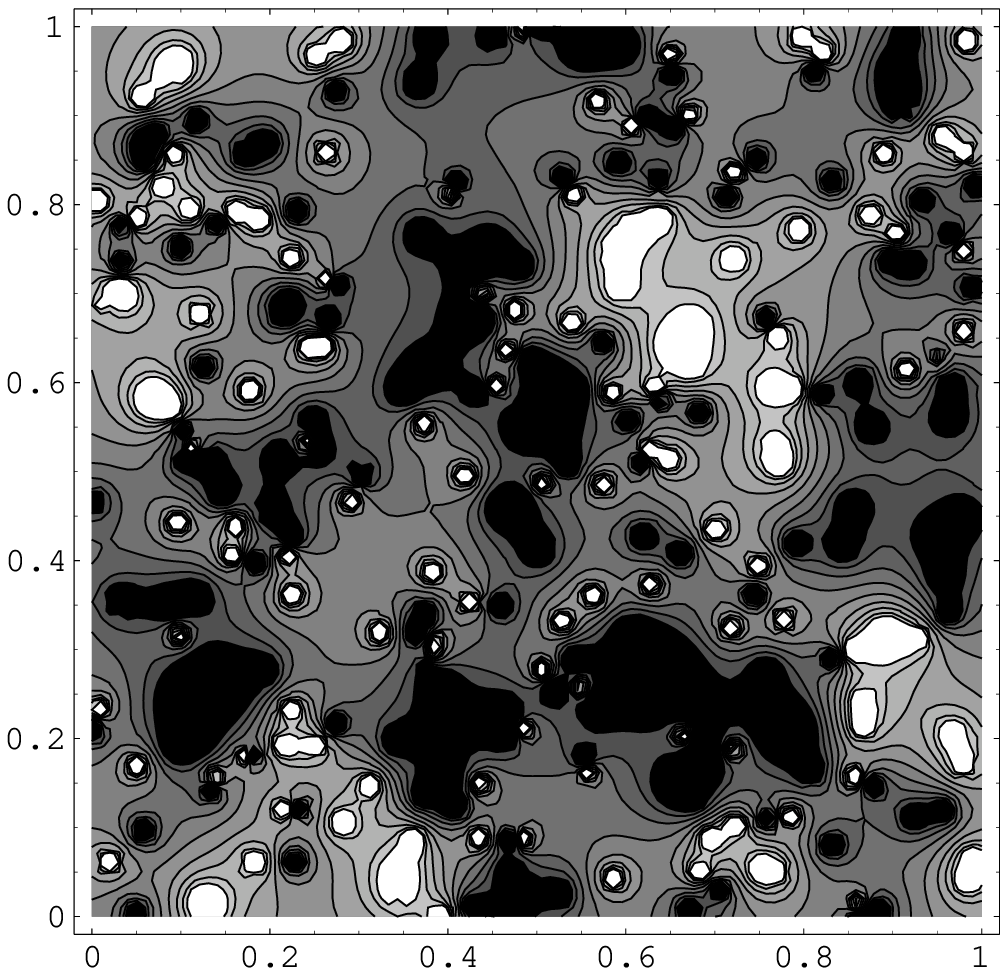}
}
}
\caption{Threshold and isodensity contour potentials for $n=30$ and $n=300$ particles of charge $\pm 1/\sqrt{n}$ and with $G$ as in Fig. \ref{1fig}. Isodensity
contours of integer values from -4 to 4 are displayed.
The fractal structure of the continuum limit (Sect. 7) becomes visible.}
\label{2fig}
\end{figure}

\noindent Here we give "local" formulations of two potentials of stochastic geometry\cite{Sa,SKM,Me}, starting with the
\underline{threshold potential}: Let $G$ and $r$ be as in Theorem \ref{4.1theo} and for $C>0$ we define the energy density
\begin{equation}
\label{5.4eqa}
v_C(t)=\left\{ \begin{array}{ll}0&\mbox{ if } t<C\\ 1&\mbox{ else}\end{array}\right.
\end{equation}
which obviously is in the class of Theorem \ref{4.1theo} (set $b=1/C$). We now get (cf. Fig. \ref{2fig})
\begin{eqnarray}
\label{5.5eqa}
U(y_1,\ldots,y_n;s_1,\ldots,s_n)&=&\int_{\R^d}v_C\left(\sum_{l=1}^ns_l\, G(x-y_l)\right)dx\nonumber \\
&=& \left| \left\{ x\in\R^d:\, \sum_{l=1}^ns_l\, G(x-y_l)\geq C\right\}\right|
\end{eqnarray}
If we, in particular, choose $G$ and $r$ as in Sect. 5.1, we get the so-called Boolean grain model of stochastic
geometry \cite{SKM}. We can also define similar energy densities $v_C^{\rm sym}(t)=v_C(|t|)$ and $v_{-C}=v_C^{\rm sym}-v_C$
to obtain related potentials which "threshold" also negative values of $X_\Lambda$.

Next we formulate the \underline{isodensity contour potential}: Let us assume that $G$ is $C^1$-differentiable in $\R^d\setminus\{0\}$ (cf. Prop. \ref{2.1prop} (ii) for examples)
and $\lim_{x\to 0}|G(x)|\linebreak =\infty$, $\lim_{|x|\to\infty}G(x)=0$. For $C>0$ we define heuristically
\begin{equation}
\label{5.6eqa}
v_C^{\rm i.d.c.}(X)=\delta(X-C)|\nabla X|
\end{equation}
or, more precisely,
\begin{eqnarray}
\label{5.7eqa}
U(y_1,\ldots,y_n;s_1,\ldots,s_n)&=&\int_{\R^d} v^{\rm i.d.c.}_C\left(\sum_{l=1}^ns_l\, G(x-y_l)\right)dx\nonumber\\
&=& \lim_{\epsilon\downarrow 0}\epsilon^{-1}\int_{\R^d}\Bigg[v_{C-\epsilon}\left(\sum_{l=1}^ns_l\, G(x-y_l)\right)\nonumber\\
&-&v_C\left(\sum_{l=1}^ns_l\, G(x-y_l)\right)\Bigg] \left|\sum_{l=1}^ns_l\nabla G(x-y_l)\right|  dx\nonumber \\
&=& \left|\left\{ x\in\R^d: \sum_{l=1}^ns_l\, G(x-y_l)=C\right\}\right|_{d-1}
\end{eqnarray}
where $|.|_{d-1}$ denotes the $d-1$-dimensional (surface) volume (cf. Fig \ref{2fig}). Clearly, $\nabla G(x-y_j)$ is well-defined on the set of points
where $v_{C-\epsilon}-v_C$ does not vanish. The last step follows from the fact that
obviously the Hausdorff dimension of the set on the right hand side is $d-1$. This also shows that $U$ is well-defined.

Potentials like $v_C$ and $v^{\rm i.d.c.}_C$ might be of particular interest in the continuum limit (see Section 7) since
they are designed to measure the fractal properties of the sample paths in that limit, see Fig. \ref{2fig}.

\

\

{\noindent \it 5.3 Particle systems with positive definite pair interactions}

\noindent Let $r=\delta_1$ (in this subsection we may thus omit the variables $s_l\equiv 1$) and $G$ as in Theorem \ref{4.1theo} reflection invariant under $x\to-x$
and let $G$ fulfill $\int_{\R^d}G\, dx\not=0$.
We set $\Phi=G*G$ and we get $\Phi\in L^1(\R^d,dx)$. $\Phi$ is positive definite in the sense that $\Phi$ is the
Fourier transform of a (not necessarily finite) non negative function on $\R^d$. We consider two separate situations:
Either $\Phi$ is the Fourier transform of a non negative $L^1(\R^d,dx)$-function and hence is continuous. Or we assume that
$\Phi$ is nonnegative, in this case possibly $\Phi(0)=\infty$. Also, we remark that choosing $G=G_\alpha$, $0<\alpha\leq 1/2$, as in Proposition \ref{2.1prop}
leads to the second case, cf. Remark \ref{2.1rem}.

Let $\chi\in C^{\infty}_0(\R^d)$ be symmetric, nonnegative such that $\int_{\R^d}\chi\, dx=1$. For $\epsilon>0$, we set
 $\chi^\epsilon(x)=\chi(x/\epsilon)/\epsilon^d$ and we introduce an ultra-violet cut-off setting
 $G^\epsilon=\chi^\epsilon*G$ and $\Phi^{\epsilon}=G^\epsilon*G^\epsilon$.

 We consider the quadratic energy density $v(t)=t^2$ for the ultra-violet regularized model, namely
 \begin{eqnarray}
 \label{5.8eqa}
 U_\epsilon(y_1,\ldots,y_n)&=&\int_{\R^d}\left(\sum_{l=1}^n G^\epsilon(x-y_l)\right)^2dx\nonumber \\
 &=&\sum_{l=1}^n\int_{\R^d}\!\! G^\epsilon(x-y_l)^2dx+\sum_{l,j=1\atop l\not=j}^n\int_{\R^d}\! \! G^\epsilon(x-y_l)\, G^\epsilon(x-y_j)\, dx\nonumber \\
 &=& n\Phi^\epsilon(0)+\sum_{l,j=1\atop l\not=j}^n\Phi^\epsilon (y_l-y_j)~.
 \end{eqnarray}
If $\Phi(0)=+\infty$, then the first term on the right hand side of (\ref{5.8eqa}) in the limit $\epsilon\downarrow 0$ gives an
infinite contribution, while the second term remains well-defined (for $y_l\not=y_j$). Since the first term is proportional
to $n$, it can be seen as a (negative) chemical potential or a self-energy which become infinite if $\epsilon\downarrow 0$ -- this is
very similar to the self-energy problem of a charged point-particle in ordinary electro dynamics. Subtracting
this infinite contribution ("self energy renormalization") gives a suitable renormalization for the
quadratic potential of the interacting CPN. We want to show that this can be done in a way preserving
the local structure of the interaction: We set (note that $\int_{\R^d} G^\epsilon dx=\int_{\R^d} G \, dx \not=0$)
\begin{equation}
\label{5.9eqa}
:t^2:^{\rm s.e.r.}_\epsilon=t^2-c_\epsilon t, ~~~~c_\epsilon=\left. \Phi^\epsilon(0)\right/ \int_{\R^d} G^\epsilon \, dx
\end{equation}
and we get that the renormalized potential
\begin{eqnarray}
\label{5.10eqa}
U_\epsilon^{\rm s.e.r.}(y_1,\ldots,y_n)&=&\int_{\R^d}:\left(\sum_{l=1}^nG^\epsilon (x-y_l)\right)^2:^{\rm s.e.r.}_\epsilon\, dx\nonumber\\
&=& \sum_{l,j=1\atop j\not=l} \Phi^\epsilon(y_l-y_j)
\end{eqnarray}
has a well-defined limit as $\epsilon\downarrow 0$, which is just given by the potential resulting from the pair
interaction $\Phi$. If $\Phi\geq 0$, stability is obvious. In the case where $\Phi$ is positive definite and continuous,
stability follows from Proposition 3.2.7 of \cite{Ru}. It can be seen in the same reference that positive definite
pair potentials play a quite special r\^ole in the theory of stability.

We would like to point out that a quadratic interaction for a CPN is obviously non-trivial, since a particle gas with
pair interactions is obviously different from a gas of noninteracting particles. However, the interaction becomes trivial in the Gaussian (continuum) limit
of Section 7 (the interacting process in that limit becomes Gaussian) as can be seen most easily by performing
the continuum limit with an ultra-violet cut-off\footnote{The continuum limit for the renormalized potential can be performed setting
$r_{1/\sqrt{z}}=\delta_{1/\sqrt{z}}$ and $a_z=-\sqrt{z}$ in (\ref{2.1eqa}) in order to avoid problems with the
stability of the potential. This is only slightly different from the techniques in Sect. 7.}.

\section{High temperature expansion}

In this section we give a construction of the infinite volume
limit $\Lambda \uparrow \R^d$ (the removal of the infra-red
cut-off, respectively) using techniques from continuous particle
systems. In particular, we give a high temperature expansion for
the correlation functional for the case of trigonometric
interaction. The main trick is to write the trigonometric
interaction as effective potential of a (formal) two-component
marked Potts model at imaginary temperature. Though imaginary
temperature might look strange, we show that it does not interfere
with the usual cluster expansion method \cite{Ru}. Once the
complex valued correlation functional has been constructed for the
Potts model, real valuedness of the interaction is restored by
restriction to one component. The construction of Gibbs measures
then follows from the general analysis of the excellent review
article \cite{KK}, see also the original article by Lennard
\cite{Le}. Obviously, here the techniques are inspired by
statistical mechanics of continuous, classical particles. For
another construction of infinite volume measures (working also
outside the LD-HT regime) with a quantum field flavor that applies
to the case where the interaction energy density $v$ is concave
 and uses FKG inequalities, cf. \cite{Go}.

\

{\noindent \it 6.1 Trigonometric interactions}

\noindent From now on we  focus on the case of
\underline{trigonometric interactions}. Let $\nu$ be a complex
valued measure on $\R$ with $\nu(A)=\overline{\nu(-A)}$ $\forall
A\in{\cal B}(\R)$. Furthermore, let $\nu$ have compact support $\subseteq
[-c',c']$, $|\nu|([-c',c'])<\infty$. Let $b=\int_{[-c',c']}|\alpha|d|\nu|(\alpha)$ and
$b'=\int_{[-c,c]}|s|dr(s)$ with $r$ as in equation (\ref{2.2eqa}).
For the definition of the modulus $|\nu|=|\Re\nu|+|\Im\nu|$ of
$\nu$ see e.g. \cite{Hal}. $\nu$ is called the
 \underline{interaction measure}.
We set
\begin{equation}
\label{6.1eqa}
v(t)=\int_\R(1-e^{i\alpha t})\,d\nu(\alpha)~,~t\in\R.
\end{equation}
Obviously $v$ is real-valued and fulfills the conditions of
Theorem \ref{4.1theo}. Suppose that also $G$ is given as in that
theorem.

For $c>0$ let $\Gamma_0^c$ ($\Gamma^c$) be the space of signed,
real-valued measures $\eta$ on $\R^d$ with (locally) finite
support such that $-c\leq\eta\{x\}\leq c$ $\forall x\in\R^d$. For
reasons that are connected with the use of Potts models in the
next section, in this section we work with an infra-red cut-off
\underline{and} a finite volume.  For $\eta\in\Gamma_0^c$ and
$\Lambda\subset \R^d$ compact we thus define the interaction
$U_\Lambda:\Gamma_0^c\to\R$ by $U_\Lambda(\eta)=\langle
v(G*\eta),1_\Lambda\rangle$. Furthermore let $F_\Lambda$ be a
Poisson noise\footnote{The associated interacting Poisson noise in
this section is denoted by $\tilde F_\Lambda$.} as in Section 4.3.
We define the \underline{correlation functional}
$\rho_\Lambda:\Gamma_0^c\to(0,\infty)$ associated with $F_\Lambda$
and $U_\Lambda$  at the inverse temperature $\beta$
\begin{equation}
\label{6.2eqa} \rho_\Lambda(\eta)={1 \over  \tilde\Xi_\Lambda}
1_{\{ {\rm supp}\eta\subseteq \Lambda
\}}(\eta)\E_{P^F}\left[e^{-\beta
U_\Lambda(\eta+F_\Lambda)}\right]~,~\forall \eta\in
\Gamma_0^c,~\tilde\Xi_\Lambda=\E_{P^F}\left[e^{-\beta
U_\Lambda(F_\Lambda)}\right].
\end{equation}
What is remarkable is that this correlation functional fulfills the following:

\begin{proposition}
\label{6.1prop}
The correlation functional $\rho_\Lambda$ fulfills the uniform (in $\Lambda$) Ruelle bound
$|\rho_\Lambda(\eta)|\leq (e^{\beta B})^{\sharp\eta}$ for all $\eta\in\Gamma_0^c$ with $B=bc\|G\|_1$.
\end{proposition}
\noindent{\bf Proof.} As $v$ is differentiable and $|v'|<b$, we get
\begin{eqnarray*}
\left|U_\Lambda(\eta+\gamma)-U_\Lambda(\gamma)\right|&=& \left|\int_0^1{d\over dt} U_\Lambda(\gamma+t\eta)\,dt\right|\\
&\leq&b\int_\Lambda|G*\eta|\, dx\leq b\|G\|_1\int_\Lambda d|\eta|\leq B\sharp \eta
\end{eqnarray*}
Combining this with the definition of $\rho_\Lambda$ in (\ref{6.2eqa}) then gives the assertion.
\kasten

The uniform Ruelle bound is crucial for the  passage from infinite
volume correlation functionls to Gibbs measures, cf. \cite{KK}. In
Section 6.4 we come back to this point.

\

{\noindent \it 6.2 Two component formal Potts model at imaginary temperature}

\noindent Let $c'>0$ be as in Section 6.1.  Clearly
$\Lambda^l\times\R^l\ni(y_1,\ldots,y_l;s_1,\ldots,s_l)\to\sum_{j=1}^ls_j\delta_{y_j}\in\Gamma_0^{c'}\subseteq{\cal
S}'$ is continuous and hence measurable. For $A\in{\cal
B}(\Gamma_0)$ we thus get that $
1_{\{\sum_{j=1}^ls_l\delta_{y_l}\in A\}}$ is measurable w.r.t.
${\cal B}(\Lambda^l)\otimes {\cal B}(\R^l)$. We set
\begin{eqnarray}
\label{6.3eqa}
P^{Z_\Lambda}(A)&=&e^{-\beta |\Lambda|\nu([-c',c'])}\delta_0(A)+e^{-\beta|\Lambda|\nu([-c',c'])}\sum_{l=1}^\infty {\beta^l\over l!}
\int_{\Lambda^{\times l}\times[-c',c']^{\times l}}\nonumber\\
&&~\times~1_{\{\sum_{j=1}^l\alpha_j\delta_{y_j}\in A\}}(y_1,\ldots,y_l;\alpha_1,\ldots,\alpha_l)\, dy_1\cdots dy_n\,d\nu(\alpha_1)\cdots d \nu(\alpha_n).\nonumber\\
\end{eqnarray}
Here $\delta_0(A)=1$ if $0\in A$ and $0$ otherwise.
\begin{lemma}
\label{6.1lem} The function $P^{Z_\Lambda}:{\cal
B}(\Gamma_0^{c'})\to\C$ is a complex valued measure on
$(\Gamma_0^{c'},{\cal B}(\Gamma^{c'}_0))$.
\end{lemma}
\noindent {\bf Proof.} This follows from the fact that $P^{Z_\Lambda}$ is a direct sum of such measures. \kasten

We note that $P^{Z_\Lambda}$ can be seen as a complex valued
generalization of a Poisson noise measure. In particular, if $\nu$
is a probability measure, $P^{Z_\Lambda}$ is the defining measure
for the marked Poisson process in the finite volume $\Lambda$ with
mark distribution $\nu$ and intensity $\sigma$.

Let $F_\Lambda$ be as in the preceding subsection with associated
measure $P^{F_\Lambda}$. For
$H:\Gamma^c_0\times\Gamma_0^{c'}\to\C$ measurable and
$L^1(P^F\otimes |P^{Z_\Lambda}|)$ integrable, we define the linear
functional
\begin{equation}
\label{6.4eqa} \E_{P^{F_\Lambda}\otimes P^{Z_\Lambda}}\left[H
\right]=\int_{\Gamma_0}\E_{P^{F_\Lambda}}[H(F_\Lambda,\gamma)]\,
dP^{Z_\Lambda}(\gamma).
\end{equation}
Let $u:\Gamma_0^c\times\Gamma_0^{c'}\to\R$ be defined by
$u(\eta,\gamma)=\langle \eta,G*\gamma\rangle$. $u$ is the
interaction of a two-component marked Potts model where the
component one interacts with the component two but there is no
interaction within either component. The \underline{formal Potts
model at imaginary temperature} that we consider here is defined
by complex valued Gibbs measure in the finite volume
\begin{equation}
\label{6.5eqa} dP^{\rm f. P.}(\eta,\gamma)={e^{iu(\eta,\gamma)}
\over \E_{P^F\otimes
P^{Z_\Lambda}}[e^{iu(F_\Lambda,Z_\Lambda)}]}dP^{F_\Lambda}\otimes
P^{Z_\Lambda}(\eta,\gamma).
\end{equation}
Here $Z_\Lambda(\gamma)=\gamma$ is the coordinate process of the
second component. This gives not an ensemble of statistical
mechanics (unless $\nu$ is a probability measure), as the second
component has a non-probability "distribution" $P^{Z_\Lambda}$.
Nevertheless, this object can be treated analogously to the
measure defining the ordinary Potts model. In particular this
applies to the correlation functional
\begin{equation}
\label{6.6eqa} \rho^{\rm f. P.}(\eta,\gamma)={1\over
\Xi_\Lambda^{\rm f. P.}}1_{\{{\rm supp}\eta\subseteq\Lambda,~{\rm
supp}\gamma\subseteq
\Lambda\}}(\eta,\gamma)\E_{P^{F_\Lambda}\otimes
P^{Z_\Lambda}}\left[e^{i
u(\eta+F_\Lambda,\gamma+Z_\Lambda)}\right],
\end{equation}
$\forall \eta\in\Gamma_0^c$,  $\gamma\in\Gamma_0^{c'}$ where
$\Xi_\Lambda^{\rm f. P.}=\E_{P^{F_\Lambda}\otimes
P^{Z_\Lambda}}\left[e^{i u(F_\Lambda,Z_\Lambda)}\right]$. The
following crucial observation implies in particular that
$\Xi_\Lambda^{\rm f.P.}>0$:

\begin{proposition}
\label{6.2prop} Let $\tilde\Xi_\Lambda$and $\rho_\Lambda$ be as in
the preceding section. The following identities hold:

\noindent (i) $\Xi^{\rm f. P.}_\Lambda=\tilde\Xi_\Lambda$;

\noindent (ii) $\rho^{\rm f. P.}_\Lambda(\eta,0)=\rho_\Lambda(\eta)$.
\end{proposition}
\noindent {\bf Proof.} By the definitions (\ref{6.2eqa}) and (\ref{6.6eqa}) and Fubini's theorem, it is sufficient to integrate out the second component and show
$$e^{-\beta U_\Lambda(\eta)}=\int_{\Gamma_0^{c'}}e^{iu(\eta,\gamma)}dP^{Z_\Lambda}(\gamma).$$
with $U_\Lambda$ the trigonometric interaction defined in Section 6.1. Using the definition (\ref{6.4eqa}) to evaluate the right hand side, we get
\begin{eqnarray*}
&&\int_{\Gamma_0^{c'}}e^{iu(\eta,\gamma)}dP^{Z_\Lambda}(\gamma)\\
&=&e^{-\beta|\Lambda|\nu([-c',c'])}\sum_{l=0}^\infty{\beta^l\over l!}\int_{\Lambda^l\times[-c',c']^{\times l}}e^{i\langle\eta,G*\sum_{j=0}^l\alpha_j\delta_{y_j}\rangle}dy_1\cdots dy_ld\nu(\alpha_1)\ldots d\nu(\alpha_l)\\
&=&e^{-\beta|\Lambda|\nu([-c',c'])}\sum_{l=0}^\infty{\beta^l\over l!}\left(\int_{\Lambda\times[-c',c']}e^{i\alpha G*\eta(y)}d\nu(\alpha)dy\right)^l\\
&=&e^{\beta\int_\Lambda [\int_{\R}e^{i\alpha G*\eta(y)}-1 d\nu(\alpha)]dy}=e^{-\beta\int_\Lambda v(G*\eta)\,dy}. \end{eqnarray*}
\kasten

Spelled out in words Proposition \ref{6.2prop} means that one can obtain the model with trigonometric interaction as the projection (or Widom-Rowlinson model) of a formal Potts model at imaginary temperature. What one has gained from this representation is that
the formal Potts model is a model with a pure two-point interaction, hence the usual cluster expansion procedure of Ruelle goes trough, cf. the following subsection. That the
formal Potts model at imaginary temperature for general $\nu$ does not possess the necessary positivity properties poses no problems, as we are only interested in the projection, where positivity holds, cf. Section 6.4.

\

{\noindent \it 6.3 The cluster expansion}

\noindent To specify the domain of convergence for our expansion, we define
\begin{equation}
\label{6.7eqa}
C_1=\sup_{y\in\R^d,s\in{\rm supp}r}\int_{\R^d}\int_\R\left|e^{is\alpha G(y-y')}-1\right|d|\nu|(\alpha)dy'\leq cb\|G\|_1<\infty
\end{equation}
and
\begin{equation}
\label{6.8eqa}
C_2=\sup_{y'\in\R^d,\alpha \in{\rm supp}\nu}\int_{\R^d}\int_\R\left|e^{is\alpha G(y-y')}-1\right|dr(\alpha)dy'\leq c'b'\|G\|_1<\infty
\end{equation}

The following theorem is based on the convergence of Ruelle's cluster expansion \cite[Chapter4.4]{Ru}:

\begin{theorem}
\label{6.1theo}
The high-temperature low-density expansion of the infinite volume limit of the correlation functional in the case of trigonometric interactions converges
for $z>0$ and $\beta\in\R$ such that $|z|<1/(eC_1)$ and $|\beta|<1/(eC_2)$. In particular

\noindent (i) $\rho(\eta)=\lim_{\Lambda\uparrow\R^d}\rho_\Lambda(\eta)$ exists for $\eta\in \Gamma_0^c$ and depends analytically on $\beta$ and $z$. In particular, for $z$ fixed, the high temperature expansion of $\rho$ converges;

\noindent (ii) $\rho$ is invariant under the action of the Euclidean group, i.e. $\rho(\eta)=\rho(\eta_{\{g,a\}})$ for $g\in O(d)$, $a\in \R^d$;

\end{theorem}

\noindent {\bf Proof.} (i) To obtain the cluster expansion for $\rho^{\rm f.P.}=\lim_{\Lambda\uparrow\R^d}\rho_\Lambda^{\rm f.P.}$, only a few modifications w.r.t. \cite[Chapter 4.4]{Ru} are necessary.

Let $\S_a=\R^d\times\R\times\{a\}$, $a=1,2$ and $\S=\S_1\dot\cup\S_2=\R^d\times\R\times\{1,2\}$. Any pair of finite marked configurations $(\eta,\gamma)\in\Gamma^c_0\times\Gamma^{c'}_0$ can then be identified with a non-marked configuration $\xi\in \hat\Gamma_0(\S)$
on $\S$ defined as follows: First, given $\eta=\sum_{j=1}^ns_j\delta _{x_j}$ we define a non-marked configuration $\tilde\eta=\sum_{j=1}^n\delta_{(x_j,s_j,1)}$ on $\S_1$ and likewise $\gamma$ defines a non-marked configuration $\tilde\gamma$ on $\S_2$.
Then we set $\xi|_{\S_1}=\tilde\eta$ and $\xi|_{\S_2}=\tilde\gamma$. Such a non-marked configuration $\xi=\sum_{j=1}^n\delta_{q_j}$, $q_j\in\S$ can be identified with the finite subset $\{q_1,\ldots,q_n\}$ of $\S$. 

Let $\sigma$ be the complex measure on $\S$ obtained by $\sigma|_{\R^d\times\R\times\{1\}}=dy\otimes r$ and $\sigma|_{\R^d\times\R\times\{2\}}=dy\otimes \nu$ with $dy$ the Lebsgue measure.
For $\Lambda \subseteq \R^d$ compact let and $\S_\Lambda=\Lambda\times\R\times\{1,2\}$. For $q\in\S$ let furthermore $\zeta(q)=z$ if $q=(y,s,1)$, $y\in\R^d$, $s\in\R$ and $\zeta(q)=\beta$ otherwise.

  Let $\chi$ be a function on $\S$ with $\supp\chi\subseteq\S_\Lambda$ for some $\Lambda\subseteq\R^d$ compact.  For $\Psi:\hat \Gamma_0(\S)\to\C$ measurable and bounded, we can define
$$
\langle\chi,\Psi\rangle(\zeta)=\Psi_0+\sum_{n=1}^\infty {1\over n!}\int_{\S^n}(\zeta\chi)(q_1)\cdots(\zeta\chi)(q_n)\Psi(\{q_1,\ldots,q_n\})\, d\sigma(q_1)\cdots d\sigma(q_n).
$$
Letting $\Psi(\xi)=e^{iu(\eta,\gamma)}$ with $(\eta,\gamma)\in\Gamma^c_0\times\Gamma^{c'}_0$ associated with $\xi$, we obtain the following representation of $\rho^{\rm f.P.}_\Lambda$:
$$
\rho_\Lambda^{\rm f.P.}(\xi)=1_{\{\xi\subseteq\S_\Lambda\}}(\xi)\langle\chi_\Lambda,\Psi\rangle(\zeta)^{-1}\langle\chi_\Lambda,D_\xi\Psi\rangle(\zeta),~~\xi\in\hat\Gamma_0(\S),
$$
where $D_\xi\Psi(\tau )=\Psi(\xi\cup\tau)$, $\xi,\tau\in\hat\Gamma_0(\S)$,  and $\chi_\Lambda=1_{\S_\Lambda}$.

Using Ruelle's $*$-product (known as $s$-product in QFT \cite{Bor})
\begin{equation}
\label{6.8aeqa}
\Psi_1*\Psi_2(\xi)=\sum_{\tau\subseteq\xi}\Psi_1(\tau)\Psi_2(\xi\setminus \tau)~,~~ \xi\in\hat\Gamma_0(\S),
\end{equation}
one obtains as in \cite[Chapter 4.4]{Ru} the following expansion
of $\rho^{\rm f.P.}_\Lambda$ in the formal parameter $\zeta$
\begin{equation}
\label{6.9eqa}
\rho^{\rm f.P.}_\Lambda(\xi)=1_{\{\xi\subseteq\S_\Lambda\}}(\xi)\langle\chi_\Lambda,\tilde\varphi_\xi\rangle(\zeta),~~\tilde\varphi_\xi=\Psi^{-1}*D_\xi\Psi.
\end{equation}
Here $\Psi^{-1}$ is the inverse of $\Psi$ w.r.t. the $*$-multiplication (as $\Psi(\emptyset)=1$ this inverse exists). We have to study the range of convergence of $\langle\chi_\Lambda,\tilde\varphi_\xi\rangle(\zeta)$ as $\Lambda\uparrow\R^d$. To this aim, we define the pair-potential
$o(q,q')=ss'G(y-y')$ for $q=(y,s,a)$, $q'=(y',s',a')$ with $a'\not=a$ and $o(q,q')=0$ otherwise. For $q\in\xi$ we set $W(q,\xi)=\sum_{q'\in\xi\setminus\{q\}}o(q,q')$ and for $q\in\S$, $ \xi\in\hat\Gamma_0(\S)$ such that $q\not\in\xi$, $K(q,\xi)=\prod_{q'\in\xi}[e^{io(q,q')}-1]$. As in Ruelle's book, we then obtain the recurrence
formula for $\tilde\varphi_\xi$ and $q\in\xi$
\begin{equation}
\label{6.10eqa}
\tilde\varphi_\xi(\tau)=e^{iW(q,\xi)}\sum_{\kappa \subseteq\tau(q)^c} K(q,\kappa)\tilde\varphi_{\xi\setminus\{q\}\cup\kappa}(\tau\setminus\kappa),
\end{equation}
where $\tau(q)^c$ for $q=(y,s,a)$ is defined as $\{q'=(y',s',a')\in\tau:a'\not=a\}$.

One obtains from (\ref{6.10eqa}) by induction over $n+m$ with $m=m_1+m_2=\sharp\xi_1+\sharp\xi_2$, $\xi_a=\xi\cap \S_a$, $a=1,2$, and $n=n_1+n_2=\sharp\tau_1+\sharp\tau_2$ that for $\theta_1,\theta_2>0$ $\exists C=C(\theta_1,\theta_2)<\infty$ such that
\begin{eqnarray}
\label{6.11eqa}
&&\sup_{(q_1,\ldots,q_m)\in\S_1^{m_1}\times \S_2^{m_2}\atop q_j\not=q_l, j\not=l}\int_{\S^{n_1}_1\times\S^{n_2}_2}\left|\tilde \varphi_{\{q_1,\ldots,q_m\}}(\{q_1',\ldots,q_n'\})\right|\,d|\sigma|(q_1')\cdots d|\sigma|(q_n')\nonumber\\
&&~~~~~~~~~~~~~~~~~~~~~\leq~ Cn_1!n_2!\theta_1^{m_1}\theta_2^{m_2}\left(e^{\theta_1 C_1}\over\theta_1\right)^{n_1+m_1}\left(e^{\theta_2 C_2}\over\theta_2\right)^{n_2+m_2}
\end{eqnarray}
This estimate for $\theta_a=C_a^{-1}$, $a=1,2$, implies that for $m=\sharp\xi$ fixed the right hand side of (\ref{6.9eqa}) converges uniformly (in $\Lambda\subseteq\R^d$ compact) if
$$
|z|<1/(eC_1)~~\mbox{and}~~|\beta|<1/(eC_2)
$$
From the uniform convergence of $\rho^{f.P.}_\Lambda(\xi)$ it follows that $\rho^{\rm f.P.}(\xi)=\lim_{\Lambda\uparrow\R^d}\rho_\Lambda^{\rm f.P.}(\xi)$ exists and is analytic in the above parameter domain. Combining this with Proposition \ref{6.2prop}, one obtains the assertion (i) of the theorem.

(ii) Note that $\rho_\Lambda(\eta_{\{g,a\}})=\rho_{g\Lambda+a}(\eta)$. Invariance of $\rho$ now follows from the equivalence of the limits $\Lambda\uparrow\R^d$ and $g\Lambda+a\uparrow\R^d$.
\kasten

\begin{remark}
\label{6.1rem} {\rm Let us briefly sketch three methods for an
analytic or numerical evaluation of $\rho$. The details can be
worked out by (more or less lengthy) straight forward
calculations.

\noindent (i) Meyer's series for $\rho^{\rm f.P.}$:  The usual
graphical methods connected with the Meyer series \cite[p. 88]{Ru}
now can be used for the explicit calculation of the expansion
coefficients in $\beta$ and $z$ of $\rho$.

\noindent (ii)  High temperature expansion: One can go back to Eq.
(\ref{6.2eqa}) and calculate the $\beta$-expansions of the
nominator and the denominator, which essentially amounts to
calculating the functional Fourier transforms of $P^F$. Taking the
quotient in the sense of formal power series then yields an
expansion, which is known to converge in the infinite volume limit
for $z$ and $\beta$ sufficiently small.

\noindent (iii) Small $G$ expansion: Finally, it is possible to
adapt \cite{DGO} to the formal Potts model and to obtain a
representation of the moments of the infinite volume Gibbs measure
with trigonometric interaction (to be constructed in the following
Section) in terms of generalized Feynman graphs\footnote{Here it
is necessary that $G$ is a regular function in order to avoid
ultra-violet singularities.}, which amounts to a formal expansion
in powers of $G$. This method has the advantage that only finitely
many moments of $\nu$ appear as parameters in the expansion up to
a finite order. Hence the form of the interaction (at least in
priciple) can be determined comparing the expansion with
experimental data. }
\end{remark}

\

{\noindent \it 6.4 Construction of Gibbs measures}

\noindent Here we want to construct the Gibbs measure associated
with the correlation functional $\rho$. Before we can do this,
some preparations are needed. Here we mostly follow \cite{KK}.

The $\sigma$-finite  Lebesgue-Poisson measure $\lambda_z$ is
defined on $(\Gamma_0^c,{\cal B}(\Gamma_0^c))$ by setting
\begin{equation}
\label{6.12eqa} \lambda_z(A)=\delta_0(A)+ \sum_{l=1}^\infty
{z^l\over l!}\int_{\R^{dl}\times[-c,c]^{\times
l}}1_A(\sum_{j=1}^ls_j\delta_{y_j})\, dy_1\cdots dy_ldr(s_1)\cdots
dr(s_l)
\end{equation}
$A\in{\cal B}(\Gamma_0^c)$.

 It is well-known, see e.g. \cite{KK}, that $\rho_\Lambda$ can be represented as
\begin{equation}
\label{6.13eqa}
\rho_\Lambda(\eta)={dP^{\tilde F_\Lambda}\over d\lambda_z}(\eta)~~~\mbox{for}~\lambda_z-a.e. ~\eta\in \Gamma_0^c
\end{equation}
Here $dP^{\tilde F_\Lambda}(\eta)={1\over \tilde \Xi_\Lambda}\,
e^{-\beta U_\Lambda(\eta)}dP^{F_\Lambda}(\eta)$.

Let $\Gamma_\Lambda^c=\{\gamma\in\Gamma^c:{\rm
supp}\gamma\subseteq \Lambda\}$ for $\Lambda\subseteq \R^d$
measurable. Note that for $\Lambda$ compact,
$\Gamma_\Lambda^{c}\subseteq \Gamma_0^c$. Suppose $\hat
\rho:\Gamma_0^c\to\R$ is a given functional that fulfills a Ruelle
bound $|\hat \rho(\eta)|\leq C^{\sharp \eta}$ for some $C>0$.
Assume furthermore that the functional is \underline{Lennard
positive} in the sense
\begin{equation}
\label{6.13aeqa} q^\Lambda(\eta)=\int_{\Gamma_\Lambda^c}
(-1)^{\sharp \eta}\hat\rho(\eta+\gamma)d\lambda_z(\gamma)\geq 0
\end{equation}
for $\lambda_z$-- a.e. $\eta$ and all $\Lambda\subseteq \R^d$
compact. One can then check that $\{P^\Lambda\}_{\Lambda\subseteq
\R^d~\rm compact}$ defined by
$dP^\Lambda(\eta)=q^\Lambda(\eta)d\lambda_z(\eta)$ is a projective
family of probability measures. Hence the inductive limit $P$ of
this family exists by Kolmogorov's theorem as a measure on
$(\Gamma^c,{\cal B}(\Gamma^c))$, cf. \cite[Prop. 4.5 and Theorem
4.5]{KK}. In particular this implies that the constructed measure
has support on tempered marked configurations.

Furthermore, let $\hat\rho_n:\Gamma_0^c\to\R$ be a sequence of
Lennard-positive correlation functionals that fulfill a uniform
Ruelle bound $|\hat\rho_n(\eta)|\leq C^{\sharp \eta}$ for some
$C>0$ independent of $n$. Let furthermore
$\hat\rho(\eta)=\lim_{n\to\infty}\hat\rho_n(\eta)$ exist
$\lambda_z$-- a.e. and suppose that for some $D>0$ sufficiently
small $\|\hat\rho-\hat\rho_n\|_D\to 0$ as $n\to\infty$ with
$\|\hat\rho-\hat\rho_n\|_{D}={\rm
esssup}_{\eta\in\Gamma_0^c}D^{\sharp
\eta}|\hat\rho(\eta)-\hat\rho_n(\eta)|$. Then the limiting
functional $\hat \rho$ is Lennard positive and fulfills the same
Ruelle bound as the $\hat\rho_n$'s, \cite[Prop.4.9]{KK}. Hence
there exists a measure $P$ associated to $\hat\rho$, as explained
in the preceding paragraph.

Let  $p_\Lambda:\Gamma_0^c\to\Gamma_\Lambda^c$ be the projection
given by $\gamma\to 1_\Lambda\gamma$. A function $H:\Gamma^c\to\R$
is $\Lambda$-measurable, if it is measurable w.r.t. ${\cal
B}_\Lambda(\Gamma^c)=p_\Lambda^{-1}(\Gamma_\Lambda^c)$. If
$\hat\rho_n$ and $\hat\rho$ are as above with associated measures
$P$ and $P_n$ on $(\Gamma^c,{\cal B}(\Gamma^c))$, then the
measures $P_n$ converge locally to $P$, i.e. for all
$H:\Gamma^c\to\R$ positive that is $\Lambda$-measurable for some
$\Lambda\subseteq\R^d$ compact, we get
$\lim_{n\to\infty}\E_{P_n}[H]=\E_P[H]$, see \cite[Cor. 4.11]{KK}.

Also, the  translation invariance of $\hat \rho$ is equivalent to
the the translation invariance of the associated measure, cf.
\cite[Prop. 3.11]{KK}.

Applying these pieces of general theory to the case of the
preceding subsection, we obtain

\begin{proposition}
\label{6.3prop} Let $P^{\tilde F_\Lambda}$ be the measures on
$(\Gamma_0,{\cal B}(\Gamma_0))$ associated with the correlation
functionals $\rho_\Lambda$ defined in Section 6.1. Then

\noindent (i) There exists a uniquely determined measure $P^{\tilde F}$ on that measurable space which is
associated with $\rho=\lim_{\Lambda\uparrow\R^d}\rho_\Lambda$;

\noindent (ii) $\lim_{\Lambda\uparrow\R^d}P^{\tilde F_\Lambda}=P^{\tilde F}$ holds in the sense of local convergence;

\noindent (iii) $P^{\tilde F}$ is translation invariant.

\end{proposition}
\noindent{\bf Proof.} The estimate (\ref{6.11eqa}) implies that $\lim_{\Lambda\uparrow\R^d}\|\rho_\Lambda-\rho\|_D=0$ for $0<D<z$. The three assertions therefore
 follow
from the general formalism.\kasten

Obviously, Proposition \ref{6.3prop} also  implies the existence
of $P^{\tilde X}=L^{-1}_*P^{\tilde F}$ as a measure on $({\cal
S}',{\cal B}({\cal S}'))$.

The interaction without infra-red cut-off  is $U(\eta)=\int_{\R^d}
v(G*\eta)\, dx$, $\eta\in\Gamma_0^c$.  Let $\eta\in\Gamma_0^c$ and
$\gamma\in\Gamma^c$. Then the mutual interaction $W$ between
$\eta$ and $\gamma$ is by definition
\begin{equation}
\label{6.14eqa}
W(\eta,\gamma)=\int_{\R^d} [v(G*(\eta+\gamma))-v(G*\eta)-v(G*\gamma)]\, dx.
\end{equation}
 As the derivative of $v$ is bounded, it is easy to see that $W(\eta,\gamma)$ is well-defined and that
$|W(\eta,\gamma)|<2B\sharp \eta$ with $B=bc\|G\|_1$ as above.

We say that the measure $P$ on $(\Gamma^c,{\cal B}(\Gamma^c))$ is a \underline{Gibbs measure} for the interaction $U$, inverse temperature $\beta$ and activity $z$ if for arbitrary $\Lambda\subseteq \R^d$ compact
and $H:\Gamma^c\to\R$ non-negative the following holds
\begin{equation}
\label{6.15eqa}
\E_{P}[H]=\int_{\Gamma^c_\Lambda}\int_{\Gamma_{\R^d\setminus\Lambda}^c}H(\eta+\gamma)e^{-\beta U(\eta)-\beta W(\eta,\gamma)} dP(\gamma)d\lambda_z(\eta).
\end{equation}
(\ref{6.15eqa}) are called \underline{Ruelle equations}. Under the
given conditions they are equivalent with other definitions of
Gibbs measures as e.g. Dobrushin-Lanford-Ruelle equations,
Georgii-Nguyen-Zessin equations and the standard definition of
Gibbs measures via conditional probablilities, cf. \cite[Theorem
3.12]{KK}. The following theorem verifies the Gibbs property for
$P^{\tilde F}$.

\begin{theorem}
\label{6.2theo} Let $z,\beta$ as in Theorem \ref{6.1theo}, $G\in
L^1(\R^d,dx)$. Then  $P^{\tilde F}$ is a Gibbs measure
w.r.t. $U,z,\beta$;

\end{theorem}
\noindent {\bf Proof.} On both sides of (\ref{6.15eqa}) the
function $H$ can be approximated from below by elementary
functions. Thus, it suffices to consider the case where $H$ is a
characteristic function $1_A$, $A\in{\cal B}(\Gamma_0)$. On both
sides of (\ref{6.14eqa}) we have to evaluate $\sigma$-finite
measures. It is therefore sufficient to consider $A$ from a
$\cap$-stable generating subsystem of ${\cal B}(\Gamma^c)$. We may
thus assume that $A\in{\cal B}_{\Lambda'}(\Gamma^c)$ for some
$\Lambda'\subseteq\R^d$ compact.

Let $\Lambda$ be fixed.  For compact $\Lambda''\subseteq \R^d$ we
consider the infra-red cut-off interaction $U_{\Lambda''}$ as in
Section 6.1 and we let
$W_{\Lambda''}(\eta,\gamma)=U_{\Lambda''}(\eta+\gamma)-U_{\Lambda''}(\eta)-U_{\Lambda''}(\gamma)$,
$\eta,\gamma\in\Gamma_0^c$.

Using (\ref{6.13eqa}) and
$$\int_{\Gamma_{\tilde\Lambda}^c}\tilde H(\eta)\,d\lambda_z(\eta)=\int_{\Gamma_{\tilde\Lambda\setminus\tilde\Lambda'}^c}\int_{\Gamma_{\tilde\Lambda'}^c} \tilde H(\eta+\gamma)\,d\lambda_z(\eta)d\lambda_z(\gamma)$$
for compact sets $\tilde\Lambda'\subseteq\tilde\Lambda\subseteq
\R^d$ and $\tilde H:\Gamma^c\to\R$ non-negative and  measurable,
it is easy to show that the Ruelle equations (\ref{6.15eqa}) hold
for $P^{\tilde F_{\Lambda''}}$ instead of $P^{\tilde F}$ and
$U_{\Lambda''}$ and $W_{\Lambda''}$ instead of $U$ and $W$.

By Proposition \ref{6.3prop} (ii) we get that
$\lim_{\Lambda''\uparrow\R^d}\E_{P^{\tilde
F_\Lambda}}[H]=\E_{P^{\tilde F}}[H]$. In order to verify the
Ruelle equations in the infinite volume limit
$\Lambda''\uparrow\R^d$ one thus has to prove that the r.h.s. of
the Ruelle equation with cut-off $\Lambda''$ converge to the
r.h.s. without that cut-off. The modulus of the difference, which
we abbreviate by $I$, can be estimated as follows
($\Lambda'''\subseteq \R^d$ is an arbitrary compact set and
$\gamma_{\Lambda'''}=1_{\Lambda'''}\gamma$):
\begin{eqnarray}
\label{6.16eqa}
I &\leq&\left|\int_{\Gamma^c_\Lambda}\int_{\Gamma_{\R^d\setminus\Lambda}^c}H(\eta+\gamma)e^{-\beta U(\eta)-\beta W(\eta,\gamma)} dP^{\tilde F}(\gamma)d\lambda_z(\eta)\right.\nonumber\\
&-&\left.\int_{\Gamma^c_\Lambda}\int_{\Gamma_{\R^d\setminus\Lambda}^c}H(\eta+\gamma)\,e^{-\beta U_{\Lambda''}(\eta)-\beta W_{\Lambda''}(\eta,\gamma)} dP^{\tilde F}(\gamma)d\lambda_z(\eta)\right|\nonumber\\
&+&\left|\int_{\Gamma^c_\Lambda}\int_{\Gamma_{\R^d\setminus\Lambda}^c}H(\eta+\gamma)\,e^{-\beta U_{\Lambda''}(\eta)-\beta W_{\Lambda''}(\eta,\gamma)} dP^{\tilde F}(\gamma)d\lambda_z(\eta)\right.\nonumber\\
&-&\left.\int_{\Gamma^c_\Lambda}\int_{\Gamma_{\R^d\setminus\Lambda}^c}H(\eta+\gamma)\,e^{-\beta U_{\Lambda''}(\eta)-\beta W_{\Lambda''}(\eta,\gamma_{\Lambda'''})} dP^{\tilde F}(\gamma)d\lambda_z(\eta)\right|\nonumber\\
&+&\left|\int_{\Gamma^c_\Lambda}\int_{\Gamma_{\R^d\setminus\Lambda}^c}H(\eta+\gamma)\,e^{-\beta U_{\Lambda''}(\eta)-\beta W_{\Lambda''}(\eta,\gamma_{\Lambda'''})} dP^{\tilde F}(\gamma)d\lambda_z(\eta)\right.\nonumber\\
&-&\left.\int_{\Gamma^c_\Lambda}\int_{\Gamma_{\R^d\setminus\Lambda}^c}H(\eta+\gamma)\,e^{-\beta U_{\Lambda''}(\eta)-\beta W_{\Lambda''}(\eta,\gamma_{\Lambda'''})} dP^{\tilde F_{\Lambda''}}(\gamma)d\lambda_z(\eta)\right|\nonumber\\
&+&\left|\int_{\Gamma^c_\Lambda}\int_{\Gamma_{\R^d\setminus\Lambda}^c}H(\eta+\gamma)\,e^{-\beta U_{\Lambda''}(\eta)-\beta W_{\Lambda''}(\eta,\gamma_{\Lambda'''})} dP^{\tilde F_{\Lambda''}}(\gamma)d\lambda_z(\eta)\right.\nonumber\\
&-&\left.\int_{\Gamma^c_\Lambda}\int_{\Gamma_{\R^d\setminus\Lambda}^c}H(\eta+\gamma)\,e^{-\beta U_{\Lambda''}(\eta)-\beta W_{\Lambda''}(\eta,\gamma)} dP^{\tilde F_{\Lambda''}}(\gamma)d\lambda_z(\eta)\right|\nonumber\\
\end{eqnarray}
Let  us call these terms
$I_1(\Lambda'',\Lambda'''),\ldots,I_4(\Lambda'',\Lambda''')$. Let
$\epsilon >0 $ be given -- we have to show that for $\Lambda''$
sufficiently large and a suitable $\Lambda'''$ we get
$I_j(\Lambda'',\Lambda''')<\epsilon$, $j=1,\ldots,4$.

$I_1(\Lambda'',\Lambda''')$  in fact only depends on $\Lambda''$.
Note that the integrand in (\ref{6.14eqa}) is dominated by
$2b|G|*|\eta|\in L^1(\R^d,dx)$, hence by dominated convergence
$U_{\Lambda''}(\eta)\to U(\eta)$ and
$W_{\Lambda''}(\eta,\gamma)\to W(\eta,\gamma)$ for
$\eta\in\Gamma_0^c$ and $\gamma\in\Gamma^c$. At the same time
$e^{2\beta B\sharp\eta}$ is an upper bound for
$|H(\eta+\gamma)e^{-\beta U_{\Lambda''}(\eta)-\beta
W_{\Lambda''}(\eta,\gamma)}|$, if $H$ is bounded by one.
Consequently, $I_1(\Lambda'',\Lambda''')\to 0$ as
$\Lambda''\uparrow\R^d$ holds by dominated convergence.

For $\Lambda'''$ fixed,  $I_3(\Lambda'',\Lambda''')\to 0$ as
$\Lambda''\uparrow\R^d$ follows from Proposition \ref{6.3prop}
(ii).

It remains to show that one can find a compact set $\Lambda'''$
such that $I_2(\Lambda'',\Lambda'''),\linebreak
I_4(\Lambda'',\Lambda''')<\epsilon$ for all compact
$\Lambda''\subseteq\R^d$. Note that this is trivial for $G$ of
finite range $R$ and
$d(\Lambda,\partial\Lambda''')=\inf\{|x-y|:x\in\Lambda,~y\in\R\setminus\Lambda'''\}>2R$
since then
$W_{\Lambda''}(\eta,\gamma_{\Lambda'''})=W_{\Lambda''}(\eta,\gamma)$
implies $I_2(\Lambda'',\Lambda''')=I_4(\Lambda'',\Lambda''')=0$.

In the next step we consider $I_4(\Lambda'',\Lambda''')$ in the
general case. Let for $R>0$
$\Lambda_R=\{x\in\R^d:\exists y\in\Lambda~\mbox{s.t.}~|x-y|\leq R\}$. We set
$\Lambda'''=\Lambda_{R'}$ for some $R'>0$ and we have to show that
$I_4(\Lambda'',\Lambda_{R'})\to 0$ uniformly in $\Lambda''$ as $R'\to\infty$. Let
us begin with the estimate
\begin{eqnarray}
\label{6.17eqa}
&&\left|e^{-\beta W_{\Lambda''}(\eta,\gamma)}-e^{-\beta W_{\Lambda''}(\eta,\gamma_{\Lambda_{R'}})}\right|\nonumber\\
&\leq& e^{2|\beta| B\sharp\eta}\left(e^{|\beta|
|W_{\Lambda''}(\eta,\gamma)-W_{\Lambda''}(\eta,\gamma_{\Lambda_{R'}})|}-1\right).
\end{eqnarray}
For $R>0$ arbitrary, we can combine (\ref{6.17eqa}) with
\begin{eqnarray}
\label{6.17aeqa}
&&|W_{\Lambda''}(\eta,\gamma)-W_{\Lambda''}(\eta,\gamma_{\Lambda_{R'}})|\nonumber\\
&&\leq 2b\left[\int_{\Lambda_R}|G|*|\gamma_{\R^d\setminus\Lambda_{R'}}|dx+\int_{\R^d\setminus\Lambda_R}|G|*|\eta|dx\right]\nonumber \\
&&=2b[\langle |G|*1_{\Lambda_R}\cdot
1_{\R^d\setminus\Lambda_{R'}},|\gamma|\rangle+\langle
|G|*1_{\R^d\setminus\Lambda_R},|\eta|\rangle]
\end{eqnarray}
and we obtain
\begin{eqnarray}
\label{6.18eqa} &&\int_{\Gamma^c_{\R^d\setminus\Lambda}}
\left|e^{-\beta W_{\Lambda''}(\eta,\gamma)} -e^{-\beta
W_{\Lambda''}(\eta,\gamma_{\Lambda_{R'}})}\right|
dP^{\tilde F_{\Lambda''}}(\gamma)\nonumber\\
&&~~~~\leq e^{2B|\beta|\sharp\eta} \left(e^{2b|\beta|\langle
|G|*1_{\R^d\setminus\Lambda_R},|\eta|\rangle}
\int_{\Gamma^c}e^{2b|\beta|\langle |G|*1_{\Lambda_R}\cdot
1_{\R^d\setminus\Lambda_{R'}},|\gamma|\rangle}
dP^{\tilde F_{\Lambda''}}(\gamma)-1\right)\nonumber\\
\end{eqnarray}
For a function $h:\R^d\times[-c,c]\to\R$ we define the
\underline{Lebsgue-Poisson coherent state}
$e_\lambda(h,\xi)=\prod_{l=1}^nh(x_l,s_l)$,
$\xi=\sum_{j=1}^ns_l\delta_{x_l}\in\Gamma_0^c$. The following
identity holds for $\tilde H:\Gamma_{\tilde\Lambda}^c\to\R$
non-negative and $\tilde\Lambda\subseteq \R^d$ an arbitrary
measurable set
\begin{equation}
\label{6.18aeqa}
\int_{\Gamma_{\tilde\Lambda}^c}\int_{\Gamma_{\tilde\Lambda}^c}\tilde H(\eta+\gamma)e_\lambda(h_1,\eta)e_{\lambda}(h_2,\gamma)d\lambda_z(\eta)d\lambda_z(\gamma)=
\int_{\Gamma_{\tilde\Lambda}^c}\tilde H(\eta)e_\lambda(h_1+h_2,\eta)d\lambda_z(\eta)
\end{equation}
and can be found in \cite[Corollary 2.5]{KK} for
$\Gamma_{\tilde\Lambda}^c$ replaced with $\Gamma_0^c$. One can
check the above identity along the same lines, cf. the proof of
Lemma 2.1 of that reference.  Using the notation
$h(x,s)=|s||G|*1_{\Lambda_R}(x) 1_{\R^d\setminus\Lambda_{R'}}(x)$
we thus get
\begin{eqnarray}
\label{6.19eqa} 1&\leq&\int_{\Gamma^c}e^{2b|\beta|\langle
|G|*1_{\Lambda_R}\cdot
1_{\R^d\setminus\Lambda_{R'}},|\gamma|\rangle}
dP^{\tilde F_{\Lambda''}}(\gamma)\nonumber\\
&=& \int_{\Gamma_{\Lambda''}^c}\int_{\Gamma_{\Lambda''}}e_\lambda(e^{2b|\beta|h},\gamma)e_\lambda(-1,\xi)\rho_{\Lambda''}(\gamma+\xi)d\lambda_z(\xi)d\lambda_z(\gamma)\nonumber\\
&=&\int_{\Gamma_{\Lambda''}^c}e_\lambda(e^{2b|\beta|h}-1,\gamma)\rho_{\Lambda''}(\gamma)d\lambda_z(\gamma)\nonumber\\
&\leq&\int_{\Gamma_{\Lambda''}^c}e_\lambda(e^{2b|\beta|h}-1,\gamma)C^{\sharp\gamma}d\lambda_z(\gamma)\nonumber\\
&\leq&e^{zC\int_{\Lambda''}[e^{2b|\beta|h(x,s)}-1]dr(s)dx}\leq
e^{2zCbc|\beta|e^{2B|\beta|}\int_{\R^d\setminus\Lambda_{R'}}|G|*1_{\Lambda_R}dx},
\end{eqnarray}
with $B=cb\|G\|_1\geq \sup_{x\in\R^d,s\in[-c,c]} h(x,s)$ and $C$ the
Ruelle constant that does not depend on $\Lambda''$, cf. Prop.
\ref{6.1prop}. Inserting (\ref{6.19eqa}) into (\ref{6.18eqa}) we
obtain for $I_4(\Lambda'',\Lambda_{R'})$
\begin{eqnarray}
\label{6.20eqa}
&&I_4(\Lambda'',\Lambda_{R'})\nonumber\\
&&\leq\int_{\Gamma_\Lambda^{c}}e^{2B|\beta|\sharp\eta}\left(e^{2b|\beta|\langle
|G|*1_{\R^d\setminus\Lambda_R},|\eta|\rangle}
e^{2zCcb|\beta|e^{2B|\beta|}\int_{\R^d\setminus\Lambda_{R'}}|G|*1_{\Lambda_R}dx}
-1\right)d\lambda_z(\eta)\nonumber\\
&&\leq
e^{2zB|\beta||\Lambda|}\left(e^{2zCcb|\beta| e^{2B|\beta|}\int_{\R^d\setminus\Lambda_{R'}}|G|*1_{\Lambda_R}dx+2zcb|\beta|\int_\Lambda|G|*1_{\R^d\setminus\Lambda_{R}}dx}
-1\right)\nonumber\\
&&=
e^{2zB|\beta||\Lambda|}\left(e^{2zbc|\beta|(Ce^{2B|\beta|}\int_{\R^d\setminus\Lambda_{R'}}|G|*1_{\Lambda_R}dx+\int_{\R^d\setminus\Lambda_R}|G|*1_\Lambda dx)}
-1\right).
\end{eqnarray}
Let $\epsilon'>0$ be arbitrary. We have to show that we can choose $R,R'>0$ such that each of the integrals in the exponent on the r.h.s. of (\ref{6.20eqa}) is smaller than $\epsilon'$. We note that
$|G|*1_\Lambda\in L^1(\R^d,dx)$, thus $\int_{\R^d\setminus \Lambda_R} |G|*1_\Lambda\,dx<\epsilon'$ for $R=R(\epsilon')>0$ large enough. Let such $R$ be fixed, we then see that $|G|*1_{\Lambda_R}\in L^1(\R^d,dx)$, hence we can
find an $R'=R'(R,\epsilon')>0$ large enough, such that $\int_{\R^d\setminus\Lambda_{R'}}|G|*1_{\Lambda_R}\, dx<\epsilon'$.
Choosing $\epsilon'=\epsilon'(\epsilon)>0$ small enough and $R,R'$ accordingly, we can finally achieve that the right hand side of (\ref{6.20eqa}) becomes
 smaller than $\epsilon$, which establishes the required estimate for $I(\Lambda'',\Lambda_{R'})$.

To estimate $I_2(\Lambda'',\Lambda_{R'})$, we remark that
(\ref{6.17aeqa}) is independent of $\Lambda''$. Thus one obtains
(\ref{6.18eqa}) with $\tilde F_{\Lambda''}$ replaced by $\tilde
F$. The integral over $\Gamma^c$ in the first line of
(\ref{6.19eqa}) with $\tilde F_{\Lambda''}$ replaced by $\tilde F$
fulfills the same uniform bound as on the right hand side, as we
have by monotone convergence and Prop \ref{6.3prop} (ii)
\begin{eqnarray}
\label{6.21eqa} &&\int_{\Gamma^c}e^{2b\beta\langle
|G|*1_{\Lambda_R}\cdot
1_{\R^d\setminus\Lambda_{R'}},|\gamma|\rangle}
dP^{\tilde F}(\gamma)\nonumber\\
&&~~~~~~~~~=\sup_{\tilde\Lambda\subseteq\R^d~{\rm
compact}}\int_{\Gamma^c}e^{2b\beta\langle |G|*1_{\Lambda_R}\cdot
1_{\R^d\setminus\Lambda_{R'}},|\gamma_{\tilde\Lambda}|\rangle}
dP^{\tilde F}(\gamma)\nonumber\\
&&~~~~~~~~~\leq\sup_{\tilde\Lambda,\Lambda''\subseteq\R^d~{\rm
compact}}\int_{\Gamma^c}e^{2b\beta\langle |G|*1_{\Lambda_R}\cdot
1_{\R^d\setminus\Lambda_{R'}},|\gamma_{\tilde\Lambda}|\rangle}
dP^{\tilde F_{\Lambda''}}(\gamma)\nonumber\\
&&~~~~~~~~~=\sup_{\Lambda''\subseteq\R^d~{\rm
compact}}\int_{\Gamma^c}e^{2b\beta\langle |G|*1_{\Lambda_R}\cdot
1_{\R^d\setminus\Lambda_{R'}},|\gamma|\rangle} dP^{\tilde
F_{\Lambda''}}(\gamma)
\end{eqnarray}
Hence the estimate (\ref{6.20eqa}) also holds for
$I_2(\Lambda'',\Lambda_{R'})$.
\kasten

\

{\noindent \it 6.5 Cluster property and extremality of the state}

\noindent The aim of this subsection is to show that $P^{\tilde
F}$ is a \underline{pure} or \underline{extremal} Gibbs state,
i.e. $P^{\tilde F}$ cannot be written as the convex combination of
two translation invariant measures on $(\Gamma^c,{\cal
B}(\Gamma^c))$.

We first prove a cluster property for the correlation functional
$\rho$.

Let $h:\R^{dn}\times[-c,c]^{n}\to\R$ and $\hat\rho:\Gamma_0^c\to\R$ a correlation functional. We define
\begin{equation}
\label{6.8beqa}
\hat\rho(h)=\int_{\R^{dn}\times[-c,c]^{\times n}}\hat\rho(\sum_{j=1}^ns_j\delta_{y_j})\,h(y_1,\ldots,y_n,s_1,\ldots,s_n)\, dy_1\cdots dy_ndr(s_1)\cdots dr(s_n)
\end{equation}
and $h_{\{g,a\}}(y_1,\ldots,y_n,s_1,\ldots,s_n)=h(gy_1+a,\ldots,gy_n+a,s_1,\ldots,s_n)$ $g\in O(d)$, $a\in\R^d$. For $\eta=\sum_{j=1}^n s_j\delta_{y_j}\in\Gamma_0^c$, $\eta_{\{g,a\}}=\sum_{j=1}^ns_j\delta_{g^{-1}(y_j-a)}$.

\begin{proposition}
\label{6.4prop}
Let $z,\beta$ as in Theorem \ref{6.1theo}. Then
 $\rho$ fulfills the cluster property
$$\lim_{\tilde \Lambda\uparrow\R^d}{1\over |\tilde \Lambda|}\int_{\tilde \Lambda}[\rho(f\otimes h_{\{1,a\}})-\rho(f)\rho(h)] da=0$$ for
$h:\R^{dn_2} \times[-c,c]^{n_2}\to\R$,
$f:\R^{dn_2}\times[-c,c]^{n_2}\to\R$ infinitely differentiable,
bounded and decreasing like a Schwartz test function in all
$\R^d$-arguments, $n_1,n_2\in\N$.
\end{proposition}
\noindent{\bf Proof.} Again, we closely  follow \cite[Section
4.4.7]{Ru}. Let $\log^*$ be the logarithm w.r.t. the $*$ product
(\ref{6.8aeqa}) and let $\rho^T=\log^*\rho:\Gamma_0^c\to\R$ be the
cluster functional associated with $\rho$. Let
$\varphi=\log^*\Psi$, with $\Psi$ as in the proof of Theorem
\ref{6.1theo}, then the following representation holds:
\begin{equation}
\label{6.11aeqa} \rho^T(\eta)=\sum_{n=0}^\infty{1\over n!}
\int_{\S^n}
\zeta(q_1)\cdots\zeta(q_n)D_{(\eta,0)}\varphi(\{q_1,\ldots,q_n\})\,d\sigma(q_1)\cdots
d\sigma(q_n).
\end{equation}
And  from (\ref{6.11eqa}) one obtains for $m=n_1+n_2$,
$q_1=(y,s,1)$ in combination with $\tilde
\varphi_{\{q\}}=\varphi_{\{q\}}$, cf. \cite[Eq. (4.24)]{Ru},
\begin{eqnarray*}
&&\int_{\S_1^{n_1-1}\times S_2^{n_2}}|\varphi(\{q_1,\ldots,q_n\})|d|\sigma|(q_2)\ldots d|\sigma|(q_n)\\
&&~~~~~~~~~~~~~~\leq C (n_1-1)! n_2! C_1^{-1} (eC_1)^{n_1} (eC_2)^{n_2}
\end{eqnarray*}
This finally gives the estimate
\begin{eqnarray*}
\label{6.11ceqa}
&&\int_{\R^{d(m-1)}\times[-c,c]^{\times(m-1)}}|\rho^T(\sum_{j=1}^ns_j\delta_{y_j})|\,dy_2\cdots d y_mdr(s_2)\cdots dr(s_m)\nonumber\\
&&~~~~~~~~~~~~~~~~~~~~~~~~~~~~~~~~~~~~~~~~~~\leq C(m-1)!C_1^{-1}{(eC_1)^m\over 1-|z|eC_1}{1\over 1-|\beta|eC_2}
\end{eqnarray*}
By a simple change of variables $\int_{\R^d}|\rho^T(f\otimes h_{\{1,a\}})|da<\infty$ follows for $f,h$ as in the assertion, which implies (cf. \cite[Section 4.4.3]{Ru}) $\int_{\R^d}|\rho(f\otimes h_{\{1,a\}})-\rho(f)\rho(h)|da<\infty$.
\kasten

One way to link the cluster  property of the correlation
functional to ergodicity of the measure is to express the moments
of the measure in terms of the correlation functional. This at the
same time gives us a formula for the infinite volume Schwinger
functions
 of the
associated random field $\tilde X$.
\begin{proposition}
\label{6.5prop}
Let $z,\beta$ as in Theorem \ref{6.1theo}. Then

\noindent (i) All moments of $P^{\tilde F}$ exist. In terms of the
correlation functional $\rho$ they are given by
\begin{equation}
 \label{6.21ceqa}
 \E_{P^{\tilde F}}\left[\prod_{j=1}^l\tilde F(f_j)\right]= \sum_{\{I_1,\ldots,I_j\}:1\leq j\leq l,I_r\subseteq \{1,\ldots,l\}\atop I_1\dot\cup\cdots\dot\cup I_j=\{1,\ldots,l\}}\rho({\bf f}_I)
 \end{equation}
 where for ${\bf f}=(f_1,\ldots,f_l)\in{\cal S}^{\times l}$ and $I=(I_1,\ldots,I_j)$ as in the sum in (\ref{6.21aeqa}) ${\bf f}_I(y_1,\ldots,y_j,s_1,\ldots,s_j)=\prod_{q=1}^js_q^{\sharp I_q}\prod_{p\in I_q}f_p(y_q)$.

\noindent (ii) The Schwinger functions of the associated
interacting CPN $\tilde X$ exist and are are given by
 $S_n(f_1\otimes \cdots\otimes f_l)=\E_{P^{\tilde X}}[\prod_{j=1}^lX(f_j)]=\E_{P^{\tilde F}}[\prod_{j=1}^l\tilde
 F(G*f_j)]$. Furthermore, the Schwinger functions are analytic in
 the coupling constant (inverse temperature) $\beta$ and the
 Feynman series converges on the indicated domain.
\end{proposition}
{\noindent \bf Proof.} (i) From the formula for the local densities $q^\Lambda$ and (\ref{6.18aeqa}), one obtains the
following formula for the Laplace transform of $P^{\tilde F}$
\begin{equation}
\label{6.21aeqa}
\E_{P^{\tilde F}}[e^{\langle f,\tilde F\rangle}]=\int_{\Gamma_\Lambda^c}e_\lambda(e^f-1,\eta)\rho(\eta)d\lambda_z(\eta)
\end{equation}
where $f\in{\cal D}(\R^d)$ with ${\rm supp}f\subseteq \Lambda$ and
$e_\lambda$ is the Lebesgue-Poisson coherent state defined in the
proof of Theorem \ref{6.1theo}. In (\ref{6.21aeqa}) we can replace
$\Gamma_\Lambda^c$ with $\Gamma_0^c$.  From the Ruelle-bound for
$\rho$ it follows that the r.h.s. is well-defined not only for $f$
with compact support, but also for $f\in{\cal S}$ and hence
(\ref{6.21aeqa}) extends by continuity. Existence of the (two
sided) Laplace-transform implies existence of moments of all
orders. Taking derivatives of the Laplace-transform at zero yields
\begin{equation}
\label{6.21deqa}
 \E_{P^{\tilde F}}\left[\prod_{j=1}^l\tilde F(f_j)\right]=
 \sum_{j=1}^l{1\over l!}\sum_{(I_1,\ldots,I_j):I_q\subseteq \{1,\ldots,l\}\atop I_1\dot\cup\cdots\dot\cup I_j=\{1,\ldots,l\}}\rho({\bf f}_I)
 \end{equation}
This gives (\ref{6.21aeqa}) is by symmetry of $\rho$.

(ii) This is a immediate corollary from (i) and $\tilde X=G*\tilde
F$. The analyticity and the range of convergence follows from the
related statements for $\rho$. \kasten

Combination of Propositions \ref{6.4prop} and \ref{6.5prop} gives us

\begin{theorem}
\label{6.3theo}
$P^{\tilde F}$ is an extremal Gibbs measure.
\end{theorem}
\noindent {\bf Proof.}
Extremality of $P^{\tilde F}$ is equivalent to the ergodicity property
\begin{equation}
\label{6.21beqa}
\lim_{\Lambda\uparrow\R^d}{1\over|\Lambda|}\int_\Lambda\E_{P^{\tilde F}}[H_1H^a_2]\,da=\E_{P^{\tilde F}}[H_1]\E_{P^{\tilde F}}[H_2] ~~~\forall H_1,H_2\in L^2(\Gamma^c,P^{\tilde F}),
\end{equation}
where $H_2^a(\eta)=H_2(\eta_{\{1,a\}})$, $\eta\in\Gamma^c$, cf. e.g. \cite[Section 3.2]{Bat}. By approximation of both sides of (\ref{6.21beqa}) it is furthermore easy to see that it suffices to check (\ref{6.21beqa}) for $H_1$ and $H_2$ in a set that has dense algebraic span in $L^2(\Gamma^c,P^{\tilde F})$. Since the
two-sided Laplace-transform exists for $P^{\tilde F}$, functions of the form $H_p=\prod_{j=1}^{l_p}\tilde F(f_j^p)$, $p=1,2$,  with $f_l^p\in{\cal S}$ form such a set. On can thus use the cluster property of $\rho$, cf. Theorem \ref{6.1theo}, and (\ref{6.21aeqa}) as follows:

We note that the right hand side of (\ref{6.21aeqa}) is a sum over all partitions of $\{1,\ldots,l\}$ into disjoint sets.  By (\ref{6.21aeqa}) (see also \cite[Section 4.4.3]{Ru}) it is therfore sufficient to check that for
${\bf f}^a=(f_1^1,\ldots,f_{l_1}^1,f^2_{1,\{1,a\}},\ldots, f^2_{l_2,\{1,a\}})$ and ${\bf f}^p=(f_1^p,\ldots,f_{l_p}^p)$, $p=1,2$ we have
$\lim_{\Lambda\uparrow\R^d}{1\over|\Lambda|}\int_\Lambda\rho({\bf f}^a_I)da=0$ if $I=\{I_1,\ldots,I_j\}$ s.t. $\exists I_q\in I$ with $I_q\cap\{1,\ldots,l_1\}\not =\emptyset$ and $I_q\cap\{l_1+1,\ldots,l_1+l_2\}\not=\emptyset$ and
$\lim_{\Lambda\uparrow\R^d}{1\over|\Lambda|}\int_\Lambda\rho({\bf f}^a_I)da=\rho({\bf f}^1_{I^1})\rho({\bf f}^2_{I^2})$ with $I^1=\{I_q\in I:I_q\subseteq\{1,\ldots,l_1\}\}$, $I^2=\{I_q-l_1:I_q\subseteq\{l_1+1,\ldots,l_1+l_2\}\}$ otherwise.

The second condition is just the cluster property of $\rho$, cf. Proposition \ref{6.4prop}. To verify the first condition, one can look into the definition of ${\bf f}_I^a$ and $\rho({\bf f}^a_I)$ to see (using also Ruelle bounds) that $\rho({\bf f}^a_I)\to 0$ 
faster than any inverse power of $|a|$, which implies the first condition. In fact, in that case at least one (non-translated) $f^1_j$ and one (translated) $f^2_{l,\{1,a\}}$ are evalated w.r.t. he same integration variable. The product of these two functions thus decreases rapidly, if $a$ gets large.   
\kasten

\

{\noindent\it 6.6 An alternative construction using 'duality' }

\noindent Here we prove an equality between correlation
functionals of interacting particles systems and characteristic
functionals\footnote{We are grateful to an anonymous referee for
pointing out to us that the 'duality transformation' discussed
here has already been considered by V. Shkripnik from Kiev in some
unpublished preprints in the 1970ies.} of interacting CPNs making
it possible to apply the results on the infinite volume for the
correlation functional from Section 6.3 to the removal of the
infra-red cut-off for the characteristic functional. The results
obtained here are somewhat weaker than those obtained by the
detour through correlation functionals, as in the previous
section. But still we hope that this new way of performing the
thermodynamic limit for the measure is of independent interest, as
it e.g. gives a direct perturbative control over the
characteristic function.

 We restrict to  trigonometric interactions given
by energy densities $v(t)=\int_{[-c',c']}(1-e^{i\alpha
t})\,d\nu(\alpha)$ where $\nu$  is a symmetric probability measure
on $[-c',c']$ for some $c'>0$. Furthermore, in this subsection we
also assume that the L\'evy measure $r$ is symmetric. Clearly, in
that case the formal Potts model of Section 6.2 becomes a real
Potts model (however still at imaginary temperature) and one
expects a specific symmetry or 'duality' depending on the
question, whether the Potts model is projected to its first or its
second component.

The following technical lemma states that there is a pointwise
definition $X(x)$ of the convoluted Poisson noise $X$ and that
also the trigonometric interactions $v(X(x))$ have a pointwise
meaning:

\begin{lemma}
\label{6.2lem} Let $X_\Lambda=G*F_\Lambda$ be a CPN with assumptions on $G$ and $F_\Lambda$ as above.
For $p\geq 1$, the mapping $\R^d\times\R\ni (x,\alpha)\to e^{i\alpha X_\Lambda(x)}\in L^p({\cal S}',P^{X_\Lambda})$ is
continuous.
\end{lemma}
\noindent {\bf Proof.} We prove that $e^{i\alpha X_\Lambda(x)}$ is well-defined for $(x,\alpha)\in\R^d\times\R$. Continuity can be proven in an
analogous manner. Let $X_{\Lambda,\epsilon} =\chi_\epsilon *X_\Lambda$ be a ultra-violet regularization of $X_\Lambda$, cf. Section 5.3. Without loss of generality
we restrict ourselves to the case $p=2n$, $n\in\N$. Then, for $\epsilon,\epsilon'>0$,
\begin{eqnarray}
\label{6.22eqa}
&&\E_{P^{X_\Lambda}}\left[ \left| e^{i\alpha X_{\Lambda,\epsilon}(x)}-e^{i\alpha X_{\Lambda,\epsilon'}(x)}\right|^p\right]\nonumber\\
&=& \E_{P^{X_\Lambda}}\left[ \left( e^{i\alpha X_{\Lambda,\epsilon}(x)}-e^{i\alpha X_{\Lambda,\epsilon'}(x)}\right)^n \left( e^{-i\alpha X_{\Lambda,\epsilon}(x)}-e^{-i\alpha X_{\Lambda,\epsilon'}(x)}\right)^n\right]\nonumber\\
&=&\sum_{j,l=0}^n\left( {n\atop j}\right)\left({n\atop l}\right)(-1)^{l+j}\E_{P^{X_\Lambda}}\left[ e^{i\alpha[(n-l)X_{\Lambda,\epsilon}(x)+lX_{\Lambda,\epsilon'}(x)-(n-j)X_{\Lambda,\epsilon}(x)-jX_{\Lambda,\epsilon'}(x)]}\right]\nonumber \\
&=&\sum_{j,l=0}^n\left( {n\atop j}\right)\left({n\atop l}\right)(-1)^{l+j}\,\, e^{\int_{\Lambda-x} \psi(\alpha(j-l)G*(\chi_\epsilon-\chi_{\epsilon'})(y))\, dy}
\end{eqnarray}
We note that $\sum_{j,l=0}^n ( {n\atop j})({n\atop j})(-1)^{l+j}=((1-1)^n)^2=0$. In order to prove that $e^{i\alpha X_{\Lambda,\epsilon}(x)}$ forms a Cauchy sequence in $L^p({\cal S}',P^{X_\Lambda})$ it is thus
sufficient to show that the integrals on the right hand side of (\ref{6.22eqa}) vanish as $\epsilon,\epsilon'\downarrow 0$. Clearly, $|\int_{\Lambda-x} \psi(\alpha(j-l)G*(\chi_\epsilon-\chi_{\epsilon'})(y))
\, dy|\leq zcn|\alpha|\int_{\R^d}|G*\chi_\epsilon-G*\chi_{\epsilon'}|\,dy$
and now the assertion of the lemma follows from $G*\chi_\epsilon\to G$ in $L^1(\R^d,dx)$ as $\epsilon\downarrow 0$ (the latter again is a consequence of the Riesz convergence theorem \cite{Ba}).
 \kasten

In particular, Lemma \ref{6.2lem} shows that the characteristic
functional ${\cal C}_X$ of $X$ can be extended to the space
$\Gamma_0=\cup_{c>0}\Gamma_0^c$ of finite configurations. Since
the measure for the interacting CPN $\bar X$, $P^{\bar
X_\Lambda}$, is absolutely continuous w.r.t. $P^X$, the same holds
for the characteristic functional ${\cal C}_{\bar X_\Lambda}$. Let
us now recall all the data entering in the definition of $\bar
X_\Lambda$ (cf. the left row of Table 1). We say that an
interacting particle system $\tilde F_\Lambda$ is
\underline{dual}\footnote{There is a conceptual difference between
this notion of "duality"
 and the notion of $\tilde F_\Lambda$ being associated to $\tilde X_\Lambda$.} to the interacting CPN $\bar X_\Lambda$ if the defining
 data for $\tilde F_\Lambda$ can be obtained from the defining data of $\bar X_\Lambda$ according to Table 1. The following theorem clarifies
 the sense of this notion:
 \begin{table}
\begin{center}
\begin{tabular}{||l|l|l||}
\hline\hline
& $\bar X_\Lambda$ interacting CPN & $\tilde F_\Lambda$ interacting particle system \\ \hline
1. & activity $z$ & inverse temperature $\beta$ \\\hline
2. & inv. temperature (coupling const.) $\beta$& activity $z$  \\ \hline
3. & IR cut-off and finite volume $\Lambda$ & finite volume and IR-cut off $\Lambda$ \\ \hline
4. & L\'evy measure $r$ & interaction measure $\nu$ \\ \hline
5. & interaction measure $\nu$ & L\'evy measure $r$\\ \hline
6. & integral kernel $G$ & integral kernel $G$ \\
\hline\hline
\end{tabular}
\end{center}
\caption{ Identifications for the characteristic--correlation
functional duality between 'Poisson' quantum fields and
interacting particle systems for the case of trigonometric
interactions}
\end{table}
\begin{theorem}
\label{6.4theo} Let $\bar X_\Lambda$ be an interacting CPN with a
 trigonometric interaction and let $\tilde F_\Lambda$ be dual to
$\bar X_\Lambda$. Then, the characteristic functional of $\bar
X_\Lambda$ and the correlation functional of $\tilde F_\Lambda$
are related via ${\cal C}_{\bar
X_\Lambda}(\eta)=\rho_\Lambda(\eta)$ $\forall \eta \in \Gamma_0$,
$\supp \eta\subseteq \Lambda$.
\end{theorem}
\noindent {\bf Proof.} By Lemma \ref{6.2lem} and the estimate $|V_\Lambda|<2|\Lambda|$ we get that all expressions in the following chain of
equations are well-defined:
\begin{eqnarray}
\label{6.23eqa}
{\cal C}_{\bar X_\Lambda}(\eta)&=&{1\over\Xi_\Lambda}\, \E_{P^X}\left[e^{iX(\eta)}e^{-\beta V_\Lambda}\right]\nonumber\\
&=& {e^{-\beta |\Lambda|}\over \Xi_\Lambda}\sum_{n=0}^\infty {(-\beta)^n\over n!}\E_{P^X}\left[e^{iX(\eta)}(V_\Lambda-|\Lambda|)^n\right]\nonumber\\
&=& {e^{-\beta|\Lambda|}\over \Xi_\Lambda} \sum_{n=0}^\infty{\beta^n\over n!}\int_{\Lambda^n\times [-b,b]^{\times n}}\nonumber \\
&\times& e^{-z\int_{\R^d}\int_{[-c,c]}[1-e^{i s(G*\eta(y)+\alpha_1G(y-y_1)+\cdots +\alpha_nG(y-y_n))}]dr(s)\, dy}\nonumber \\
&\times& dy_1\cdots dy_n\, d\nu(\alpha_1)\cdots d\nu(\alpha_n)
\end{eqnarray}
Defining $U_\Lambda(\eta)=\int_{\Lambda}v_r(G*\eta)\, dx$ with $v_r(t)=\int_{[-c,c]}(1-e^{its})dr(s)$ we now get the statement of the theorem
comparing the right hand side of (\ref{6.23eqa}) and
the defining equation (\ref{6.2eqa}), cf. also (\ref{3.2eqa}). That here the partition function $\Xi_\Lambda$ of the interacting CPN is equal to the partition function
$\tilde \Xi_\Lambda$ can be seen by an analogous argument. \kasten

Theorem \ref{6.4theo} generalizes the well-known connection of
trigonometric interactions and particle systems with certain pair
interactions (equivalence of massive / massless sine-Gordon model
and Yukawa / Coulomb gas, respectively), see e.g.
\cite{AHK1,AHK2,F} and references therein. In fact, in the
ultra-violet regularized case one can obtain this classical
'duality' from Theorem \ref{6.3theo} by a scaling in the spirit of
Corollary \ref{7.1cor} below, see also Section 7.4.

From Theorem \ref{6.4theo} we get that for interacting CPNs with negative trigonometric interactions the high-temperature ('Feynman') expansion of the characteristic
functional is equivalent to the low activity expansion of the dual particle system and vice versa. Hence the results of Section 6.3 carry over to characteristic functionals of the
fields in the following way:

\begin{corollary}
\label{6.1cor}
Let $\beta$ and $z$ be the coupling constant and the activity of the interacting CPN $\tilde X_\Lambda$. If $|\beta|<1/(eC_1)$ and $|z|<1/(eC_2)$ with $C_1,C_2$ as in (\ref{6.7eqa}) and (\ref{6.8eqa}). Then

\noindent (i) ${\cal C}_{\tilde X}(\eta)=\lim_{\Lambda\uparrow\R^d}{\cal C}_{\tilde X_\lambda}(\eta)$ exists for $\eta\in\Gamma_0$ and is analytic in $z$ and $\beta$;

\noindent (ii) ${\cal C}_{\tilde X}(\eta)$ is continuous at zero
in the sense that ${\cal C}_{\tilde
X}(\sum_{l=1}^n\alpha_l\delta_{x_l})\to 1$ if
$\alpha_1,\ldots,\alpha_l\linebreak \to 0$;

\noindent (iii) ${\cal C}_{\tilde X}:\Gamma_0\to\C$ hence defines
a projective family of measures $(P^{\tilde X}_J)_{J\subseteq\R^d
\mbox{ \rm finite}}$;

\noindent (iv) There exists a canonical measure $P^{\tilde X}_{\rm can.}$ on the space of functions $\omega:\R^d\to\R$ equipped with the sigma-algebra generated by pointwise evaluation $f\to f(x)$.
The infinite volume interacting CPN $\tilde X(x)(\omega)=\omega(x)$ can be seen as the canonical process of $P^{\tilde X}_{\rm can.}$ in the above sense.
\end{corollary}
\noindent{\bf Proof.} (i) This follows from Theorem \ref{6.1theo} and Theorem \ref{6.3theo}.

(ii) One can use the representation through the dual correlation functional and one obtains the following uniform estimate
\begin{eqnarray*}
|\rho_\Lambda(\eta)-1|&=&\left|{1\over\Xi_\Lambda}\E_{P^F}\left[e^{-\beta U_\Lambda(\eta+F_\Lambda)}-e^{-\beta U_\Lambda(F_\Lambda)}\right]\right|\\
&\leq&{1\over\Xi_\Lambda}\E_{P^F}\left[\left|e^{-\beta U_\Lambda(\eta+F_\Lambda)-\beta U_\Lambda(F_\Lambda)}-1\right|e^{-\beta U_\Lambda(F_\Lambda)}\right]\\
&\leq&\sup_{\gamma\in\Gamma_0}\left|e^{-\beta U_\Lambda(\eta+\gamma)-\beta U_\Lambda(\gamma)}-1\right|\leq e^{\beta b \|G\|_1\sum_{l=1}^n|\alpha_l|}-1
 \end{eqnarray*}
 (iii) Thus, ${\cal C}_{\tilde X}$ defines a family of positive definite (as the limit of positive definite functions) and continuous functions that obviously generates a projective
 family of finite dimensional distributions $P_J^{\tilde X}$ of the random vectors $(X(x_1),\ldots,X(x_n))$ for $J=\{x_1,\ldots,x_n\}$ (again, the projectivity property is evident for $\Lambda\subseteq\R^d$ compact and it survives the limit as the vectors converge in
distribution by L\'evy's theorem).

(iv) follows from (iii) and Kolmogorov's theorem on the existence of the inductive limit $P_{\rm can.}^{\tilde X}$ of the family of finite dimensional distributions.
\kasten

\section{The continuum limit}
In this section we discuss the continuum scaling limit of interacting particle systems
with infra-red cut-off\footnote{Working with finite volume instead of an infra-red cut-off would lead to
Gaussian tail fields outside this volume, which would lead to misleading "tail-effects" in the scaling.}. On the level of interacting CPNs, this scaling can be seen as a kind of
implementation of the renormalization group.

\

{\noindent \it 7.1 Scaling limits}

\noindent Here we first consider the situation of a gas of
charged, noninteracting particles. The number of positiveandnegative charges is assumed to be equal in average, hence the gas macroscopically is neutral.
If we let the number of particles per unit volume (the activity $z$) go to infinity
$s$ scale charges with a factor $s\to s/\sqrt{z}$, we obtain the so-called \underline{continuum limit}.
See e.g. \cite{DMP} for an overview over the scaling of particle systems.

Let $r$ be a probability measure on $[-c,c]$ s.t. $r\{0\}=0$. For $0<\lambda<\infty$ and $A\subseteq \R$ measurable,
we define $r_\lambda(A)=r(\lambda A)$. Let $F$ be the Poisson noise determined by the L\'evy measure $r$ and the activity $z=1$,
cf. Eqs. (\ref{2.1eqa}) and (\ref{2.2eqa}). We then denote\footnote{The superscript $z$ in this section is used in a different sense than in
Section 6, since there the charges remained unscaled.} the Poisson noise determined by the L\'evy measure
$r_{1/\sqrt{z}}$ and activity $z\geq 1$ by $F^z$. Throughout the section we assume $\int_{[-c,c]}s\,dr(s)=0$. We also set
$\sigma^2=2\int_{[-c,c]}s^2\,dr(s)$ and $\psi_z(t)=z\int_{[-c/\sqrt{z},c/\sqrt{z}]}(e^{ist}-1)\, dr_{1/\sqrt{z}}(s)$. Finally, by $F_g^\sigma$ we denote the Gaussian noise with intensity $\sigma>0$ (cf. (\ref{2.1eqa})--(\ref{2.2eqa}))
and we write $X^z=G*F^z$, $X_g^\sigma=G*F_g^\sigma$ for the associated convoluted Poisson and Gaussian noise, respectively.
The basic facts on the continuum limit are given by the following proposition:

\begin{proposition}
\label{7.1prop}
With definitions as above we get

\noindent (i) $F^z\stackrel{\cal L}{\to}F_g^\sigma$ as $z\to\infty$;

\noindent (ii) $X^z\stackrel{\cal L}{\to}X_g^\sigma$ as $z\to\infty$.
\end{proposition}
{\bf Proof.}  We have (i) $\Leftrightarrow$ (ii) and it therefore suffices to prove the first statement. By L\'evy's theorem
convergence in law is equivalent with the convergence of characteristic functionals. It is thus sufficient to prove (cf. (\ref{2.1eqa})--(\ref{2.2eqa}))
$\int_{\R^d}\psi_z(f)\, dx\to\int_{\R^d}(-{\sigma^2\over 2}f^2)\, dx$ as $z\to\infty$ $\forall f\in{\cal S}$. Since
$\psi_z(t)\to-{\sigma^2\over 2}t^2$ as $z\to\infty$ we have pointwise convergence and since $|\psi_z(f)|\leq {\sigma^2\over 2}f^2$ $\forall z\geq 1$ the
statement follows by dominated convergence.
\kasten

We recall from Section 2.3 that Proposition \ref{7.1prop} is of particular interest in the case $G=G_{\alpha,m_0}$, cf. Prop. \ref{2.1prop},
since then $X_{\alpha,m_0,g}^\sigma=G_{\alpha,m_0}*F_g^\sigma$ is a generalized free field for $0<\alpha<1/2$ and is Nelson's free field
of mass $m_0>0$ for $\alpha=1/2$.

Next we investigate the effect of a length scale transformation $x\to\lambda x$, $x\in\R^d$, $0<\lambda<\infty $, on
the Poisson noise $F$ and the CPN $X_{\alpha,m_0}=G_{\alpha,m_0}*F$, respectively.
\parbox{7cm}{
The basic observation is that
increasing the activity can be performed by a scaling of the length ($z\sim\lambda^d$), cf. Fig. 3.

Also one has to take into account that for a locally finite marked configuration we have a scaling dimension $\lambda^{-d}$, since $\delta(\lambda x)=\lambda^{-d}\delta(x)$.
To obtain the same scaling as in Proposition \ref{7.1prop}, we thus have to define
\begin{equation}
\label{7.1eqa}
F_\lambda(x)=\lambda^{d/2}F(\lambda x),
\end{equation}
$0<\lambda <\infty$, where this scaling relation has to be understood in the sense of distributions.
}
\hspace{1cm}
\parbox{6cm}{
\begin{center}
\framebox(5,5){
\begin{picture}(5,5)
\put(1.25,1.25){\framebox(2.5,2.5){$~$}}
\put(1.45,3){$\bullet$}
\put(0.25,0.73){$\bullet$}
\put(4.1,2.6){$\bullet$}
\put(1.27,3.78){$\bullet$}
\put(4.49,4.2){$\bullet$}
\put(2.1,4.8){$\bullet$}
\put(1.1,2.5){$\bullet$}
\put(2.25,3.12){$\bullet$}
\put(4.49,1){$\bullet$}
\put(0.44,2.13){$\bullet$}
\put(0.77,3.3){$\bullet$}
\put(2.92,4.01){$\bullet$}
\put(.91,0.18){$\bullet$}
\put(3.51,3.33){$\bullet$}
\put(4.1,2.21){$\bullet$}
\put(2.9,0.4){$\bullet$}
\put(2.41,1.73){$\bullet$}
\put(0.82,4.13){$\bullet$}
\put(1.75,1,75){$\Lambda$}
\put(1.75,0.5){$\lambda\cdot\Lambda$}
\end{picture}
}
\end{center}
{\small Figure 3: The average number of particles in a region $\Lambda$ is proportional to $|\Lambda|$. A scaling $\Lambda\to\lambda\Lambda$
thus scales the activity by a factor $\lambda^d$. Here $d=2$, $\lambda=2$.}
}
\setcounter{figure}{3}
\begin{proposition}
\label{7.2prop}
With definitions as above

\noindent (i) $F_\lambda\stackrel{\cal L}{=}F^z$ for $z=\lambda^d$;

\noindent (ii) For $X_{\alpha,m_0,\lambda}(x)=\lambda^{(d-4\alpha)/2}X_{\alpha, m_0/\lambda}(\lambda x)$ we get
$X_{\alpha,m_0,\lambda}\stackrel{\cal L}{=}G_{\alpha,m_0}*F_\lambda$;

\noindent (iii) For $0<\alpha\leq 1/2$, $X_{\alpha,m_0,\lambda}\stackrel{\cal L}{\to}X_{\alpha,m_0,g}^\sigma$ as $\lambda\to\infty$ where
the latter is a (generalized) free field.
\end{proposition}
\noindent {\bf Proof.} (ii) follows from (i) and Proposition \ref{2.1prop} (v). (iii) follows from (i), (ii) and
Proposition \ref{7.1prop}. To prove (i) let $f^\lambda(x)=f(x/\lambda)$, $f\in{\cal S}$, $x\in\R^d$, $\lambda>0$.
Then, by (\ref{7.1eqa}), $\langle F_\lambda,f\rangle= \lambda^{-d/2}\langle F,f^\lambda\rangle$. Thus,
\begin{eqnarray}
\label{7.2eqa}
{\cal C}_{F_\lambda}(f)&=&e^{\int_{\R^d}\int_{[-c,c]}(e^{is\lambda^{-d/2}f^{\lambda}(y)}-1)\, dr(s)\, dy}\nonumber\\
&=&e^{\lambda^d\int_{\R^d}\int_{[-\lambda^{-d/2}c,\lambda^{-d/2}c]}(e^{isf(y)}-1)\,dr_{\lambda^{-d/2}}(s)\, dy}, ~~f\in{\cal S},
\end{eqnarray}
and the claim follows from Eqs. (\ref{2.1eqa}) and (\ref{2.2eqa}).
\kasten

The scaling \ref{7.2prop} (iii) is of the same form as for block-spin transformations implementing the
renormalization group for lattice systems, \cite{FFS}. In the general sense, that the renormalization group
is a scaling limit adding more and more "microstructures" to a given region, we can say that the continuum limit
for the models studied in this article is a suitable formulation of the renormalization group.

\begin{remark}
\label{7.1rem}
{\rm
(i) It is an interesting fact that it is just property (ii) of Remark \ref{2.1rem} which prevents us
from taking a pointwise continuum limit: If we have $G\in L^2(\R^d,dx)$, then the i.i.d. variables
$Z_j(x)=S_jG(x-Y_j)$ with $S_j,Y_j$ distributed as $r\otimes dx|_\Lambda/|\Lambda|$ for a finite volume $\Lambda\subseteq\R^d$
have finite variance ${\sigma^2\over 2}\int_\Lambda|G(x-y)|^2\, dy<\infty$ and therefore fulfill the requirements of
the central limit theorem. Under such conditions, the quantity $X_{\Lambda}^z(x)=\sum_{j=1}^{N^z_\Lambda}Z_j(x)/\sqrt{z}$ ($N_\Lambda^z$
being a Poisson random variable with intensity $z|\Lambda|$) converges
in law to a Gaussian random variable, and one can thus expect a pointwise definition of the process
$X_{\Lambda,g}^\sigma(x)$. If however $G\not\in L^2(\R^d,dx)$, as it is the case in for the examples relevant for QFT,
then the variance of $Z_j(x)$ is infinite. Heuristically speaking, $X_\Lambda^z(x)$ then converges to
a "Gaussian random variable with infinite fluctuations" -- thus there is no pointwise limit. Ultra-violet divergences
and renormalization in these cases have to be taken into account. In the case $d=2$, $G=G_{1/2}$, the
variance of $Z_j(x)$ only diverges logarithmically, which already gives a hint that ultra-violet divergences in this
specific case\footnote{This is the standard case considered usually in constructive QFT in two dimensions, see \cite{AFHKL,GJ,Si}.} will be rather mild.

\noindent (ii) From the above discussion it clear that the Gaussian (continuum) limit can also be taken in the canonical
ensemble (CE) by replacing $N_\Lambda$ with it's expectation $|\Lambda|$. Interactions for the CE can be defined as in Section 4.
 It is however open, whether also the analytic continuation
\cite{AGW1} can be performed in the CE. On the other hand, the CE is of advantage if one wants to work with potentials
which might not be stable.
}
\end{remark}

{\noindent \it 7.2 The continuum limit for trigonometric interactions with ultra-violet cut-off}

\noindent Here we study the continuum limit of  CPNs with ultra-violet and infra-red regularized
bounded interactions and we show convergence in law to the corresponding
perturbed Gaussian models.

Let $G^\epsilon$ be a ultra-violet regularization
of the kernel $G$ (cf. Section 5.3). $\Lambda\subseteq\R^d$ is assumed to be compact. Let $X^{z}_\epsilon=G^\epsilon *F^z$ and
$X_{\epsilon}^\sigma=G^\epsilon*F^\sigma$. Here we dropped the superscripts $g$ for notational simplicity and we adopt the convention
that a (convoluted) noise with superscript $\sigma$ is Gaussian. It is clear that $X_\epsilon^\sigma=\chi^\epsilon*X^\sigma$ has paths in the set $C^\infty(\R^d,\R)$.
For $v:\R\to\R$ being measurable and bounded (by a constant $a>0$) we can thus define the potentials
\begin{equation}
\label{7.3eqa}
V_{\Lambda,\epsilon}^{z/\sigma}=\left\langle v(X_\epsilon^{z/\sigma}),1_\Lambda\right\rangle.
\end{equation}
and measures (for $\beta >0$, we also note that $\left|V_{\Lambda,\epsilon}^{z/\sigma}\right|\leq a|\Lambda|$ a.s.)
\begin{equation}
\label{7.4eqa}
P^{\bar X^{z/\sigma}_{\epsilon,\Lambda}}={e^{-\beta V_{\Lambda,\epsilon}^{z/\sigma}}\over \Xi_{\Lambda,\epsilon}^{z/\sigma}}\, P^{X^{z/\sigma}_\epsilon},
~~\Xi_{\Lambda,\epsilon}^{z/\sigma}=\E_{P^{X_\epsilon^{z/\sigma}}}\left[ e^{-\beta V_{\Lambda,\epsilon}^{z/\sigma}}\right].
\end{equation}
Let $\bar X_{\epsilon,\Lambda}^{z/\sigma}$ be the associated coordinate processes. We now obtain the same result as Proposition \ref{7.1prop} for
the perturbed models:

\begin{theorem}
\label{7.1theo}
 $\bar X_{\epsilon,\Lambda}^{z}\stackrel{\cal L}{\to}\bar X_{\epsilon,\Lambda}^{\sigma}$ as $z\to\infty$.
\end{theorem}
\noindent {\bf Proof.} As convergence in law is equivalent with the convergence of characteristic functionals, we have
to prove ${\cal C}_{\bar X_{\epsilon,\Lambda}^{z}}(f)\to{\cal C}_{\bar X_{\epsilon,\Lambda}^{\sigma}}(f)$ $\forall f\in{\cal S}$.
Since $V^z_{\Lambda,\epsilon}$ is an uniformly (in $z$) bounded random variable, we get that
the expression
\begin{equation}
\label{7.5eqa}
{\cal C}_{\bar X_{\epsilon,\Lambda}^{z}}(f)={\sum_{n=0}^\infty{(-\beta)^n\over n!}\E_{P^{X_\epsilon^z}}\left[e^{iX_\epsilon^z(f)}\left(V_{\Lambda,\epsilon}^{z}\right)^n\right]
\over \sum_{n=0}^\infty{(-\beta)^n\over n!}\E_{P^{X_\epsilon^z}}\left[\left(V_{\Lambda,\epsilon}^{z}\right)^n\right]}
\end{equation}
converges to the related expression with $z$ replaced with $\sigma$ if all terms in the numerator and denominator converge separately.
Using Fubini's theorem we get for a term in the numerator
\begin{eqnarray}
\label{7.6eqa}
\E_{P^{X_\epsilon^z}}\left[e^{i X_\epsilon^z(f)}\left(V_{\Lambda,\epsilon}^{z}\right)^n\right]&=&
\int_{\Lambda^{\times n}}\E_{P^{X^z}}\bigg[ e^{i X^z(\chi_\epsilon*f)}\nonumber \\
&\times&
v(X^z(\chi_{\epsilon,y_1}))\cdots v( X^z(\chi_{\epsilon,y_n}))\bigg] dy_1\cdots dy_n
\end{eqnarray}
and the corresponding term in the denominator is obtained setting $f=0$. Here $\chi_\epsilon$ is the ultra-violet cut-off
function (cf. Section 5.3) and $\chi_{\epsilon,y}(x)=\chi_\epsilon(x-y)$. Since $\chi_{\epsilon,y}\in{\cal S}$ we now get
the pointwise convergence of the integrand on the right hand side of (\ref{7.6eqa}) to the related integrand with $z$ replaced with $\sigma$ from
the convergence in law of $X^z$, cf. Proposition \ref{7.1prop} (ii). Since the integrand is uniformly bounded by $(a|\Lambda|)^n$, convergence
of the right hand side of (\ref{7.6eqa}) then follows from dominated convergence.
\kasten

We want to modify this result in the following way: We replace the functions $v(t)$ in (\ref{7.3eqa}) with functions
$:v(t):^{z/\sigma}_{\epsilon}=\int_{[-b,b]}:\cos(\alpha t):^{z/\sigma}_\epsilon\, d\nu(\alpha)$ where
\begin{equation}
\label{7.7eqa}
:\cos(\alpha t):_\epsilon^{z/\sigma}=\cos(\alpha t)\left/
\E_{P^{X^{z/\sigma}_\epsilon}}[\cos(\alpha X_\epsilon^{z/\sigma}(x))]\right. ,~~x\in\R^d.
\end{equation}
Here $\nu$ is a finite, complex measure on $[-b,b]$ s.t. $\nu(A)=\overline{\nu(-A)}$ for $A\subseteq [-b,b]$ measurable.
These energy densities define the (ultra-violet regularized) trigonometric interactions \cite{AHK1,AHK2}. It is not difficult to prove that under the given
conditions $:v(t):^z_\epsilon$ is uniformly bounded (in $z$ and $t$) and $:v(t):_\epsilon^z\to :v(t):_\epsilon^\sigma$ uniformly in $t$ as
$z\to\infty$.  Thus, the proof of Theorem \ref{7.1theo} carries over to the modified interactions:

\begin{corollary}
\label{7.1cor}
Let $\bar X_{\Lambda,\epsilon}^z$ be the ultra-violet regularized interacting CPN with trigonometric interaction
specified as above and let $\bar X_{\Lambda,\epsilon}^\sigma$ be the related perturbed Gaussian model. Then the statement of
Theorem \ref{7.1theo} still holds.
\end{corollary}

{\noindent \it 7.3 Triviality for trigonometric potentials without renormalization}

\noindent We now want to consider the continuum limit without ultra-violet cut-off in the case of
trigonometric potentials without renormalization "$:~:_0^z$", i.e. we set the denominator in (\ref{7.7eqa})
equal to one: Let $v(t)=\int_{\R} \cos(\alpha t)\, d\nu(\alpha)$ for some finite, complex measure $\nu$ on $\R$
such that $\nu\{0\}=0$ and $\nu(A)=\overline{\nu(-A)}$ for $A\subseteq \R$ measurable. Let furthermore $r$ be symmetric,
$r(A)=r(-A)$, $A\subseteq [-c,c]$ measurable. In this case $\psi_z$ is real and $\psi_z(t)\leq 0$. We chose $G\in L^1(\R^d,dx)$
such that $G\not \in L^2(\R^d,dx)$, cf. Remarks \ref{2.1rem} (ii) and \ref{7.1rem} (i) for the motivation. Finally,
we define $V_\Lambda^z$ as in Eq. (\ref{7.3eqa}) with $\epsilon=0$ and by Theorem \ref{3.2theo} we get that this is well-defined (since
$\nu$ is finite, $v$ is bounded). We get the following lemmma:

\begin{lemma}
\label{7.1lem}
$\|V_\Lambda^z\|_{L^2({\cal S}',P^{X^z})}\to0$ as $z\to\infty$.
\end{lemma}
\noindent {\bf Proof.} We get by Fubini's theorem for bounded functions
\begin{equation}
\label{7.8eqa}
\E_{P^{X^z}}\left[\left|V_\Lambda^z\right|^2\right]=\int_{\Lambda^{\times 2}\times \R^2}\E_{P^{X^z}}\left[ e^{i(\alpha_1 X^z(y_1)+\alpha_2X^z(y_2))}\right]dy_1dy_2d\nu(\alpha_1)d\nu(\alpha_2)
\end{equation}
with (cf. Eq. (\ref{2.1eqa}) and Lemma \ref{6.2lem})
\begin{equation}
\label{7.9eqa}
\E_{P^{X^z}}\left[ e^{i(\alpha_1 X^z(y_1)+\alpha_2X^z(y_2))}\right]=e^{\int_{\R^d}\psi_z(\alpha_1G(x-y_1)+\alpha_2G(x-y_2))dx}
\end{equation}
and the integral in the exponent on the right hand side exists for $0<z<\infty$, since $|\psi_z(t)|\leq c\sqrt{z}|t|$.

If we can show that the right hand side of (\ref{7.9eqa}) vanishes $dy_1dy_2d\nu(\alpha_1)d\nu(\alpha_2)$ a.e., we get the statement of the lemma by dominated convergence (since $\psi_z\leq 0$).
Let $g(t)=\psi_z(\sqrt{z}t)/zt^2$. One easily verifies that $g$ is continuous and $g(0)=-\sigma^2/2$. By Fatou's lemma we get for $\alpha_1,\alpha_2\not=0$ and $y_1\not=y_2$
\begin{eqnarray}
\label{7.10eqa}
&&\limsup_{z\to\infty}\int_{\R^d}\psi_z(\alpha_1G(x-y_1)+\alpha_2G(x-y_2))dx\nonumber \\
&&\hspace{2cm}=\limsup_{z\to\infty}\int_{\R^d}g([\alpha_1G(x-y_1)+\alpha_2G(x-y_2)]/\sqrt{z})\nonumber \\
&&\hspace{5cm}\times (\alpha_1G(x-y_1)+\alpha_2G(x-y_2))^2 dx\nonumber \\
&&\hspace{2cm}\leq \int_{\R^d}\limsup_{z\to\infty} g([\alpha_1G(x-y_1)+\alpha_2G(x-y_2)]/\sqrt{z})\nonumber\\
&&\hspace{5cm}\times(\alpha_1G(x-y_1)+\alpha_2G(x-y_2))^2\,dx \nonumber \\
&&\hspace{2cm}= -{\sigma^2\over 2}\int_{\R^d}(\alpha_1G(x-y_1)+\alpha_2G(x-y_2))^2\, dx=-\infty
\end{eqnarray}
This concludes the proof.
\kasten

Let $\bar X_\Lambda^z$ be the interacting CPN with infra-red cut-off $\Lambda$ associated to $V_\Lambda^z$. We then get:

\begin{theorem}
\label{7.2theo}
$\bar X_{\Lambda}^z\stackrel{\cal L}{\to}X^\sigma$ as $z\to\infty$, i.e. the limit is trivial (Gaussian).
\end{theorem}
\noindent {\bf Proof.} Again we have to show convergence of characteristic functionals. Let $f\in {\cal S}$, then
\begin{equation}
\label{7.11eqa}
\E_{P^{X^z}}\left[e^{iX^z(f)}e^{-\beta V_\Lambda^z}\right]={\cal C}_{X^z}(f)-\E_{P^{X^z}}\left[e^{iX^z(f)}\left( 1-e^{-\beta V_{\Lambda}^z}\right)\right]
\end{equation}
and from Lemma \ref{7.1lem} we get
\begin{eqnarray}
\label{7.12eqa}
\left|\E_{P^{X^z}}\left[e^{iX^z(f)}\left( 1-e^{-\beta V_{\Lambda}^z}\right)\right]\right|&\leq& \E_{P^{X^z}}\left[\left|1-e^{-\beta V_\Lambda^z}\right|^2\right]^{1/2}\nonumber \\
&\leq& C \E_{P^{X^z}}\left[\left|V_\Lambda^z\right|^2\right]^{1/2}\to 0 \mbox{ as }z\to\infty
\end{eqnarray}
where $C=\beta|\Lambda||\nu|(\R)e^{\beta|\Lambda||\nu|(\R)}$. Likewise one can show that $\Xi^z_\Lambda\to 1$ as $z\to\infty$. The statement now follows from Proposition \ref{7.1prop} (ii).
\kasten

\begin{figure}[htb]
\centerline{
\includegraphics[width=8cm]{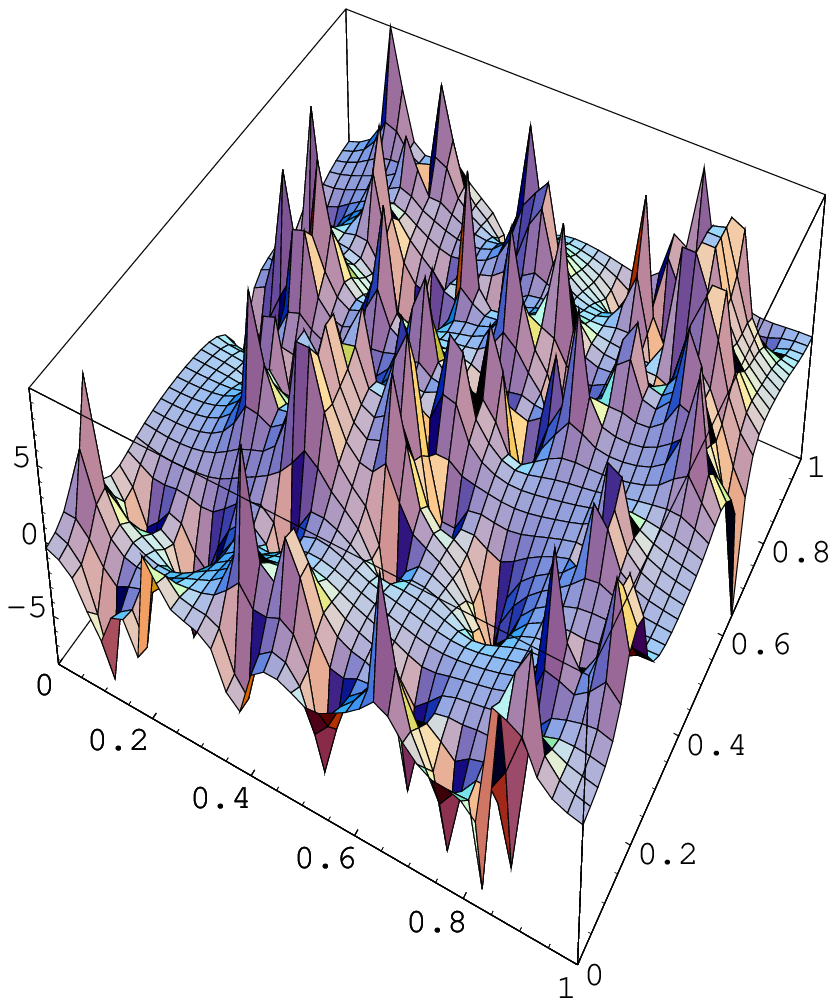}
}
\caption{Sample paths of $X^z$ in the unit cube for $z=100$ (see also Fig. \ref{1fig}). }
\label{4fig}
\end{figure}

\begin{remark}
\label{7.2rem}
{\rm
(i) Clearly, Theorem \ref{7.2theo} is what one would expect from the analysis of the sine-Gordon model \cite{F,FP,FS}:
The normal ordering $:\cos(\alpha t):_\epsilon^\sigma$, cf. (\ref{7.7eqa}), in this case can be understood as a renormalization
of the coupling constant, i.e. we chose the energy density $\cos(\alpha t)$ with coupling constant $\beta_\epsilon^\sigma=\beta_0/\E_{P^{X^\sigma_\epsilon}}[\cos(\alpha X^\sigma_\epsilon(x))]$
and one can easily check that $\beta_\epsilon^\sigma\uparrow \infty$ as $\epsilon\downarrow 0$. Since this coupling constant renormalization
leads to a well-defined limit potential, it is natural to expect that without renormalization of $\beta$ the limit is trivial. This is the same statement as in Theorem \ref{7.2theo},
where we however use the continuum limit $z\to\infty$ without ultra-violet cut-off instead of the limit $\epsilon\downarrow 0$. We will continue this discussion in the following subsection.

\noindent (ii) Even though Theorem \ref{7.2theo} does not come as a surprise,
it's interpretation is of some interest: If $z\to\infty$ the spatial fluctuation of sample paths of $X^z$ increase rapidly, cf. Fig. \ref{4fig}.
This leads to increasing oscillations of the function $\cos(\alpha X^z(x))$ and thus
$\int_\Lambda\cos(\alpha X^z(x))\, dx$  integrates out to zero as $z\to\infty$.

\noindent (iii) For a different approach to the triviality of the sine-Gordon model without renormalization, based on random Colombeau distributions, see \cite{AHR}.
}
\end{remark}

Another intresting approach to triviality in quantum field theory that is less motivated by trigonometric interactions than 
Remark \ref{7.2rem} (ii) but probably works for all bounded non-renormalized interaction densities $v$ is to look at the "spatial" properties of the sample paths as depending on the strength of the ultra-violet singularity. Plots as in
Fig. 3 at high scaling parameter are appropriate, cf. Fig. 5. Already for $z=1000$ one can see that in the ultra-violet finite case (Fig. 5a) long range "Gaussian tails" dominate the sample path. 
Hence the fuctuations of the potential energy prevail in the scaling limit. In contrast to this, the ultra-violet divergent case exposes a strong "localization" of the path properties due to
the "volume" of the singularities. Thus, each of the one hundred little squares with in average 10 particles is "approximately independent" from its neighbors and contributes an amount proportional to
the covered volume $\sim 1/100$. One thus recognizes the regime of the law of large numers and the convergence of the potential to a constant (i.e. triviality of the interaction) is expected. Again, the uv-critical case (Fig. 5b) is just the
uv-singularity strength of constructive quantum field theoryin $d=2$ dimensions.
  
\begin{figure}
\label{5fig}
\centerline{
\includegraphics[width=4.3cm]{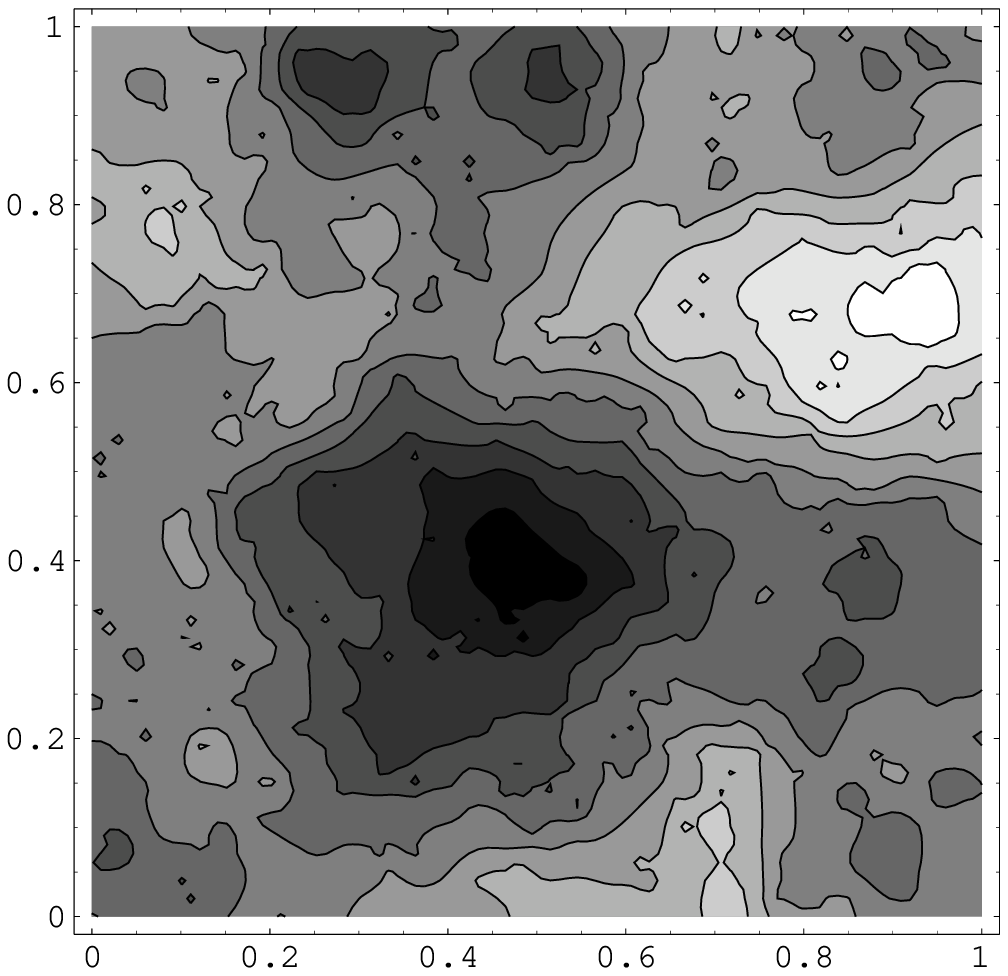}
\hspace{.3cm}
\includegraphics[width=4.3cm]{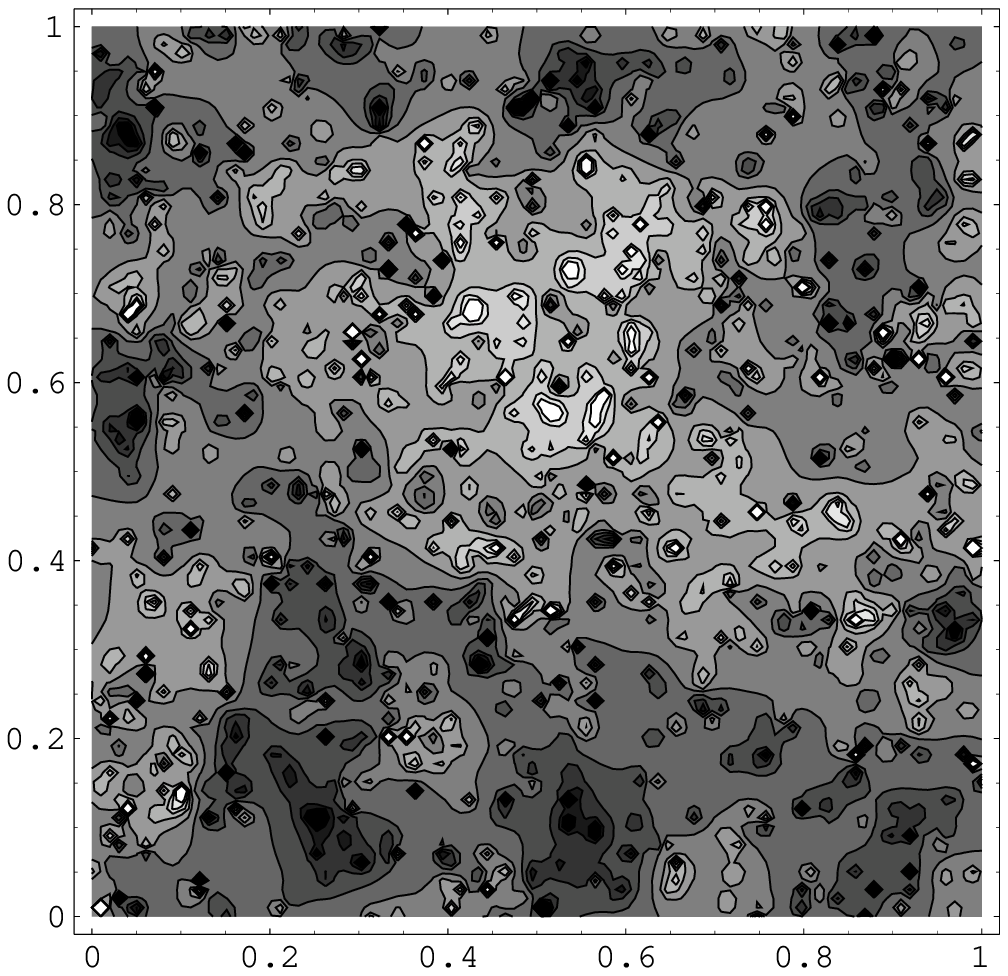}
\hspace{.3cm}
\includegraphics[width=4.3cm]{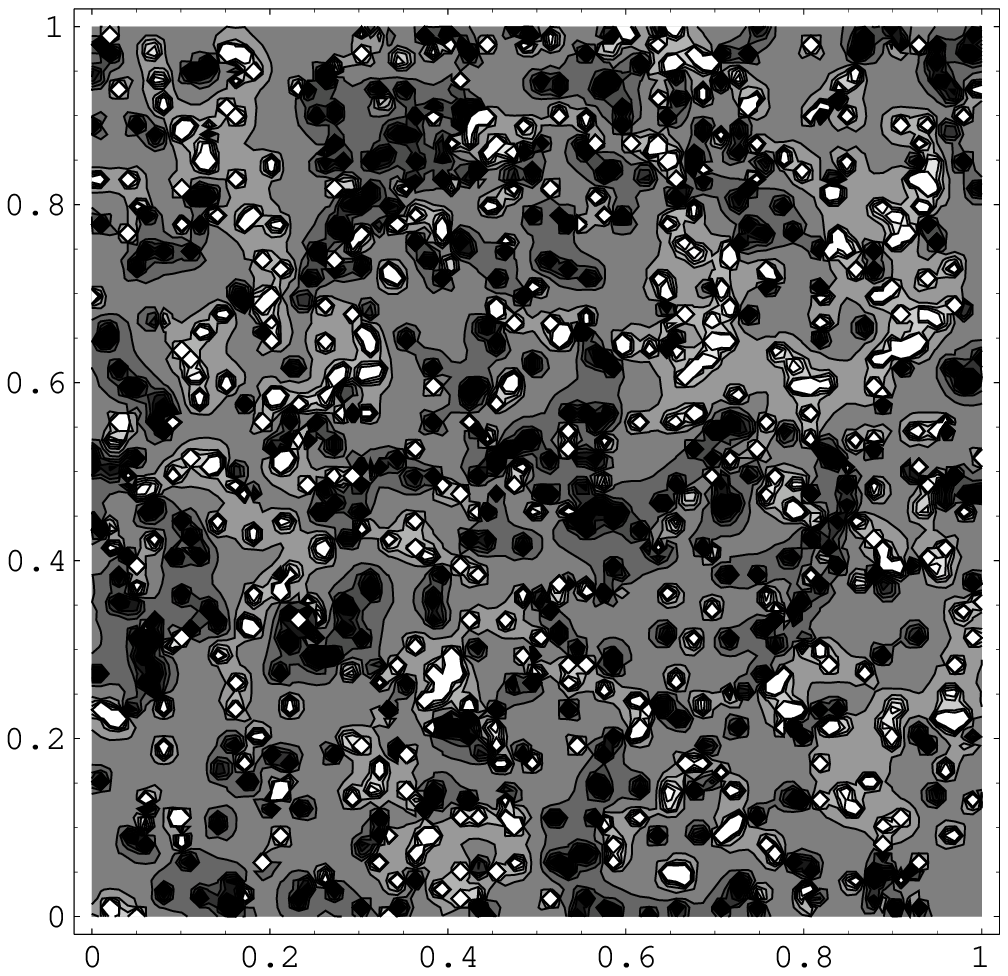}
}
 \caption{Density plot of the static field $\sim |x|^{-\alpha}$ of a two
dimensional system of $1000$ non-interacting charged particles a) for $\alpha =0.2$
(uv-finite scaling limit); b) $\alpha =1$ (uv-critical); c) $\alpha
=2$ (uv-divergent).}
\end{figure}

\

{\noindent \it 7.4 Some remarks on the continuum limit for the sine-Gordon model}

\noindent Here we give some remarks on the continuum limit for the sine-Gordon (sG) model in $d=2$ dimensions with coupling constant renormalization, namely we show that the Boltzmann weights
of the dual particle system converge to those of the Yukawa gas, which is dual the sine-Gordon model, see e.g. \cite{F}. We also comment on a simultanous expansion in the coupling constant
$\beta$ and a re-scaled activity $\zeta$ and we show that the continuum limit yields convergence in the expansion's coefficients. A treatment which goes beyond these
very preliminary results and investigates convergence in law of the 'Poissonian' sine-Gordon models under the continuum limit would be desirable. But the technical details
of such a treatment seem to be rather complicated,  as it is case for the proof of the ultra-violet stability of the classical sine-Gordon model \cite{DL,F,FP,FS}. It therefore goes beyond the scope of
the present article.

 We fix $G=G_{1/2,m_0}$, $G_1=G*G$, cf. Proposition \ref{2.1prop}. We consider the interacting CPN
with energy density $:\cos(\alpha t):^z=:\cos(\alpha t):^z_0$, cf. (\ref{7.7eqa}). Using the language of
particle systems, we define the potential for the dual particle system with external source $f$ as
\begin{eqnarray}
\label{7.13eqa}
U^{\rm sG}_z(f;y_1,\ldots,y_n;\alpha_1,\ldots,\alpha_n)&=&\int_{\R^d}-\psi_z(G*f(x)+\sum_{l=1}^n\alpha_lG(x-y_l))\nonumber\\
&+&\sum_{l=1}^n\psi_z(\alpha_lG(x-y_l))\, dx
\end{eqnarray}
where $f\in{\cal S}$, $y_1,\ldots,y_n\in\R^d$, $y_j\not=y_l$, $j\not=l$ and $\alpha_1,\ldots,\alpha_n\in\supp \nu\subseteq [-b,b]$.
Here the integrals of the second term in (\ref{7.13eqa}) do not depend on the $y_l$ and these terms arise from the coupling constant
renormalization (\ref{7.7eqa}). We also define
\begin{eqnarray}
\label{7.14eqa}
U^{\rm sG}_\sigma(f;y_1,\ldots,y_n;\alpha_1,\ldots,\alpha_n)&=&{\sigma^2\over 2}\bigg[f*G_1*f(0)+2\sum_{l=1}^n\alpha_lG_1*f(y_l)\nonumber\\
&+&\sum_{l,j=1\atop j\not=l}^n \alpha_j\alpha_l G_1(y_j-y_l)\bigg]
\end{eqnarray}
which for $f=0$ gives the Yukawa potential for particles with charges $\alpha_l$. We consider the \underline{$f$-dependent Boltzmann weights}
$e^{-\zeta U^{\rm sG}_z(f;y_1,\ldots,y_n;\alpha_1,\ldots,\alpha_n)}$ for the dual particle system of the interacting CPN and $e^{-\zeta U^{\rm sG}_\sigma(f;y_1,\ldots,y_n;\alpha_1,\ldots,\alpha_n)}$
for the Yukawa gas. Here $\zeta>0$ is an inverse temperature for the dual particle systems and hence is a scaling factor for the activity (the intensity $\sigma$, respectively) for the quantum field systems, cf. Theorem \ref{6.3theo}.
We get the following expansion in $\beta$ and $\zeta$ for the characteristic functional of $\bar X^{(z,\zeta)}_\Lambda$, defined as the interacting CPN with sG--interaction and L\'evy-characteristic $\zeta \psi_z$:
\begin{eqnarray}
\label{7.15eqa}
{\cal C}_{\bar X^{(z,\zeta)}_\Lambda}(f)&=&{1\over \Xi^{(z,\zeta)}_\Lambda}\sum_{l,n=0}^\infty{(-\zeta)^l(-\beta)^n\over l!\, n!}\int_{\Lambda^{\times n}\times[-b,b]^{\times n}}\nonumber \\
&\times&\left[ U_z^{\rm sG}(f;y_1,\ldots,y_n;\alpha_1,\ldots,\alpha_n)\right]^ldy_1\cdots dy_n\, d\nu(\alpha_1)\cdots d\nu(\alpha_n)\,.\nonumber \\
\end{eqnarray}
The related expansion for the partition function is obtained from the expansion of the
 numerator by setting $f=0$. From the fact that
$|U_z^{\rm sG}(f;y_1,\ldots,y_n;\alpha_1,\ldots,\linebreak \alpha_n)|\leq C(n,z,c)$
 where $C(n,z,c)$ is linearly bounded in $n$, we get that the expansion
 (\ref{7.15eqa}) converges absolutely for any fixed $z<\infty$, independently of the dimension $d$.

For $d=2$, the related expansion for the characteristic functional of the Gaussian sine-Gordon model exists term by term, which can be deduced from (\ref{7.14eqa}) and the fact that
$G_1(x)\sim -\ln|x|/2\pi$ for $|x|$ small. It is known for the special case $\nu=(\delta_b+\delta_{-b})/2$
that if we sum up over $l$ under the integral, then the series converges absolutely for any $\beta$ provided $0<\zeta<2/\sigma^2\sqrt{4\pi b}$, cf. \cite{F}. From the analysis of that model it seems to us that
after summing up $n$,
at most asymptotic convergence in $l$ can be expected, since ultra-violet divergences for $\zeta<0$ are more severe than in the case $\zeta>0$.
This can be explained from the fact that the Yukawa gas at negative temperatures becomes unstable.
Here we ignore the question of convergence and consider (\ref{7.15eqa}) as a formal power series in $\beta$ and $\zeta$.

\begin{proposition}
\label{7.3prop} With definitions as above

\noindent (i) The $f$-dependent Boltzmann weights (potentials $U_z^{\rm sG}$) of the dual particle system of the interacting CPN converge pointwisely to
the $f$-dependent Boltzmann weights (potentials $U_\sigma^{\rm sG}$) of the Yukawa gas as $z\to\infty$.

\noindent (ii) For $d=2$ the expansion (\ref{7.15eqa}) converges to the related expansion of the classical ('Gaussian') sine-Gordon model, where
$U_z^{\rm sG}$ is replaced by $U_\sigma^{\rm sG}$, in the sense of convergence of formal power series.
\end{proposition}
{\bf Proof.} (i) Using $\psi_z(t) \to -\sigma^2t^2/2$ as $z\to\infty$, it is elementary to show
\begin{equation}
\label{7.16eqa}
-\psi_z(\sum_{l=1}^nt_l+s)+\sum_{l=1}^n\psi_z(t_l)\to {\sigma^2\over 2}\left[\sum_{j,l=1\atop l\not=j}t_lt_j+2s\sum_{l=1}^nt_l+s^2\right] \mbox{ as } z\to\infty
\end{equation}
where $t_1,\ldots,t_n,s\in\R$. If we replace $t_l=\alpha_lG(x-y_l)$ and $s=G*f(x)$ we thus get the convergence of the left hand side of (\ref{7.16eqa}) to the right hand side whenever
$x\not=y_l$, $l=1,\ldots,n$. We note that under this replacement, the right hand side of (\ref{7.16eqa}) integrated over $\R^d$ w.r.t. $dx$ is just the
right hand side of (\ref{7.14eqa}). To prove the convergence of the right hand side of (\ref{7.13eqa}) to the right hand side of (\ref{7.14eqa}) for $y_j\not=y_l$, $l\not=j$, it is thus
sufficient to show that the integrand in (\ref{7.13eqa}) has an uniform (in $z$) $L^1(\R^d,dx)$-bound.

We note that $|\psi_z(t)|\leq \sigma^2t^2/2$ and $|\psi_z'(t)|\leq \sigma^2 |t|$ for all $z>0$. For $j=1,\ldots, n$ we thus get
that the modulus of the left hand side of (\ref{7.16eqa}) can be estimated as follows:
\begin{eqnarray}
\label{7.17eqa}
\ldots &=& \left| \int_0^1\psi'\big( \big[\sum_{l=1\atop l\not=j}^nt_l+s\big]u+t_j\big)\big[\sum_{l=1\atop l\not=j}^nt_l+s\big]-\sum_{l=1\atop l\not=j}\psi_z(t_j)\, du\right| \nonumber\\
&\leq& \sigma^2\sum_{l,p=1\atop l,p\not=j}^n|t_lt_p|+2\sigma^2\sum_{l=1\atop l\not= j}^n|t_ls|+s^2+{3\sigma^2\over 2}\sum_{l=1\atop l\not=j}^nt_l^2
\end{eqnarray}
If one replaces on the right hand side $t_l$ with $\alpha_l G(x-y_l)$ and $s$ with $G*f(x)$ one apparently gets a function of fast decay which is
locally integrable on $\R^d\setminus\cup_{l=1,l\not=j}^nB_{R_l^j}(y_l)$ with $R_l^j=|y_l-y_j|/2$ by our assumption
 $y_j\not=y_l$, $j\not=l$. A point $x\in\R^d$ is contained in such a set for $j$ s.t. the $|x-y_j|=\min\{|x-y_l|:l=1,\ldots,n\}$. Therefore, the union over $j=1,\ldots,n$ of all such
 sets gives $\R^d$ and there is a global $L^1(\R^d,dx)$-majorant.

 (ii) To obtain the convergence in terms of formal power series in (\ref{7.15eqa}) it suffices to prove the convergence of each expansion coefficient in the numerator
 and in the denominator (i.e. in the expansion of $\Xi^{(z,\zeta)}_\Lambda$), since the coefficients of the expansion of the fraction can be calculated from those of the numerator and denominator
 via a finite combinatorial expression (note that the zero order coefficient of the partition function is one). Furthermore, the calculation for the partition function is a special case of the calculation
 for the numerator, namely $f=0$, we only have to consider the latter.

 By (i) we have pointwise convergence of the integrands in (\ref{7.15eqa}). For $d=2$, $n$ and $l$ fixed, we can find a $L^l(\Lambda^{\times n}\times [-b,b]^{\times n},d^{2n}x\otimes r^{\otimes n})$-majorant by
 integrating the majorant constructed in (i) over $\R^2$ w.r.t. $dx$.  The fist term on the right hand side of (\ref{7.17eqa}) then gives rise to a term
 $\sum_{l,j=1,l\not=j}|\alpha_l\alpha_j|G_1(y_l-y_j)$ which is $L^p$-integrable in the variables $y_1,\ldots,y_l$ for any $p\geq 1$ since $G_1(x)\sim -\ln|x|/2\pi$ for small $x$. The terms involving $s$ and $s^2$ in
 (\ref{7.15eqa}) trivially have the same property, since under the replacements as above the integration over $dx$ can be estimated by $\sum_{l=1}^n|\alpha_l|G_1*|f|(y_l)$ and $|f|*G_1*|f|(0)$ which are manifestly
 bounded. Hence, the only really problematic term in (\ref{7.17eqa}) is the last one.

 This term, $\sum_{l=1,l\not=j}^n\alpha_l^2G(x-y_l)^2$, by the construction of the $dx$-majorant is integrated (in $x$) over $\R^2 \setminus\cup_{l=1,l\not=j}^nB_{R_l^j}(y_l)$. By Proposition \ref{2.1prop} (vi) applied
 to the case $d=2$, $\alpha=1/2$, one gets $|G(x)|<c_{1/2}(2)/|x|$.
 We can thus dominate this term by $-C_1\sum_{j,l=1j\not=l}^n\ln (|y_j-y_l|)1_{\{|y_j-y_l|<1\}}+n^2C_2$ for $C_1(\sigma,b),C_2(\sigma,b)>0$ sufficiently large.
 This establishes $L^p$, $p\geq 1$, integrability
 also for this last term and we can thus use the $dx$ integral of the majorant found in (i)
 as an $L^l$-majorant needed to prove dominated convergence in each term of (\ref{7.15eqa}).
 \kasten

 \

\small
\noindent {\bf Acknowledgments.} Discussions with Klaus R. Mecke on Section 4.2 and
Tobias Kuna on Section 6 were very helpful for the indicated parts of the article. We also thank Martin Grothaus, Armin Seyfried and Jiang-Lun Wu for interesting discussions and an anonymous referee
 for reading of the typoscript very carefully.
Financial support for the second named author via DFG projects "Stochastic analysis and systems with infinitely many degrees of freedom" and "Stochastic methods in QFT"
and for the third named author by the Grant--in--Aid Science Research No. 12640159 (Ministery of Education and Sciences, Japan) is gratefully acknowledged.

\end{document}